\documentclass[a4paper,11pt]{article}
\usepackage{jheppub}
\usepackage[T1]{fontenc}
\usepackage{amssymb}    %
\usepackage{pifont}
\usepackage{verbatim}   
\usepackage{color}      
\usepackage{subfigure}  
\usepackage{multirow}
\usepackage{dsfont}
\usepackage{comment}
\usepackage{pdflscape}
\usepackage{rotating}
\usepackage{feynmp} 
\usepackage{float}	

\renewcommand{\S}{{\mathcal{S}}}
\newcommand{\X}{{{\Xi}}}
\newcommand{\XX}{{{\Xi}_1}}
\newcommand{\V}{{\mathcal{L}_1}}
\newcommand{\J}{{\mathcal{J}}}
\newcommand{\cw}{c_W}
\newcommand{\sw}{s_W}

\newcommand{\dd}[1]{D_{#1}\Phi}
\newcommand{\du}[1]{D^{#1}\Phi}
\newcommand{\dg}[1]{(#1)^\dagger}
\newcommand{\wh}{\hat{W}}
\newcommand{\bh}{\hat{B}}
\newcommand{\tr}[1]{\mathrm{Tr}\left[{#1}\right]}

\allowdisplaybreaks[3]

\def \bal#1\eal  {\begin{align} #1 \end{align}}
\def\({\left(}
\def\){\right)}
\def\[{\left[}
\def\]{\right]}
\def\d{\mathrm{d}}
\newcommand{\bim} {\begin{itemize}[noitemsep]}

\newcommand{\eim} {\end{itemize}}
\newcommand{\be} {\begin{equation}}
\newcommand{\ee} {\end{equation}}
\newcommand{\bc}{\begin{center}}
\newcommand{\ec}{\end{center}}

\newcommand{\nn} {\nonumber\\}

\newcommand{\ud} {\mathrm{d}}

\newcommand{\epi}{\epsilon}

\begin{document}

\title{Positivity constraints on aQGC: carving out the physical parameter space}

\author[a]{Qi Bi,}
\author[a]{Cen Zhang,}
\author[b]{and Shuang-Yong Zhou}
\affiliation[a]{
Institute of High Energy Physics, and School of Physical Sciences,
University of Chinese Academy of Sciences, Beijing 100049, China
}
\affiliation[b]{Interdisciplinary Center for Theoretical Study, University of Science and Technology of China, Hefei, Anhui 230026, China}

\emailAdd{biqi@ihep.ac.cn}
\emailAdd{cenzhang@ihep.ac.cn}
\emailAdd{zhoushy@ustc.edu.cn}
\preprint{\small USTC-ICTS-19-01}
\abstract{
Searching for deviations in quartic gauge boson couplings (QGCs) is one of the
main goals of the electroweak program at the LHC. We consider positivity
bounds adapted to the Standard Model, and show that a set of positivity constraints
on 18 anomalous QGC couplings can be derived, by requiring that the vector
boson scattering amplitudes of specific channels and polarisations satisfy the
fundamental principles of quantum field theory.
We explicitly solve the positivity inequalities to remove their dependence on
the polarisations of the external particles, and obtain 19 linear inequalities, 3
quadratic inequalities, and 1 quartic inequality that only involve the QGC
parameters and the weak angle.  These inequalities constrain the possible
directions in which deviations from the standard QGC can occur, and can be used
to guide future experimental searches.
We study the morphology of the positivity bounds in the parameter
space, and find that the allowed parameter space is carved out by the
intersection of pyramids, prisms, and (approximately) cones.  Altogether, they
reduce the volume of the allowed parameter space to only 2.1\% of the total.
We also show the bounds for some benchmark cases, where one, two, or three
operators, respectively, are turned on at a time, so as to facilitate a quick
comparison with the experimental results.
}
\maketitle

\section{Introduction}

Vector boson scattering (VBS) and triboson production processes at the LHC are
among the processes most sensitive to the mechanism of electroweak symmetry
breaking.  In the Standard Model (SM), Feynman amplitudes involving four
longitudinally polarized weak bosons individually grow with energy, but at
large energy, cancellations among diagrams which involve quartic gauge boson
couplings (QGC), trilinear gauge boson couplings (TGC), and Higgs exchange
always occur, which lead to a total amplitude that does not grow as required by
unitarity.  However, effects beyond the Standard Model (BSM) could potentially
spoil these cancellations and leave a trace in these channels, which can be
observed at the LHC.  These channels are also the unique ones to probe the QGC
couplings, which are conventionally parametrised by 18 dim-8 effective
operators \cite{Eboli:2006wa,Degrande:2013rea,Eboli:2016kko}.  Both ATLAS and
CMS experiments have extensively searched for possible deviations in these
couplings. A summary of current limits can be found in \cite{qgclimits}.
The most recent Run-II analyses on VBS processes, in the $W^\pm
W^\pm jj$, $W^\pm Zjj$ and $ZZjj$ final states, have improved the precision
reach on these channel \cite{ATLAS:2018ogo,Sirunyan:2017ret,ATLAS:2018ucv,
CMS:2018ysc,Sirunyan:2017fvv}. In particular, the CMS analyses
\cite{Sirunyan:2017ret,CMS:2018ysc,Sirunyan:2017fvv} have pushed the limits on
potential BSM effects up to around TeV scale.  Further improvement with future
high luminosity upgrade can be expected. 

Even though the QGC couplings, or the coefficients of the corresponding dim-8
operators, are free parameters from the effective field theory (EFT) point of
view, to admit a UV completion they cannot simply take any values. Recently,
some of the authors have shown that dim-8 QGC operator coefficients must
satisfy a set of positivity bounds \cite{Zhang:2018shp}.  These bounds require
that certain linear combinations of these operator coefficients must be
positive. They are derived from the fundamental principles of quantum field
theory \cite{Adams:2006sv}, namely that any $VV\to VV$ scattering amplitude has
to satisfy analiticity, unitarity, and Lorentz invariance. For example, from
$WW\to WW$ scattering we can derive a condition of the form
\begin{flalign}
	\sum_iF_i x_{i,WW}(\vec a,\vec b)>0
\end{flalign}
where $F_i=F_{S,0},F_{S,1},\cdots$ are all dim-8 QGC
operator coefficients, to be defined in the next section; $\vec a,\vec b$ are
the polarisation vectors of the two $W$ bosons being scattered, and $x_{i,WW}$
only depend on the weak angle $\theta_W$, and $\vec
a,\vec b$.  For a given set of values for QGC coefficients $\{F_i\}$, if this
condition is violated for any possible value of $\vec a$ and $\vec b$, then the
$WW\to WW$ amplitude cannot be UV completed in a theory that satisfies
analiticity, unitarity and Lorentz invariance, and therefore it does not make
sense to study these coefficients. We will review the derivation of the
positivity bounds adapted to the context of SMEFT in Section~\ref{sec:setup}. Here
it suffices to emphasize that the positivity bounds are very reliable bounds
as their existence can be proven by simply assuming that some of the most basic
properties of quantum field theory are respected in the UV.

\begin{figure}
	\begin{center}
		\hfill
		\includegraphics[width=.435\linewidth]{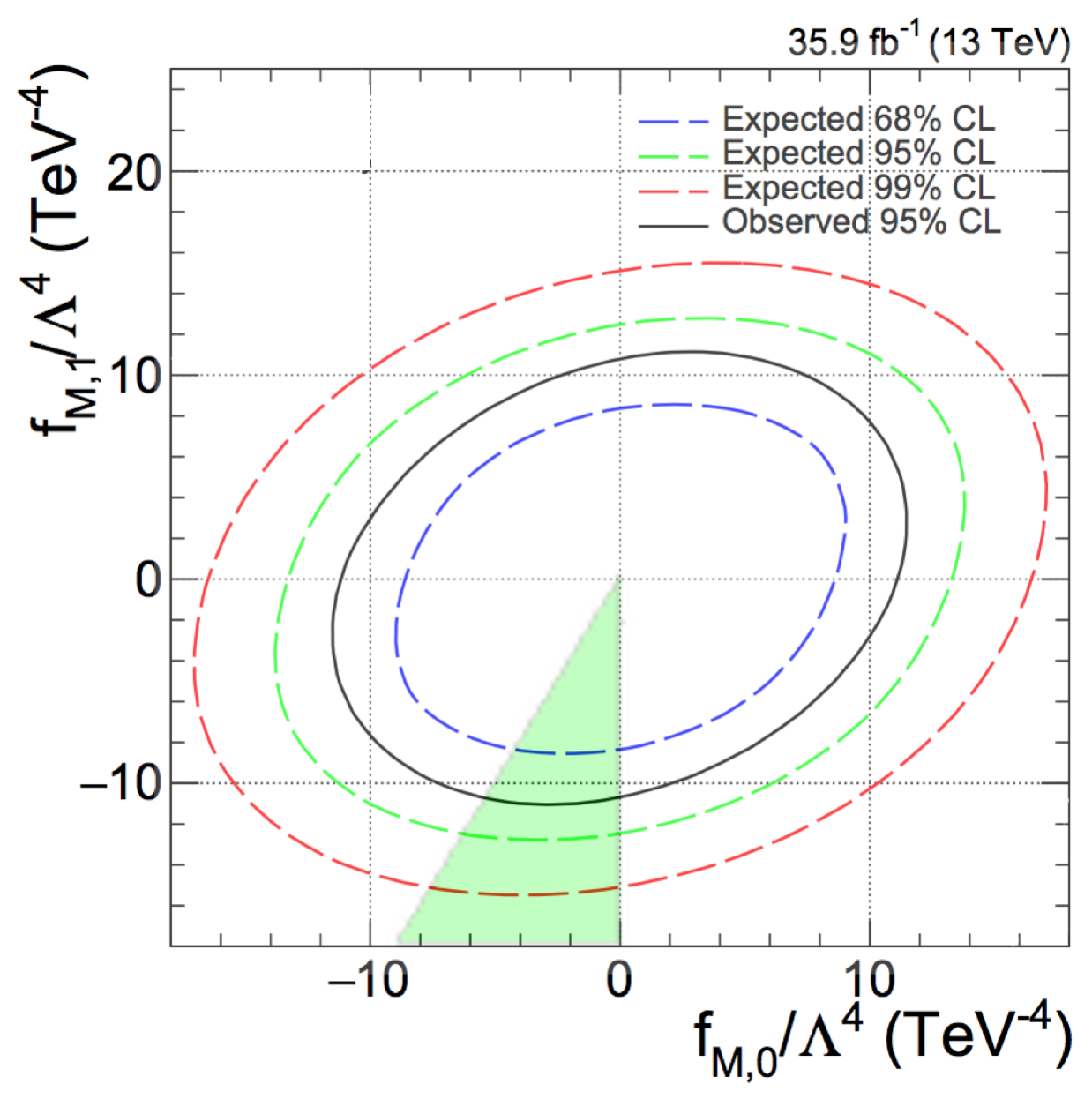}
		\hfill
		\includegraphics[width=.45\linewidth]{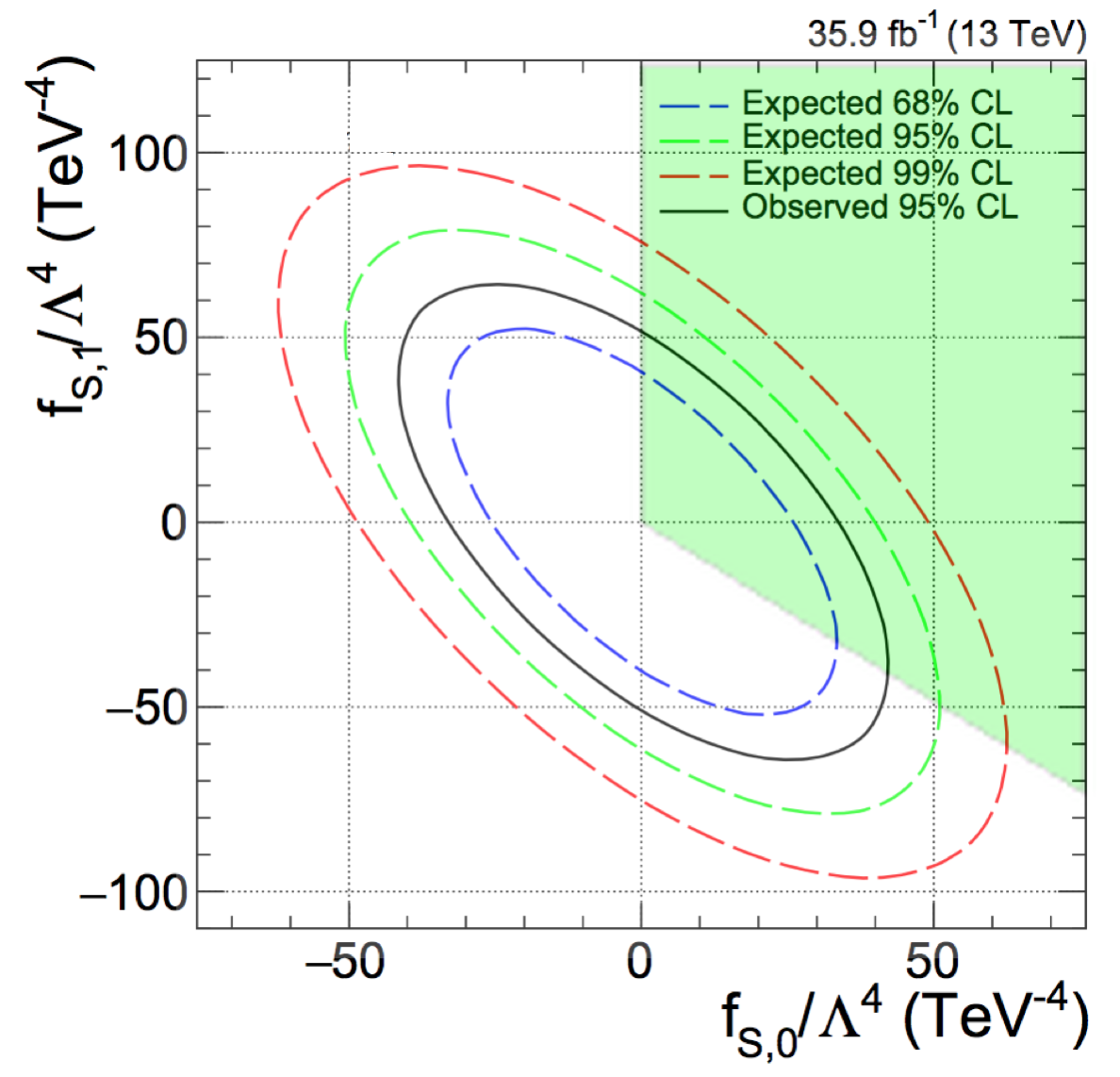}
		\hfill
	\end{center}
	\caption{Limits on pairs of QGC operator coefficients from CMS
		$WZjj$ analysis \cite{CMS:2018ysc}.  Green shaded area represents
	positivity bounds.}
	\label{fig:cms}
\end{figure}

These positivity bounds can have significant implication on experimental
measurement.  In Figure~\ref{fig:cms} we show the experimental constraints on
two pairs of
operator coefficients, from the recent CMS analysis on $WZjj$ final state,
together with the corresponding positivity bounds, shown as green shaded areas.
The parameter space outside of the green area violates positivity bounds, and
does not correspond to any real UV model.  We can see that while the measurement
constrains the magnitudes of deviations from the SM, positivity bounds
constrain the possible directions in which the deviations are possible at all.
In the two-parameter case they carve out triangle areas in the parameter space,
which reduce the physical parameter space that needs to be studied.
If all QGC parameters are turned on, the total parameter space will be reduced
by two orders of magnitude.

These results are important in several ways.  
First of all, positivity constraints lead to additional physical insight of the
multidimensional parameter space spanned by 18 QGC coefficients, and could
provide guidance as to where one should look for deviations in QGC couplings.
For example, in Figure~\ref{fig:cms} left the possible parameter space is
reduced to a very narrow range, and the experimental analysis could have been
focused only within this range, as BSM deviation cannot be observed elsewhere.
Second, theoretical studies on QGC often follow a bottom-up approach, and in
that case they are advised to keep the positivity constraints satisfied and
avoid choosing unphysical benchmark parameters.  Finally, as a more practical
aspect, experimental limits like those in Figure~\ref{fig:cms} are often
obtained by scanning the parameter space, and comparing the theory prediction at
each point with the observed events.  Therefore in a global SMEFT analysis
where all relevant operators are taken into account, positivity constraints
greatly reduce the area of parameter space that needs to be scanned, and thus
improving the efficiency of experimental analysis. 

In our previous work, these bounds are derived for each $VV\to VV$
amplitude, with arbitrary polarisation vectors $\vec a$ and $\vec b$. The
derivation will be given in Section~\ref{sec:setup} but with more details.  
The resulting positivity bounds on VBS consist of 7 linear inequalities of the
form $\sum_iF_i x_{i,j}(\vec a,\vec b)>0$, where $j=W^\pm W^\pm,W^\pm
W^\mp,W^\pm Z,ZZ, W^\pm\gamma,Z\gamma,\gamma\gamma$ denotes the scattering
channel.  While this set of conditions in principle encodes all information of
the bounds, they are difficult to use because the polarisation vectors $\vec a$
and $\vec b$ show up as free parameters.  Each of them is a complex linear
combination of three helicity states, and so in total they correspond to 12
real degrees of freedom.  The positivity bounds must be satisfied not only for
the basis helicities, but also for their arbitrary linear combinations.
Therefore, to determine if a set of QGC coefficients $\{F_i\}$ are excluded by
positivity conditions, one has to scan a 12-dimensional space of $\vec a,\vec
b$ and check if any point that violates the conditions exists.  This can be
done in a numerical way, but it takes time to efficiently scan a
multidimensional space and is sometimes inaccurate.  In
Ref.~\cite{Zhang:2018shp}, for simplicity we have only considered real
polarisation vectors. This restriction limited the constraining power of
positivity bounds.

The main purpose of this work is to find the analytical form that describes
exactly the allowed parameter space for all $\vec a$ and $\vec b$, by removing
the polarisation dependence in positivity bounds.  This is done by going
through all possible complex values for $\vec a,\vec b$, and combining all the
corresponding bounds.  We will then obtain a set of analytical
inequalities independent of $\vec a,\vec b$, and can be directly used in any
experimental or theoretical studies.  They consist of 19 linear inequalities, 3
quadratic inequalities and 1 quartic inequality.  These bounds carve out higher
dimensional pyramids, prisms, and cones in the parameter space spanned by the
18 QGC parameters.  Their intersection gives the physically allowed parameter
space. We will then discuss the shapes and volumes of these
constraints.  To better understand the physics origin of these constraints, we
will also consider a simplified model, in which various QGC couplings can be
generated by integrating out several heavy resonant states.  We will show
explicitly that in these models the resulting QGC couplings will naturally
satisfy positivity constraints.

Finally, to help future studies on QGC couplings, we provide a number of
benchmark scenarios, where we turn on only 1, 2 or 3 operators at a time.
Currently experimental analyses only consider individual operators or
at most operator pairs, and
therefore these benchmark scenarios could provide useful information for
similar analyses.  It should however be kept in mind that the
most useful way to identify the possible deviations from the SM is to perform
global analyses, by combining different channels and turning on all QGC
parameters.  In that case our results are still directly applicable, and in
fact the impact is stronger --- we will show that the volume of the full
parameter space can be reduced to about $2.1\%$.

The paper is organized as follows:
in Section~\ref{sec:setup} we introduce the effective Lagrangian and the QGC
operators, and adapt the positivity approach to the context of SMEFT.
The resulting positivity constraints are obtained and depend on the
polarisation vectors.
In Section~\ref{sec:solve} we ``solve'' these conditions by removing the 
polarisation dependence, and obtain the full set of analytical inequalities
that describes the allowed parameter space.  
We first illustrate the approach by considering two toy cases in 
Section~\ref{subsec:toy}, and then we solve the problem for each channel
in Section~\ref{subsec:general}.
The final results are collected, further simplified, and matched to specific
polarisation states in Section~\ref{subsec:polarisation}.
These are the main results of this work.
The readers who are only interested in the actual form of the positivity
constraints can directly go to Eqs.~(\ref{eq:result1starts})-(\ref{eq:MT}) and
Eqs.~(\ref{eq:result2starts})-(\ref{eq:result2ends}).

The rest of the paper is devoted to a better understanding of the constraints.
In Section~\ref{sec:morphology} we examine the shapes of the physical parameter
space constrained by positivity bounds.
In Section~\ref{sec:volume} we discuss the volume of the space allowed by
various positivity conditions, which reflect their respective constraining
strengths and indicate their relative importance.
A collection of benchmark scenarios, where only 1, 2 or 3 operators are turned on
at a time, is provided in Section~\ref{sec:benchmark}, with exact description
of the bounds in Appendices~\ref{subsec:A11} and \ref{subsec:A12}.
Finally, we summarise the most important results of this work in
Section~\ref{sec:summary}.
A simplified model with explicit BSM resonances is given in
Appendix~\ref{sec:models}, to illustrate how positivity bounds are satisfied
once these resonances are integrated out.

\section{Positivity bounds}
\label{sec:setup}

\subsection{Effective operators}
Let us first define the effective operators relevant in this study.  The QGC
couplings are parameterised in a bottom-up effective field theory (EFT)
approach---the SMEFT \cite{Weinberg:1978kz, Buchmuller:1985jz,Leung:1984ni}
\begin{flalign}
	\mathcal{L}_{\rm EFT}=\mathcal{L}_\mathrm{SM}
	+\sum_i\frac{f_i^{(6)}O_i^{(6)}}{\Lambda^2}
	+\sum_i\frac{f_i^{(8)}O_i^{(8)}}{\Lambda^4}+\cdots
\end{flalign}
Unlike the usual SMEFT truncation at the level of dim-6, a useful
parametrisation of anomalous QGC requires dim-8 operators.  This is mostly
because the QGC couplings at dim-6 are fully correlated with 3 TGC couplings,
which are better constrained by other channels such as diboson production.  The
dim-8 effective operators allow us to study potential QGC deviations
independent of existing constraints on TGC couplings.  They also parametrise
the four vector vertices with arbitrary helicities of the four gauge bosons. 
There are additional motivations for studying these operators.  Dim-6 TGC
couplings $\kappa_\gamma, \kappa_Z, \lambda_\gamma, \lambda_Z$ are only
generated by loop-induced operators \cite{Elias-Miro:2013mua}, so it is likely
that multi-boson interactions first show up in dim-8 QGC operators which are
generated at the tree level.  In some strongly coupled scenarios, dim-8 QGC
operators can be enhanced by more powers of BSM couplings and give larger
contributions than those of the dim-6 operators, see Ref.~\cite{Liu:2016idz}.
Also, in VBS scattering with at least one transversely polarised vector, dim-6
operators cannot interfere with the SM amplitude without any mass suppression
\cite{Azatov:2016sqh}, unlike dim-8 operators. Dim-6 contributions are then
further suppressed by factors of $\mathcal{O}(M_{W,Z}^2/E^2)$. Since
the dominant sensitivity in these measurements come from the energy growing
terms, dim-8 contributions are as important as dim-6 squared terms. Finally,
since some of the individual limits on dim-8 QGC operators
have already reached the TeV scale, demonstrating the LHC sensitivity to
these higher-dimensional operators, including them allows us to make the most
use of the data. We refer to Ref.~\cite{Green:2016trm} for a more detailed
review of multi-boson interaction at the LHC.

The dim-8 QGC operators are parametrised by three types of effective operators
\cite{Eboli:2006wa,Degrande:2013rea,Eboli:2016kko}.  The $S$-type operators
involve only covariant derivatives of the Higgs, which correspond to the
longitudinal components of the electroweak bosons.  The $M$-type operators
include a mix of field strengths and covariant derivatives of the Higgs, which
sources both transversal and longitudinal modes.  The $T$-type operators only
include field strengths, and thus only give rise to transversal components.  
Here, we use the convention of \cite{Degrande:2013rea}, which has become standard
in this community.  Defining
\begin{flalign}
	\wh^{\mu\nu}\equiv ig\frac{\sigma^I}{2}W^{I,\mu\nu}\,,\qquad
	\bh^{\mu\nu}\equiv ig'\frac{1}{2}B^{\mu\nu}\,.
\end{flalign}
the 18 dimension-8 QGC operators are given by
\begin{flalign}
	\begin{aligned}
	\begin{array}{l}
O_{S,0}=[\dg{\dd{\mu}}\dd{\nu}]\times[\dg{\du{\mu}}\du{\nu}]
\\
O_{S,1}=[\dg{\dd{\mu}}\du{\mu}]\times[\dg{\dd{\nu}}\du{\nu}]
\\
O_{S,2}=[\dg{\dd{\mu}}\dd{\nu}]\times[\dg{\du{\nu}}\du{\mu}]
\\
O_{M,0}=\tr{\wh_{\mu\nu}\wh^{\mu\nu}}\times\left[\dg{\dd{\beta}}\du{\beta}\right]
\\
O_{M,1}=\tr{\wh_{\mu\nu}\wh^{\nu\beta}}\times\left[\dg{\dd{\beta}}\du{\mu}\right]
\\
O_{M,2}=\left[\bh_{\mu\nu}\bh^{\mu\nu}\right]\times\left[\dg{\dd{\beta}}\du{\beta}\right]
\\
O_{M,3}=\left[\bh_{\mu\nu}\bh^{\nu\beta}\right]\times\left[\dg{\dd{\beta}}\du{\mu}\right]
\\
O_{M,4}=\left[\dg{\dd{\mu}}\wh_{\beta\nu}\du{\mu}\right]\times\bh^{\beta\nu}
\\
O_{M,5}=\frac{1}{2}\left[\dg{\dd{\mu}}\wh_{\beta\nu}\du{\nu}\right]\times\bh^{\beta\mu}+h.c.
\\
O_{M,7}=\left[\dg{\dd{\mu}}\wh_{\beta\nu}\wh^{\beta\mu}\du{\nu}\right]
	\end{array}
\end{aligned}\quad
\begin{aligned}
	\begin{array}{l}
O_{T,0}=\tr{\wh_{\mu\nu}\wh^{\mu\nu}}\times\tr{\wh_{\alpha\beta}\wh^{\alpha\beta}}
\\
O_{T,1}=\tr{\wh_{\alpha\nu}\wh^{\mu\beta}}\times\tr{\wh_{\mu\beta}\wh^{\alpha\nu}}
\\
O_{T,2}=\tr{\wh_{\alpha\mu}\wh^{\mu\beta}}\times\tr{\wh_{\beta\nu}\wh^{\nu\alpha}}
\\
O_{T,5}=\tr{\wh_{\mu\nu}\wh^{\mu\nu}}\times\bh_{\alpha\beta}\bh^{\alpha\beta}
\\
O_{T,6}=\tr{\wh_{\alpha\nu}\wh^{\mu\beta}}\times\bh_{\mu\beta}\bh^{\alpha\nu}
\\
O_{T,7}=\tr{\wh_{\alpha\mu}\wh^{\mu\beta}}\times\bh_{\beta\nu}\bh^{\nu\alpha}
\\
O_{T,8}=\bh_{\mu\nu}\bh^{\mu\nu}\times\bh_{\alpha\beta}\bh^{\alpha\beta}
\\
O_{T,9}=\bh_{\alpha\mu}\bh^{\mu\beta}\times\bh_{\beta\nu}\bh^{\nu\alpha}   ,
	\end{array}
\end{aligned}
\end{flalign}
and the QGC interactions are described by
\begin{flalign}
	\mathcal{L}_\mathrm{QGC}=\sum_i\frac{f_iO_i}{\Lambda^4}
\end{flalign}
where $f_i$ are the corresponding coefficients of the 18 operators.
Note that $O_{M,6}$ is redundant \cite{Sekulla} and is not included.
We have also added a Hermitian conjugate of $O_{M,5}$ in its definition
\cite{Perez:2018kav}.  For later convenience, we redefine the coefficients as
follows
\begin{flalign}
F_{S,i}\equiv f_{S,i}, \quad
F_{M,i}\equiv e^2f_{M,i},\quad
F_{T,i}\equiv e^4f_{T,i}.
\end{flalign}
By doing this, the resulting positivity constraints are free of additional
couplings constants.  That is, they will only involve the Weinberg angle
$\theta_W$ and the polarisation of the vector bosons.  As we will show
in Section~\ref{subsec:polarisation}, this redefinition will not change the
final form of the allowed parameter space, i.e.~we can directly replace
$F_{X,i}\to f_{X,i}$ in the final results.

In the following subsections, we briefly review the positivity bounds,
and adapt it to the context of SMEFT.

\subsection{Generic UV completion}

Positivity bounds are a powerful tool to constrain the Wilson coefficients 
of EFTs. An EFT is a valid description of the underlying
theory below a cutoff scale, beyond which the EFT needs to be UV completed,
that is, replaced with a theory valid up to arbitrary high energies. As
mentioned in the introduction, the existence of positivity bounds are
guaranteed by some fundamental properties of S-matrix such as Lorentz
invariance, unitarity, crossing symmetry, analyticity and polynomial
boundedness, making them very reliable theoretical bounds. Analyticity is the
statement that scattering amplitudes are analytical functions of the Mandelstam
variables ($s$ and $t$) in the complex domain, apart from certain poles and
branch cuts. It is tied to the concepts of causality and locality and can be
proven to any order in perturbation theory. By analyticity, one can connect the
UV behavior of the theory with the IR via deformation of the integration
contour in the complex $s$ plane. Polynomial boundedness states that when
complex momenta increase the scattering amplitudes are bounded by a certain
polynomial (or for some cases a linear exponential), which is also linked to
locality, arising from the fact that the amplitudes should be well defined in
real space after Fourier transforms. With the help of unitarity, the more
restrictive Froissart bound can be proven, which states that when $|s|\to
\infty$ the amplitude $A^{q_1q_2}_{ab}(s)$ is bounded by $C_F s \ln^2 s$, $C_F$
being a constant. These are the essential ingredients needed to derive the
dispersion relations that are used to derive the positivity bounds. In this
paper, we have only made use of the forward limit positivity bound, which we
will review in the following, with some adaptation to the context of
SMEFT.\footnote{An infinite number of generalized positivity bounds have also
been derived recently, which are valid away from the forward limit and
can involve $t$ derivatives \cite{deRham:2017avq, deRham:2017zjm}. They
are useful for high order contributions to the amplitude
\cite{deRham:2017avq, deRham:2017imi, deRham:2018qqo} or for some
``soft'' amplitudes such as scattering amplitudes in massive Galileon.
However, for the leading tree level aQGC amplitudes at ${\cal
O}(\Lambda^{-4})$, these generalized positivity bounds do not give rise
to anything nontrivial.}

\begin{figure}[ht]
\centering
\includegraphics[width=.4\linewidth]{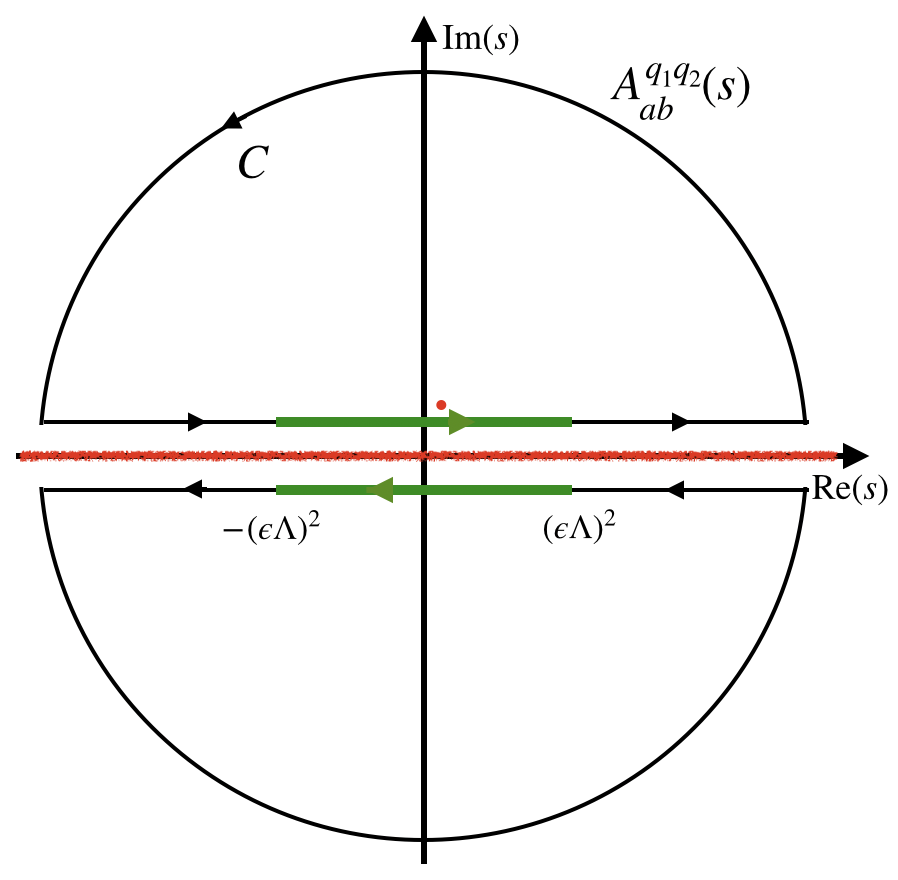}
\includegraphics[width=.4\linewidth]{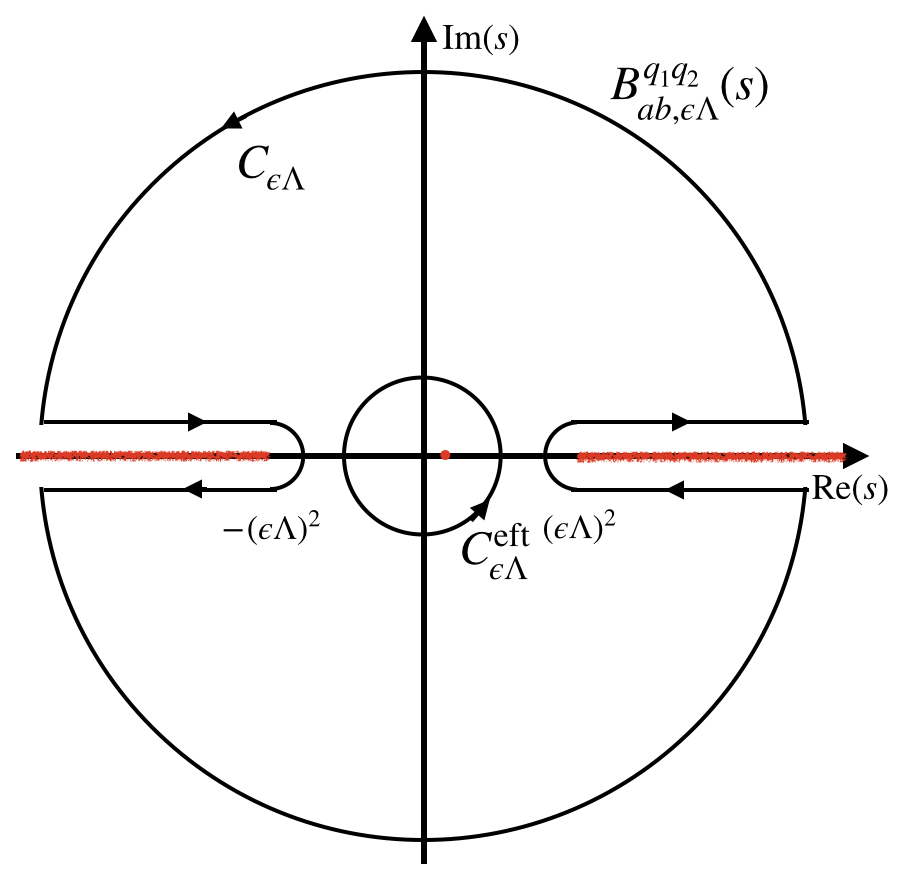}
\caption{Dispersion relation contours in the forward limit.}
\label{fig:DRplot}
\end{figure}

Let us consider elastic scattering of two particles (ie, 2 to 2 scattering)
labeled by $(m_1,S_1, a, q_1)$ and $(m_2,S_2,b, q_2)$, which denote the mass,
spin, polarisation and remaining quantum numbers of the respective particle.
For example, for $WW$ scattering $q$ is the electric charge of $W$ bosons. In
the complex $s$ plane, from perturbation theory, we can infer that there are
various simple poles arising from the tree level exchanges of particles and
also branch cuts from internal multi-particle production at loop levels. The
analyticity of the scattering amplitude requires that for fixed $t$ (we will
focus on $t=0$ here: $A^{q_1q_2}_{ab}(s)=A^{q_1q_2}_{ab}(s,t=0)$) the amplitude
is analytic in the complex $s$ plane apart from these singularities that
already appear in perturbation theory. 

In a field theory such as the SMEFT which contains massive as well as massless
particles, the multi-particle productions of the massless particles give rise
to branch cuts that cover the whole real axis in the $s$ plane from $-\infty$
to $+\infty$. However, below the scale  $\epi \Lambda$ with $\epi\lesssim1$,
the amplitude can be computed to a desired precision in perturbation theory, so
we can subtract out the simple poles and the branch cut contributions within
$-(\epi\Lambda)^2<s<+(\epi\Lambda)^2$. To this end, we define the following
modified amplitude \cite{deRham:2017imi, Bellazzini:2016xrt, deRham:2017xox}
\bal
\label{eq:BepiLam}
B^{q_1q_2}_{ab,\epi\Lambda}(s) &\equiv A^{q_1q_2}_{ab}(s)  -  \frac{1}{2\pi i}  \int^{+(\epi\Lambda)^2}_{-(\epi\Lambda)^2}  \ud s' \frac{ {\rm Disc} A^{q_1q_2}_{ab}(s')}{s'-s} 
\\
&= \frac{1}{2\pi i} \oint_{\rm C}   \ud s' \frac{A^{q_1q_2}_{ab}(s')}{s'-s} -  \frac{1}{2\pi i}  \int^{+(\epi\Lambda)^2}_{-(\epi\Lambda)^2}  \ud s' \frac{ {\rm Disc} A^{q_1q_2}_{ab}(s')}{s'-s} 
\\
& = \frac{1}{2\pi i}  \int_{\rm C'_{\epi\Lambda}} \ud s' \frac{A^{q_1q_2}_{ab}(s')}{s'-s}
= \frac{1}{2\pi i}  \oint_{\rm C_{\epi\Lambda}} \ud s' \frac{B^{q_1q_2}_{ab,\epi\Lambda}(s')}{s'-s}   ,
\eal
where the discontinuity is defined as ${\rm Disc}A^{q_1q_2}_{ab}(s') =
A^{q_1q_2}_{ab}(s'+i\varepsilon)- A^{q_1q_2}_{ab}(s'-i\varepsilon)$
($\varepsilon$ being an infinitesimal positive number) and the last equality is
the use of Cauchy's integral formula for $B^{q_1q_2}_{ab,\epi\Lambda}(s)$. In
the above, $s$ is defined away from the real axis for $A^{q_1q_2}_{ab}(s)$, but
can be analytically continued to the real axis between $-\epi\Lambda$ and
$+\epi\Lambda$ for $B^{q_1q_2}_{ab,\epi\Lambda}(s)$. As depicted in Fig
\ref{fig:DRplot}, the left diagram is the singularity structure for
$A^{q_1q_2}_{ab}(s)$ and the contour $C$ includes the upper and lower
semi-circles and the full discontinuity along the real axis; the right diagram
is the singularity structure for the modified amplitude
$B^{q_1q_2}_{ab,\epi\Lambda}(s)$ and the contour $C'_{\epi\Lambda}$ has the
discontinuity from $-(\epi\Lambda)^2<s<+(\epi\Lambda)^2$ (the green solid
contour) subtracted ($C_{\epi\Lambda}$ being the closed contour for
$B^{q_1q_2}_{ab,\epi\Lambda}$).
The amplitude $B^{q_1q_2}_{ab,\epi\Lambda}$ defined in this way
has exactly the same discontinuities as $A_{ab}^{q_1q_2}$ on the real
axis in $(-\infty,(\epi\Lambda)^2] \cup  [(\epi\Lambda)^2,+\infty)$,
but the branch cuts and poles within $[-(\epi\Lambda)^2,(\epi\Lambda)^2]$
are subtracted.  It also has the same asymptotic behavior as $A_{ab}^{q_1q_2}$
as $s\to \infty$ and satisfies the Froissart bound.
With the modified amplitude, we can define
\bal
\label{eq:fBdef}
f^{q_1q_2}_{ab,\epi\Lambda}(s) &\equiv \frac12 \frac{\ud^2 B^{q_1q_2}_{ab,\epi\Lambda}(s)}{\ud s^2} 
\\
& =  \frac{1}{2\pi i}  \(  \int_{-\infty}^{-(\epi\Lambda)^2} + \int^{\infty}_{+(\epi\Lambda)^2}  \)  \ud s' \frac{ {\rm Disc} A^{q_1q_2}_{ab}(s')}{(s'-s)^3}   ,
\label{eq:fdef2}
\eal
where we have used the Froissart bound and thus the infinite semi-circles parts of the contour integral vanish. Changing the integration variable for the left hand cut $s'\to M^2 - s'$ ($M^2\equiv 2m_1^2+2m_2^2$) leads to
\be
f^{q_1q_2}_{ab,\epi\Lambda}(s) = \frac{1}{2\pi i}  \( \int^{\infty}_{(\epi\Lambda)^2+M^2} \d s'  \frac{ {\rm Disc} A^{q_1q_2}_{ab}(M^2-s')}{(M^2 - s'-s)^3} + \int^{\infty}_{(\epi\Lambda)^2} \d s' \frac{ {\rm Disc} A^{q_1q_2}_{ab}(s')}{(s'-s)^3}  \) .
\ee
Then we can make use of the crossing relation in the forward limit
$A^{q_1q_2}_{ab}(M^2-s')=A^{q_1\bar{q_2}}_{\bar{a}\bar{b}}(s')$, where
$\bar{q_2}$ denotes the quantum numbers of the anti-particle 2 and $\bar{a}$
and $\bar{b}$ denote possible change for the polarisations.\footnote{For
example, if $q_2$ is the electric charge, then $\bar{q_2}=-q_2$. If $a$
and $b$ are helicities, then $\bar{a}=a$ and $\bar{b}=-b$; if $a$ and
$b$ are transversities, then $\bar{a}=-a$ and $\bar{b}=-b$. For
particles with spin, the crossing relations in the non-forward limit
are highly nontrivial, which makes the proof of the positivity bounds
away from the forward limit much more complicated
\cite{deRham:2017zjm}.}  For the discontinuity in $|s'|>(\epi \Lambda)^2$,
there is an extra minus sign: 
\bal
{\rm Disc}A^{q_1q_2}_{ab}(M^2-s')&= A^{q_1q_2}_{ab}(M^2-s'+i\varepsilon)-A^{q_1q_2}_{ab}(M^2-s'-i\varepsilon) 
\\
&=  A^{q_1\bar{q_2}}_{\bar{a}\bar{b}}(s'-i\varepsilon)-A^{q_1\bar{q_2}}_{\bar{a}\bar{b}}(s'+i\varepsilon) 
 =-{\rm Disc}A^{q_1\bar{q_2}}_{\bar{a}\bar{b}}(s')  .
\eal
Utilizing this, we get
\be
\label{eq:fdisc2}
f^{q_1q_2}_{ab,\epi\Lambda}(s) = \frac{1}{2\pi i} \( \int^{\infty}_{(\epi\Lambda)^2+M^2} \d s'   \frac{ {\rm Disc}A^{q_1\bar{q_2}}_{\bar{a}\bar{b}}(s')}{( s'-M^2+s)^3} +  \int^{\infty}_{(\epi\Lambda)^2} \d s'  \frac{ {\rm Disc} A^{q_1q_2}_{ab}(s')}{(s'-s)^3}  \)   .
\ee
We then convert the discontinuities of the amplitudes in the integrand to their imaginary parts via ${\rm Disc}=2i\,{\rm Im}$ and make use of the optical theorem which tells us that 
\be
{\rm Im}A^{q_1q_2}_{ab}(s') = \sqrt{(s'-M_+^2)(s'-M_-^2)}\sigma^{q_1q_2}_{ab}(s')>0  ,  ~~ s'>(\epi\Lambda)^2  ,
\ee
where $M_\pm = m_1\pm m_2$ and $\sigma^{q_1q_2}_{ab}(s')$ is the total cross section for the $s$ channel. Similarly, we have  
\bal
 {\rm Im}A^{q_1\bar{q_2}}_{\bar{a}\bar{b}}(s')  = \sqrt{(s'-M_+^2)(s'-M_-^2)}\sigma^{q_1\bar{q_2}}_{\bar{a}\bar{b}}(s')>0    , ~~     s'>(\epi\Lambda)^2   .
\eal
where $\sigma^{q_1\bar{q_2}}_{\bar{a}\bar{b}}(s')$ is the total cross section for the $u$ channel. Putting all these together, we have
\be
f^{q_1q_2}_{ab,\epi\Lambda}(s) >0 ,~~~~~   -(\epi\Lambda)^2 < s < (\epi\Lambda)^2   .
\ee
The power of analyticity is that in the complex plane the contour $C_{\epi\Lambda}$
can be deformed to $C^{\rm eft}_{\epi\Lambda}$ which can be used to compute
$f^{q_1q_2}_{ab,\epi\Lambda}(s)$ within the EFT framework to desired accuracy
below scale $\epi\Lambda$. 

In this work we are going to assume that the contributions from the higher
dimensional operators are well approximated by the tree level, which is a
reasonable assumption given that perturbativity in EFT is always required in
any realistic analysis. 
Under this assumption $f^{q_1q_2}_{ab,\epi\Lambda}(s)$ can be computed 
as a linear combination of dim-8 coefficients plus a quadratic form
of dim-6 coefficients.  We will show that the latter can be removed.
The SM makes no contribution at the tree level,
because in the r.h.s.~of Eq.~(\ref{eq:fdef2}) there are no poles nor branch
points above $\epi\Lambda$. Its leading contribution arises at the one loop
level, and could lead to small positive constant terms that weaken the
positivity bound. However, in the following we will show that they are
suppressed by inverse powers of $\epi\Lambda$, and can be safely ignored.  We
will also show that they can be completely removed for weakly coupled UV
completion.  As a result, $f^{q_1q_2}_{ab,\epi\Lambda}(s) >0 $ directly leads
to positivity constraints on Wilson coefficients.

\subsection{SM loop contribution}

To see that the SM contribution to $f^{q_1q_2}_{ab,\epi\Lambda}(s)$ is
negligible, we note that the SM and EFT contributions to
$f^{q_1q_2}_{ab,\epi\Lambda}$ are distributed differently on the real axis.
While the EFT contributions stay completely above $\Lambda$, the SM
contributions are dominated by the discontinuity below scale $\epi\Lambda$.
Therefore, in the ``improved positivity bounds'' approach
\cite{deRham:2017imi, Bellazzini:2016xrt, deRham:2017xox} that we are using, 
we can choose an $\epi\Lambda$ to subtract the SM loop contributions 
as much as possible, without losing positivity. More specifically, the second
term in the r.h.s.~of Eq.~(\ref{eq:BepiLam}) serves to subtract the SM
contribution without modifying the tree level EFT contribution. The remaining
SM contamination is suppressed at least by $(\epi\Lambda)^{-2}$.  This is
because the integrand in the r.h.s.~of Eq.~(\ref{eq:fdisc2}) goes like $s^{-3}$
for fermionic loops and $s^{-2}$ for bosonic loops. 

As an example, consider the $\gamma\gamma$ scattering channel.  The SM loop
contribution to $f^{00}_{ab,\epi\Lambda}$ can be
computed either by using Eq.~(\ref{eq:fBdef}) and subtracting the
lower energy discontinuity, or equivalently by
using the Eq.~(\ref{eq:fdisc2}).  The latter is more convenient because
in the physical regime we can use the optical theorem, so that
\begin{flalign}
f^{00}_{ab,\epi\Lambda}(0) = &\frac{1}{2\pi i} \( \int^{\infty}_{(\epi\Lambda)^2} \d s'   \frac{ {\rm Disc}A^{00}_{\bar{a}\bar{b}}(s')}{ s'^3} +  \int^{\infty}_{(\epi\Lambda)^2} \d s'  \frac{ {\rm Disc} A^{00}_{ab}(s')}{s'^3}  \)   
\nonumber\\
=&
\frac{2}{\pi}\int^{\infty}_{(\epi\Lambda)^2} \d
s'\frac{1}{s'^3}\sqrt{(s'-M_+^2)(s'-M_-^2)}
\sum_{X}\sigma^{00}_{ab}(\gamma\gamma\to X)(s')   ,
\end{flalign}
where for simplicity we have restricted to crossing symmetric amplitudes, so
$A^{00}_{\bar{a}\bar{b}}=A^{00}_{ab}$. (We will see in Section~\ref{sec:solve}
that, for the $\gamma\gamma$ scattering, linear polarisation gives rise to the
most crucial positivity bounds.)
Now $f^{00}_{ab,\epi\Lambda}(0)$ can be computed by evaluating the total cross
sections of $\gamma\gamma\to X$, but with only the SM contributions.  For one loop
contribution we need to consider $\gamma\gamma\to f\bar f$ and $\gamma\gamma\to
W^+W^-$.  Since the integration is for $s>(\epi\Lambda)^2$, we take the leading
contribution at large $s$.  The dominant contribution comes from
$\gamma\gamma\to W^+W^-$:
\begin{flalign}
	\sigma^{00}_{ab}(\gamma\gamma\to W^+W^-)(s)
	=\frac{8\pi\alpha^2}{m_W^2}
	\left[
		\left(a_1^2+a_2^2\right) \left(b_1^2+b_2^2\right)
	\right]+\mathcal{O}(s^{-1})
\end{flalign}
where $a_{1,2}$ and $b_{1,2}$ are the transversal polarisation components
of the two photons. This gives rise to an
$\mathcal{O}\left[(\epi\Lambda)^{-2}\right]$ contribution to
$f^{00}_{ab,\epi\Lambda}(0)$:
\begin{flalign}
	f^{00,WW}_{ab,\epi\Lambda}(0)
	=\frac{16\alpha^2}{(\epi\Lambda)^2m_W^2}
	\left[
		\left(a_1^2+a_2^2\right) \left(b_1^2+b_2^2\right)
	\right]+\mathcal{O}\left[(\epi\Lambda)^{-4}\right]
\end{flalign}
The $1/m_W^2$ factor is from the $t$- and $u$-channel $W$-boson propagators.
The cross section of $\gamma\gamma\to f\bar f$ does not have this behavior
and is suppressed by one extra power of $s$:
\begin{flalign}
	\sigma^{00}_{ab}(\gamma\gamma\to f\bar f)(s)
	=N_cQ^4\frac{4\pi\alpha^2}{s}
	\left[
		\left(a_1^2+a_2^2\right) \left(b_1^2+b_2^2\right)
		\log\frac{s}{m_f^2}
		-2(a_1b_1-a_2b_2)^2
	\right]+\mathcal{O}(s^{-2})
\end{flalign}
which gives rise to an $\mathcal{O}\left[(\epi\Lambda)^{-4}\right]$
contribution to $f^{00}_{ab,\epi\Lambda}(0)$:
\begin{flalign}
	f^{00,ff}_{ab,\epi\Lambda}(0)
	=&N_cQ^4\frac{2\alpha^2}{(\epi\Lambda)^4}
	\left[
		2\left(a_1^2+a_2^2\right) \left(b_1^2+b_2^2\right)
		\log\frac{(\epi\Lambda)^2}{m_f^2}
		+(a_1^2+a_2^2)(b_1^2+b_2^2)-4(a_1b_1-a_2b_2)^2
	\right]\nonumber\\&+\mathcal{O}\left[(\epi\Lambda)^{-6}\right]
\end{flalign}
Now we can estimate the size of the SM loop contamination.  The most
constraining experimental limits come from scattering of high mass $VV$
pairs, up to $1.5\sim 2$ TeV \cite{CMS:2018ysc,Sirunyan:2017fvv}, so we have
to assume that the EFT is well-behaved within this range, and take
$\epi\Lambda=2$ TeV.  Note that a large $\epi\Lambda$ does not affect the EFT
contribution: at the tree level there is no branch cut below $\Lambda$, while
loop corrections in EFT are always further suppressed by additional loop
factors.  With this value the SM loop contribution is completely dominated by
$\gamma\gamma\to WW$. We find
\begin{flalign}
	f^{00,WW}_{\epi\Lambda}=0.038\ \mathrm{TeV}^{-4}
	\label{eq:smcontamination}
\end{flalign}
for both the $a\parallel b$ and $a\perp b$ cases.

This contribution is much smaller than a typical EFT contribution.  The
current limits on $f_i$ span several orders of magnitude \cite{qgclimits}, but
even the tightest ones are around $\mathcal{O}(1)(\Lambda/\mathrm{TeV})^4$.
This is in a slightly different set of conventions by
Refs.~\cite{Eboli:2006wa,Eboli:2016kko}, whose coefficients will be denoted
with superscript EGM in the following.  
Taking into account the relation between the two sets of conventions, a typical
EFT contribution to $f^{00}_{\epi\Lambda}$ is simply of order
$f_i^{EGM}/\Lambda^4$, meaning that each operator is expected to contribute
about $\mathcal{O}(1)$ TeV$^{-4}$, much larger than
Eq.~(\ref{eq:smcontamination}).  In practice the these contributions could vary
by numerical factors.  For example, we find that the contribution from
$f_{T,7}^{EGM}$ in the $a\parallel b$ case is
\begin{flalign}
	4\sw^2\cw^2\frac{F_{T,7}^{EGM}}{\Lambda^4}=0.69\ \mathrm{TeV}^{-4}
	\frac{F_{T,7}^{EGM} \mathrm{TeV}^{4}}{\Lambda^4}
\end{flalign}
which is still more than one order of magnitude larger than the SM loops.
$F_{T,8}^{EGM}$ gives a much larger contribution, which is $9.7\
\mathrm{TeV}^{-4}$, while the smallest contribution is from $F_{T,2}^{EGM}$,
which gives $0.1\ \mathrm{TeV}^{-4}$.  In any case, we see that the SM loop
contribution cannot qualitatively affect the positivity bounds.

If future experimental sensitivity could improve significantly, it is possible
that the SM contribution should be taken into account as a constant correction
in $f^{00}_{ab,\epi\Lambda}(0)$.  Its size is not difficult to compute as we
have shown in the $\gamma\gamma$ case.  However, in the next subsection
we will show that if the UV completion is weakly coupled, this SM contribution
can be completely removed, without leaving any impact on the positivity condition.

\subsection{Weakly coupled UV completion}

The standard Wilsonian UV completion usually postulates the existence of extra
weak couplings beyond the SM, which can be used to order the loop expansion,
and the leading UV amplitude already satisfies the required properties
(particularly the Froissart bound) that are needed to derive the dispersion
relation.  In this case we can derive positivity for the specific BSM amplitude
at the leading tree or loop order, and the results are free of contaminations
from the SM.

\begin{figure}[ht]
\centering
\includegraphics[width=.45\linewidth]{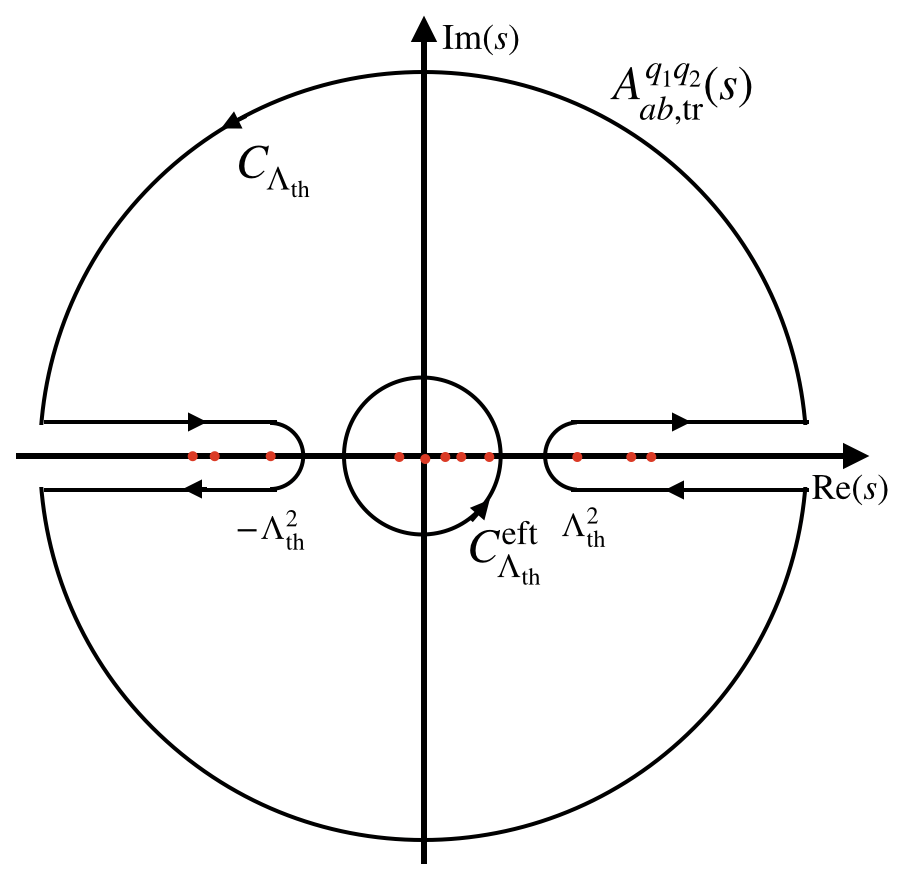}
\caption{Tree level dispersion relation contour in the forward limit.}
\label{fig:DRplot3}
\end{figure}

If the BSM amplitude arises at the tree level by one heavy particle
exchange between two SM vector boson currents, we can simply derive the
positivity bound for the tree level amplitude
$A^{q_1q_2}_{ab,{\rm tr}}$ as follows
\bal
f^{q_1q_2}_{ab,{\rm tr}}(s) &\equiv \frac12 \frac{\ud^2 B^{q_1q_2}_{ab,{\rm tr}}(s)}{\ud s^2} 
\\
& =    \frac{1}{2\pi i}  \(  \int_{-\infty}^{-\Lambda_{\rm th}^2} + \int^{\infty}_{+\Lambda_{\rm th}^2}  \)  \ud s' \frac{ {\rm Disc} A^{q_1q_2}_{ab,{\rm tr}}(s')}{(s'-s)^3}  ,
\eal
where $B^{q_1q_2}_{ab,{\rm tr}}(s)$ is the tree level amplitude with the
	lower energy poles subtracted from $A^{q_1q_2}_{ab,{\rm tr}}$,
and $\Lambda_{\rm th}$ is the energy scale of the lowest massive state that
lies beyond the EFT cutoff; see Fig \ref{fig:DRplot3}. All the
mass poles that lie above $\Lambda$ will contribute to the discontinuity from
$-\infty$ to $-\Lambda_{\rm th}^2$ and from $\Lambda_{\rm th}^2$ to $+\infty$. (The tree level poles contribute to the imaginary part thanks to the Feynman $i\varepsilon$.)
Similar to the previous subsection, we can obtain
\be
f^{q_1q_2}_{ab, {\rm tr}}(s) =  \int^{\infty}_{\Lambda_{\rm th}^2+M^2} \d s'   \frac{ {\rm Im}A^{q_1\bar{q_2}}_{\bar{a}\bar{b}, {\rm tr}}(s')}{( s'-M^2+s)^3} +  \int^{\infty}_{\Lambda_{\rm th}^2} \d s'  \frac{ {\rm Im} A^{q_1q_2}_{ab, {\rm tr} }(s')}{(s'-s)^3}   .
\ee

At the tree level, the contributions to ${\rm Im} A^{q_1q_2}_{ab, {\rm tr} }$
comes from diagrams with a heavy particle propagator $X$. By the cutting rules,
${\rm Im} A^{q_1q_2}_{ab, {\rm tr} }$ can be written as a positive sum of
complete squares of the two-to-$X$ amplitudes, meaning ${\rm Im}
A^{q_1q_2}_{ab, {\rm tr} }(s')>0$ and similarly ${\rm
Im}A^{q_1\bar{q_2}}_{\bar{a}\bar{b}, {\rm tr}}(s')>0$ for $s'>\Lambda_{\rm
th}^2$.  This leads to the leading tree level positivity bound
\be
f^{q_1q_2}_{ab,{\rm tr}}(s) >0 ,~~~~~   -\Lambda_{\rm th}^2 < s <\Lambda_{\rm th}^2  .
\label{eq:f}
\ee
without any possible contribution from the SM.

\begin{figure}[ht]
\centering
\includegraphics[width=.6\linewidth]{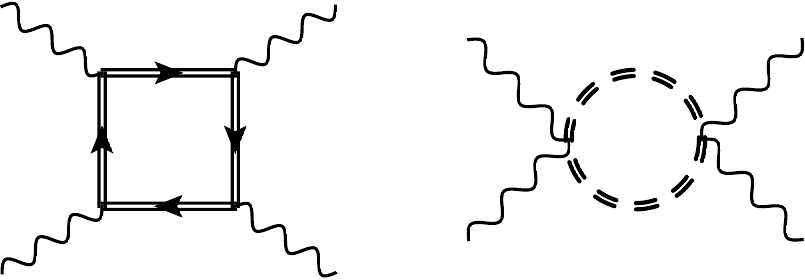}
\caption{Examples of BSM contribution arising from
heavy particle loops. Double lines represent BSM particles.}
\label{fig:heavyloops}
\end{figure}

\begin{figure}[ht]
\centering
\includegraphics[width=.6\linewidth]{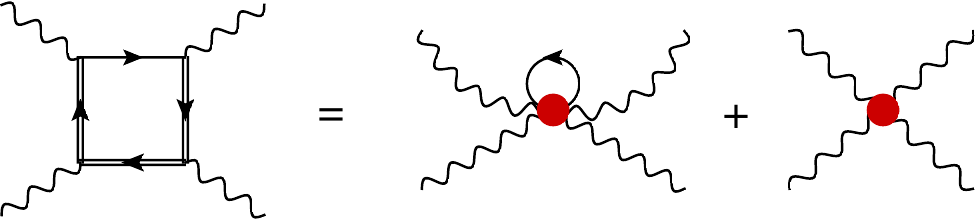}
\caption{If the leading BSM loop contribution contains SM fields
in the loop, the diagram needs to be matched to both tree-level and loop-level
diagrams in the EFT. Red blobs represent higher-dimensional vertices.}
\label{fig:loopmatch}
\end{figure}

More generally, the leading BSM contribution in VBS may arise from loop
diagrams.  For example, the EFT operator may come from 
integrating out heavy particle loops in diagrams shown in
Figure~\ref{fig:heavyloops}.  Only BSM particles can run in the loop,
otherwise the contribution needs to be matched to both tree level
and loop level EFT diagrams, contradicting with our initial assumption that the
higher dimensional contributions are well approximated by the tree level.
An example is shown in Figure~\ref{fig:loopmatch}, where the operator
that enters at the tree level is loop-induced, while the one enters at the loop level
is generated by a tree level graph.
If all loop particles are heavy, we can simply consider the leading BSM loop
contributions, which also satisfies the required properties to derive
positivity.  More specifically, we define
\be
f^{q_1q_2}_{ab, {\rm lp}}(s) \equiv \frac12 \frac{\ud^2 B^{q_1q_2}_{ab,{\rm lp}}(s)}{\ud s^2} 
 =  \int^{\infty}_{\Lambda_{\rm th}^2+M^2} \d s'   \frac{ {\rm Im}A^{q_1\bar{q_2}}_{\bar{a}\bar{b}, {\rm lp}}(s')}{( s'-M^2+s)^3} +  \int^{\infty}_{\Lambda_{\rm th}^2} \d s'  \frac{ {\rm Im} A^{q_1q_2}_{ab, {\rm lp} }(s')}{(s'-s)^3}   .
\ee
where we consider the leading BSM loop contributions to the discontinuity of
the amplitudes, which arise at $n$-loop, and now $\Lambda_{\rm th}$ is the
threshold for multi-particle production of the BSM states. By using cutting
rules, again ${\rm Im} A^{q_1q_2}_{ab, {\rm lp} }(s')$ and ${\rm
Im}A^{q_1\bar{q_2}}_{\bar{a}\bar{b}, {\rm lp}}(s')$ can be written as a sum of
products of two 2-to-$X$ amplitudes at $s'>\Lambda_{\rm th}^2$,
where $X$ is a $n+1$ heavy particle final state. The two amplitudes in each
product must be the same order, otherwise BSM loop contribution could have arose
from a lower order, so each product must be a complete square, and thus
\be
f^{q_1q_2}_{ab, {\rm lp}}(s)>0 ,~~~~~   -\Lambda_{\rm th}^2 < s <\Lambda_{\rm th}^2   .
\ee

To sum up, if the UV completion is weakly coupled, we can derive the positivity
bounds for the leading order (which may be the tree level or any loop order)
BSM contributions to VBS, and the results are not affected by any SM loop
contaminations. This means the resulting positivity bounds will continue
to be relevant, no matter how much the experimental precision will improve in
the future.

\subsection{Results}
Now we are ready to apply the positivity condition $f^{q_1q_2}_{ab}>0$ to
VBS and set constraints on QGC operators.  To this end, consider
the two-by-two scattering of the SM electroweak gauge boson fields
\begin{equation}
V_1V_2\to V_1V_2,~~~~~~ V_i=Z,W^\pm,\gamma ,
\end{equation}
with the polarisation vectors
\begin{align}
	\epsilon^\mu(V_1)&= \sum_{i=1}^3 a_i \epsilon^\mu_{(i)}
	                             =  \left(a_3\frac{p_1}{m_{V_1}},a_1,a_2,a_3\frac{E_1}{m_{V_1}}\right)    ,
	\\
	\epsilon^\mu(V_2)&=\sum_{i=1}^3 b_i \epsilon^\mu_{(i)}
	                             =\left(b_3\frac{p_2}{m_{V_2}},b_1,b_2,b_3\frac{E_2}{m_{V_2}}\right)    ,
\end{align}
where the linear polarisation basis is given by
\be
\epsilon^\mu_{(1)}(V) = (0,1,0,0), ~~~\epsilon^\mu_{(2)}(V)  = (0,0,1,0),~~~\epsilon^\mu_{(3)}(V)  = \frac{1}{m_V}(p_V,0,0,E_V),
\ee
and $a_i$, $b_i$ are arbitrary complex numbers (with non-vanishing $a_3$, $b_3$
only for massive vectors).  We will use $\vec a\equiv (a_1,a_2,a_3)$ and $\vec
b\equiv (b_1,b_2,b_3)$ to denote the polarisation state.  The amplitude can be
computed at the tree level using standard tools
\cite{Alloul:2013bka,Hahn:2000kx,Hahn:1998yk}.  We find that $f^{q_1q_2}_{ab}>0$
for all 7 $VV\to VV$ channels leads to the following conditions:
\begin{align}
&ZZ:\nonumber\\
& 8 A_1(4 (2 F_{T,0} + 2 F_{T,1} + F_{T,2}) c_W^8 + 
   2 (2 F_{T,5} + 2 F_{T,6} + F_{T,7}) c_W^4 s_W^4 + (2 F_{T,8} + F_{T,9}) s_W^8) 
\nonumber\\
&  +4A_2 (4 F_{T,2} c_W^8 + 2 F_{T,7} c_W^4 s_W^4 + F_{T,9} s_W^8) +(A_3+A'_3) ((-4 F_{M,1} + 2 F_{M,7}) c_W^6 + 2 F_{M,5} c_W^4 s_W^2 - 2 F_{M,3} c_W^2 s_W^4)
\nonumber\\
&+(A_4+A'_4) (8 (4 F_{T,0} + 4 F_{T,1} + F_{T,2}) c_W^8 + 4 (4 F_{T,5} + 4 F_{T,6} + F_{T,7}) c_W^4 s_W^4 + 
 2 (4 F_{T,8} + F_{T,9}) s_W^8)
 \nonumber\\
 &+(A_5+A'_5)((8 F_{M,0} - 2 F_{M,1} + F_{M,7}) c_W^6 + (4 F_{M,4} + F_{M,5}) c_W^4 s_W^2 + (4 F_{M,2} - 
    F_{M,3}) c_W^2 s_W^4) 
  \nonumber\\
&+    16 A_6 (F_{S,0} + F_{S,1} + F_{S,2}) c_W^4>0
\label{eq:starts}\\
\nonumber\\
&W^\pm W^\pm:\nonumber\\
& 2 A_1(8 F_{T,0} + 12 F_{T,1} + 5 F_{T,2}) +6 A_2F_{T,2}+(A_3+A_3')(-2 F_{M,1} + F_{M,7})
\nn
&   +2A_4(8 F_{T,1} +  F_{T,2}) + 2A_4' (8 F_{T,0} + 4 F_{T,1} + F_{T,2})+ A'_5(4 F_{M,0} - F_{M,1} +F_{M,7})
\nn
& + 4 A_6 (2 F_{S,0} + F_{S,1} + F_{S,2})>0
\label{eq:WW}
\\
\nonumber\\
&W^\pm W^\mp:\nonumber\\
&  2 A_1(8 F_{T,0} + 12 F_{T,1} + 5 F_{T,2}) +6 A_2F_{T,2}+(A_3+A_3')(-2 F_{M,1} + F_{M,7})
\nn
&   +2A'_4(8 F_{T,1} +  F_{T,2}) + 2A_4 (8 F_{T,0} + 4 F_{T,1} + F_{T,2})+ A_5(4 F_{M,0} - F_{M,1} + F_{M,7})
\nn
& + 4 A_6 (2 F_{S,0} + F_{S,1} + F_{S,2})>0
\\
\nonumber\\
&W^\pm Z:\nonumber\\
& 4 A_1 (4 (4 F_{T,1} + F_{T,2}) c_W^5 + (4 F_{T,6} + F_{T,7}) c_W s_W^4) + 4A_2 c_W (4 F_{T,2} c_W^4 + F_{T,7} s_W^4)
\nn
&  - 4 A_3 ((2 F_{M,1} - F_{M,7}) c_W^5 + F_{M,5} c_W^3 s_W^2 + F_{M,3} c_W s_W^4)
\nn
&+4A_3'  (-2 F_{M,1} + F_{M,7}) c_W^3 + 8 (A_4+A_4')c_W (4 F_{T,1} c_W^4 + F_{T,6} s_W^4)
\nn
& + (A_5+A_5') (2 F_{M,7} c_W^4 + F_{M,5} s_W^2 + 4 F_{M,4} c_W^2 s_W^2 - F_{M,5} s_W^4)
\nn
&+16 A_6 (F_{S,0} + F_{S,2}) c_W^3  >0
\\
\nonumber\\
&Z\gamma:\nonumber\\
& A_1 (F_{T,7} + 16 (2 F_{T,0} + 2 F_{T,1} + F_{T,2}) c_W^4 - 8 F_{T,7} s_W^2 - 
16 F_{T,5} c_W^2 s_W^2 + 8 F_{T,7} s_W^4 + 8 F_{T,8} s_W^4 
\nn
& + 4 F_{T,9} s_W^4 + 
4 F_{T,6} (1 - 2 s_W^2)^2) +A_2(8 F_{T,2} c_W^4 + 2 F_{T,9} s_W^4 + F_{T,7} (1 - 2 s_W^2)^2)
\nn
& +(A_3+A_3')(4 (4 F_{T,0} + 4 F_{T,1} + F_{T,2}) c_W^4 - 
 2 (4 F_{T,5} + F_{T,7}) c_W^2 s_W^2 + (4 F_{T,8} + F_{T,9}) s_W^4 + 
 2 F_{T,6} (1 - 2 s_W^2)^2)
\nn
& + A''_3 (F_{M,7} - (2 F_{M,1} + F_{M,3} + F_{M,5}) c_W^2 - F_{M,7} s_W^2)  > 0
\\
\nonumber\\
&W^\pm \gamma:\nonumber\\
& A_1 (16 F_{T,1} + 4 F_{T,2} + 4 F_{T,6} + F_{T,7}) + A_2 (4 F_{T,2} + F_{T,7})+A_3(-2 F_{M,1} - F_{M,3} + F_{M,5} + F_{M,7})  
\\
& + 2(A_4+A'_4) (4 F_{T,1} + F_{T,6}) >0
\\
\nonumber\\
&\gamma\gamma:\nonumber\\
&4 A_1 (8 F_{T,0} + 8 F_{T,1} + 4 F_{T,2} + 4 F_{T,5} + 4 F_{T,6} + 2 F_{T,7} + 2 F_{T,8} + F_{T,9}) + 2A_2 (4 F_{T,2} + 2 F_{T,7} + F_{T,9})
\nn
& (A_4+A'_4)( 16 F_{T,0} + 16 F_{T,1} + 4 F_{T,2} + 8 F_{T,5} + 8 F_{T,6} + 2 F_{T,7} + 4 F_{T,8} + F_{T,9}) > 0
\label{eq:ends}
\end{align}
where the arbitrary complex parameters $a_i$ and $b_i$ form the following
combinations 
\begin{align}
\begin{aligned}
&A_1\equiv |a_1|^2 |b_1|^2 + |a_2|^2 |b_2|^2,
\\
&A_2\equiv |a_1|^2 |b_2|^2 + |a_2|^2 |b_1|^2,
\\
&A_3\equiv ( |b_1|^2+ |b_2|^2)|a_3|^2,
\\
&A'_3\equiv (|a_1|^2  +|a_2|^2) |b_3|^2 ,
\\
&A''_3\equiv |b_1|^2  |a_3|^2
\end{aligned}
\qquad~~
\begin{aligned}
&A_4\equiv  a_1a_2^*  b_1 b_2^*+c.c. ,
\\
&A'_4\equiv a_1a_2^*  b_1^* b_2+c.c.  ,
\\
&A_5\equiv  (a_1 b_1 +a_2 b_2) a_3^* b_3^*  + c.c.,
\\
&A'_5 \equiv - (a_1 b_1^* +a_2 b_2^*) a_3^*b_3+c.c.
\\
&A_6\equiv |a_3|^2 |b_3|^2 ,
\end{aligned}
\end{align}
and we have also defined
\begin{align}
	s_W\equiv \sin\theta_W,\quad
	c_W\equiv \cos\theta_W,\quad
	t_W\equiv \tan\theta_W ,~~~~~\theta_W \text{~being weak angle}.
\end{align}
In Section~\ref{subsec:general} we will see that the $W^\pm W^\pm$ channel and
the $W^\pm W^\mp$ channel give equivalent bounds.  For the rest of the paper we
will only consider $W^\pm W^\pm$, and omit the superscript $\pm$.

We have neglected the contribution from dim-6 operators. In
Ref.~\cite{Zhang:2018shp} we have shown by explicit calculations that their
contributions to the l.h.s.~of the inequalities are always negative-definite,
which means that dropping these terms will only make the positivity conditions
more conservative. For completeness, here we give these contributions in the
Warsaw basis \cite{Grzadkowski:2010es}, but for the rest of the paper we will
drop them.  They read
\begin{align}
	&WZ:\nonumber\\
	&-a_3^2b_3^2s_W^4c_W^2\left(c_WC_{\varphi D}+4s_WC_{\varphi
	WB}\right)^2
	-36(a_1b_1+a_2b_2)^2e^2s_W^2c_W^2C_{W}^2
	+\mbox{dim-8 terms}
	>0
	\\
	&WW:\nonumber\\
	&-a_3^2b_3^2s_W^2c_W^4C_{\varphi D}^2
	-36 (a_1b_1+a_2b_2)^2e^2s_W^2c_W^2 C_W^2
	+\mbox{dim-8 terms}
	>0
	\\
	&W\gamma:\nonumber\\
	&-(a_1b_1+a_2b_2)^2C_W^2
	+\mbox{dim-8 terms}
	>0
\end{align}
and there are no contribution in the other four channels.

We could have considered amplitudes that involve the Higgs boson, such as
$VH\to VH$, to derive more bounds.  However, the dim-8 parametrisation of QGC
has only included the relevant operators that involve four-gauge-boson
vertices and are independent of TGCs.  In principle, other operators might also
enter the $VH\to VH$ vertex, and this will lead to results that depend on
non-QGC operators.  For this reason we have focused on the VBS channels, where
all relevant operators are included in the QGC parameterisation.

In our previous work \cite{Zhang:2018shp}, for computational simplicity we have
restricted ourselves to positivity bounds for real $a_i$ and $b_i$
corresponding to crossing symmetric amplitudes with linear polarisations.  In
this work we will consider arbitrary complex polarisations. 
We will see that
even though the vast majority of the constraining power comes from real
polarisations, including complex polarisations does bring improvements, which
could be significant if only one or several operators are turned on.

\section{Solving the positivity conditions}
\label{sec:solve}

The positivity conditions given in the previous section can be schematically
written as
\begin{flalign}
	\sum_iF_i x_{i,j}(\vec a,\vec b)>0
\end{flalign}
where $F_i$ are all coefficients, $j$ denotes the scattering channel,
and the $x_i$'s are complex quartic homogeneous polynomial functions of $a_i$
and $b_i$.  These conditions needs to be satisfied for arbitrary values
of the polarisation vectors $\vec a$ and $\vec b$.  As we have mentioned, these
conditions are inconvenient because they involve $\vec a$ and $\vec b$ as free
parameters.  
The goal of this section is to solve these inequalities and remove the $\vec a$
and $\vec b$ dependence.  This is done by going through all possible complex
values for $\vec a,\vec b$, which span a 6-dimensional complex space, obtaining
the bounds for each polarisation, and combining all the resulting bounds.  By
doing this, we will obtain the simplest description for the physical parameter
space that is independent of any free parameters, which can be used directly in
future experimental as well as theoretical studies.

\subsection{Toy cases}
\label{subsec:toy}

Before solving the positivity conditions in the most general case, it is
illustrative to consider some toy cases, to explain our approach and
to develop physics intuition.

\subsubsection{The pyramid case}

Let us consider a case where we only allow three operator coefficients,
$F_{M,0}$, $F_{M,1}$ and $F_{M,5}$ to be nonzero. We also restrict $\vec a$
and $\vec b$ to take real values only. The problem is reduced to a
3-dimensional one.  We can write the positivity conditions in the form of
the inner product of two vectors:
\begin{flalign}
\vec{F}\cdot \vec x_i(\vec a,\vec b)>0
\label{eq:cond}
\end{flalign}
where $\vec{F}=(F_{M,0},F_{M,1},F_{M,5})$ is a coefficient vector, and
$i=WW,WZ,\cdots$ are different channels. The $\vec x_i$ vectors are:
\begin{flalign}
	&\vec x_{WW}=\left(-4T_aT_b\cos\phi,-T_a^2+T_aT_b\cos\phi-T_b^2,0\right)
	\\
	&\vec x_{ZZ}=(0,-2\cw^2,\sw^2)
	\\
	&\vec x_{WZ}=\left(0,-2\cw^2T_a^2-2T_b^2,-\sw^2T_a^2\right)
	\\
	&\vec x_{W\gamma}=(0,-2,1)
	\\
	&\vec x_{Z\gamma}=(0,-2,-1)
\end{flalign}
and $\vec x_{\gamma\gamma}$ does not give any nontrivial condition.  Here
$T_a$, $T_b$, and $\cos\phi$ are defined as
\begin{flalign}
	&T_a=\frac{a_3}{\sqrt{a_1^2+a_2^2}},
	\quad
	T_b=\frac{b_3}{\sqrt{b_1^2+b_2^2}}\,.
	\\
	&\cos\phi=\frac{a_1b_1+a_2b_2}{\sqrt{a_1^2+a_2^2}
	\sqrt{b_1^2+b_2^2}}
\end{flalign}
Positivity conditions simply state that the coefficients, $\vec F$, must be 
positive once projected onto any of the $\vec x_i(\vec a,\vec b)$ vectors.

The set of vectors $X=\{\vec x_i(\vec a,\vec b)\}$ encodes all the information
to describe the actual bounds.  The problem now is to identify the simplest way
to describe this vector set, or one that gives exactly the same bounds,
without referring to the polarisation vector
$\vec{a},\vec{b}$.  A useful fact is the following: if a specific vector
$\vec x$ can be written as a positive linear combination of some other vectors
in the set, specified by channel $i$ and polarisation $\vec a_j,\vec b_j$, i.e.
\begin{flalign}
	\vec x=\sum_{i,j}c_{i,j}\vec{x_i}(\vec{a}_j,\vec{b}_j),\quad
	c_{i,j}\ge0
	\label{eq:positivelinear}
\end{flalign}
then this vector $\vec x$ can be removed form the set $X$, without affecting
the allowed parameter space. This is because
\begin{flalign}
	\vec F\cdot \vec x=\sum_{i,j}c_{i,j}\vec F\cdot
	\vec{x_i}(\vec{a}_j,\vec{b}_j)
\end{flalign}
but each term on the r.h.s is already positive, so $\vec F\cdot \vec x>0$ does not
lead to new exclusion.  Making use of this, we can keep removing
redundant $\vec{x}$ vectors from the set $X$, until it reaches the simplest form.

\begin{figure}
	\begin{center}
		\hfill
		\includegraphics[width=.5\linewidth]{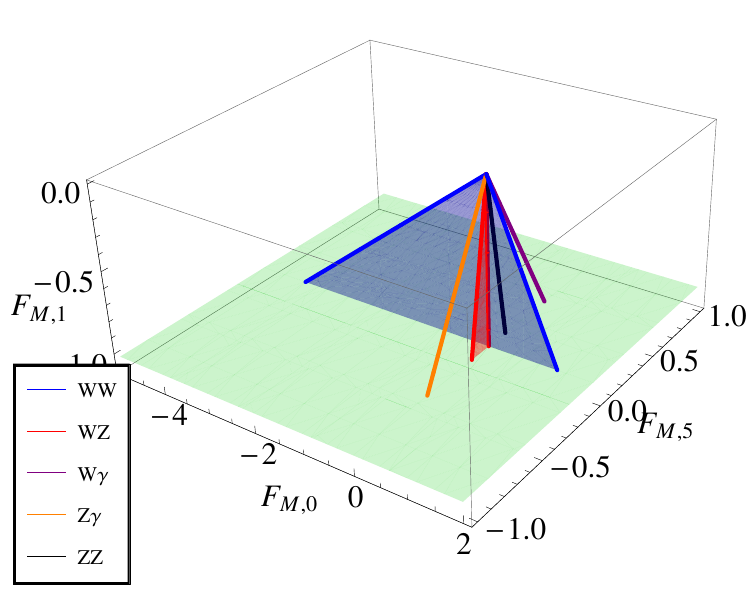}
		\hfill
		\includegraphics[width=.45\linewidth]{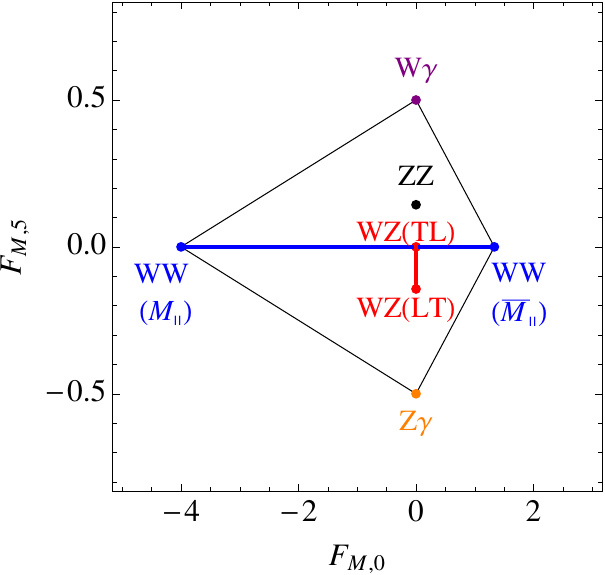}
		\hfill
	\end{center}
	\caption{Left: the vectors $\vec{x}_i(\vec{a},\vec{b})$ intersect the
plane $(u,-1,v)$, leaving a set of intersection points.
Right: the intersection points $\vec P_i(\vec{a},\vec{b})=(u,v)$ on the plane.
The colors represent different scattering channels. The $M_{\parallel}$, $\bar
M_{\parallel}$, $LT$ and $TL$ in the brackets denote the corresponding polarisation
configuration that gives the endpoints of the line segments, see
Table~\ref{tab:polar} in Section~\ref{subsec:polarisation}.}
	\label{fig:inter1}
\end{figure}

More specifically, we proceed as follows.  Notice that in this example
the second component of the $\vec{x}_i$ vectors is always negative.  We
rescale all $\vec{x}_i$ with a positive factor such that their second components
are all equal to -1, and then we focus on the other two components.  In other
words, we find the intersection points of $\vec{x}_i$ on the plane $(u,-1,v)$,
as shown in Figure~\ref{fig:inter1} left.

Note that the $\vec x_{WW}$ and $\vec{x}_{WZ}$ are functions of $\vec{a},\vec{b}$
and they each span a triangle that intersects the plane with a line segment.
Now we focus on the intersection points of $\vec{x}_i$, which we denote by
$\vec P_i(\vec{a},\vec{b})=(u,v)$.  They are
\begin{flalign}
	&\vec P_{WW}(\vec a,\vec b)=\left(-\frac{4T_aT_b\cos\phi}{T_a^2-T_aT_b\cos\phi+T_b^2},0\right),
	\quad
	\vec P_{ZZ}(\vec a,\vec b)=\left(0,\frac{\sw^2}{2\cw^2}\right),
	\\
	&\vec P_{WZ}(\vec a,\vec b)=\left(0,-\frac{\sw^2T_a^2}{2\cw^2T_a^2+2T_b^2}\right),
	\quad
	\vec P_{W\gamma}(\vec a,\vec b)=\left(0,\frac{1}{2}\right),
	\\
	&\vec P_{Z\gamma}(\vec a,\vec b)=\left(0,-\frac{1}{2}\right).
\end{flalign}
These points are shown in Figure~\ref{fig:inter1} right. From the above expressions
we can see that, in this simplified case, depending on the polarisation, the
$WW$ channel leads to intersection points on the horizontal axis, varying from
-4 to 4/3, while the $WZ$ channel leads to intersection points on the vertical
axis, varying from 0 to $-t_W^2/2$.  In more general cases, the
intersection points could form a higher dimensional point set, instead of just
line segments.

\begin{figure}[h!]
	\begin{center}
		\includegraphics[width=.5\linewidth]{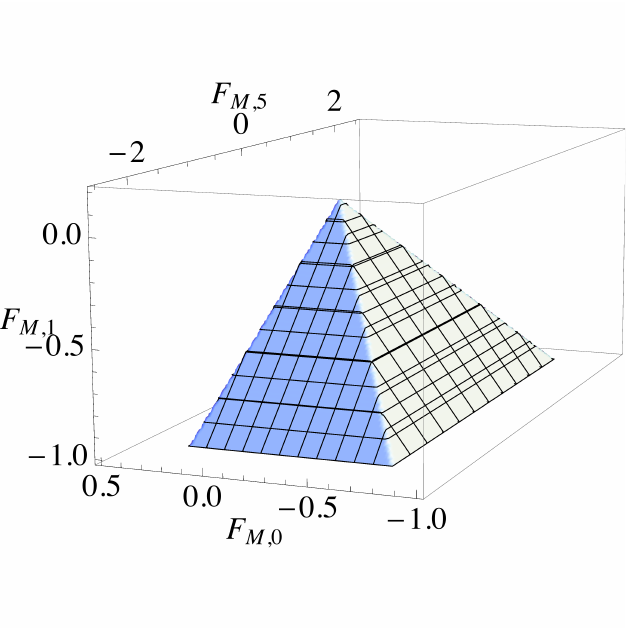}
	\end{center}
	\caption{In a subspace spanned by three coefficients, $F_{M,0}$,
	$F_{M,1}$ and $F_{M,5}$, allowed parameter space is described by a
pyramid.}
	\label{fig:p1}
\end{figure}

Now we can use Eq.~(\ref{eq:positivelinear}) to remove redundant conditions.
When this equation is satisfied, the corresponding intersection point $\vec P$
of $\vec x$ must satisfy 
\begin{flalign}
	&\vec P=\sum_{i,j}c_{i,j}\vec P_i(\vec a_j,\vec b_j)
	\\
	&\sum_{i,j}c_{i,j}=1
\end{flalign}
i.e.~the $\vec P$ is a positive linear combination of other intersection points,
and that the sum of the coefficients is one.  The latter condition comes from the fact
that the second component of $\vec x$ is always -1.  This implies that
$\vec P$ stays within a polygon formed by other $\vec P_i(\vec a_j,\vec b_j)$.
Therefore, in Figure~\ref{fig:inter1} right, we are allowed to remove any points
that are inside a polygon formed by other points.  Obviously, the largest
polygon is the convex hull of all $\vec P_i$, displayed by the black lines in
the figure.  It has four vertices, given respectively by the $W\gamma$ channel,
the $WZ$ channel, and the $WW$ channel with two different polarisation states,
which we denote as $M_\parallel$ and $\bar M_\parallel$.  The polarisation will
be better explained in Section \ref{subsec:polarisation}. Here we simply mention
that $M_\parallel$ means that $\vec a$ and $\vec b$ are parallel to each other,
while $M_\parallel$ means that $\vec a$ and $\vec b'$ are parallel to each
other, where $\vec b'=(b_1,b_2,-b_3)$.  With these four vertices, the
positivity constraints are minimally described by four linear inequalities:
\begin{flalign}
	\begin{aligned}
	&-2F_{M,1}+F_{M,5}>0,
	\\
	&-2F_{M,1}-F_{M,5}>0,
	\end{aligned}
	\quad
	\begin{aligned}
	&-4F_{M,0}-F_{M,1}>0,
	\\
	&+4F_{M,0}-3F_{M,1}>0.
	\end{aligned}
\end{flalign}
These conditions solely are equivalent to the statement that
Eq.~(\ref{eq:cond}) holds for any and all $\vec a$ and $\vec b$.  We can
rewrite the above solution in a form that is more convenient for a parameter
scan,
\begin{flalign}
	&F_{M,1}\in\left(-\infty,0\right)
	\\
	&F_{M,0}\in \left(\frac{3}{4}F_{M,1},-\frac{1}{4}F_{M,1}\right)
	\\
	&F_{M,5}\in \left(2F_{M,1},-2F_{M,1}\right)
\end{flalign}
It's easy to see that these conditions carve out a pyramid in the 3-dimensional
parameter space spanned by $F_{M,0}$, $F_{M,1}$ and $F_{M,5}$.  We show this in
Figure~\ref{fig:p1}.  Note that if we take $F_{M,5}=0$, the bounds in the
$F_{M,0}-F_{M,1}$ plane are not as tight as we have shown in Figure~\ref{fig:cms}
left.  This is because here we only considered real polarisation vectors.  If
we allow $\vec a$ and $\vec b$ to take complex values, the $ZZ$ point will
expand to a horizontal line that goes from $-\infty$ to $+2$, which leads to a
stronger constraint.

In principle, the above procedure should apply to the most general case.
However, the large dimension of the problem (18 coefficients)
and the complex values of $\vec a$ and $\vec b$ components (6 degrees of
freedom, after removing 4 free phases and 2 overall scaling factors),
make the problem more complicated.  We will solve the problem in 
two steps.  In the first step, we focus on each channel, and remove the
polarisation dependence in each channel. This corresponds to identifying
the endpoints of the $WW$ and $WZ$ line segments in Figure~\ref{fig:inter1} right.
As we have mentioned, in a more general case, the $\vec P_{i}(\vec a,\vec b)$
points could form a higher dimensional object. In that case we will identify
the boundary of this point set.  In the second step, we combine the results
of all channels, and simplify further.  This is similar to removing the
$ZZ$, $WZ(LT)$ and $WZ(TL)$ points in Figure~\ref{fig:inter1} right by taking
a convex hull of all points (here $(LT)$ and $(TL)$ denote two
polarisation configurations which correspond to $W_LZ_T$
scattering and $W_TZ_L$ scattering respectively, see Table~\ref{tab:polar} in
Section~\ref{subsec:polarisation} for more details).

\subsubsection{The cone case}
\begin{figure}
	\begin{center}
		\hfill
		\includegraphics[width=.45\linewidth]{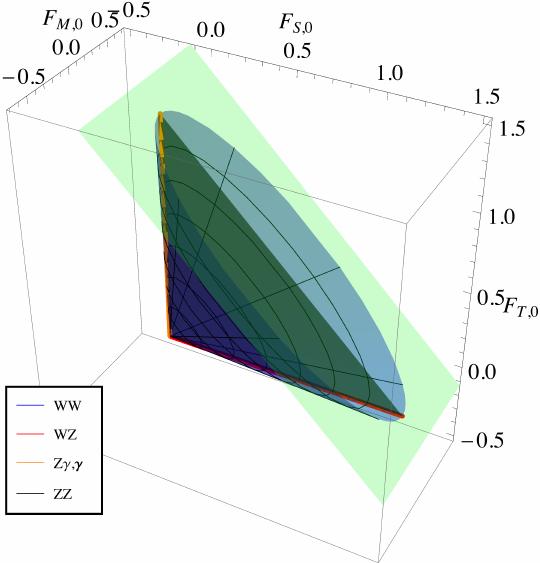}
		\hfill
		\includegraphics[width=.45\linewidth]{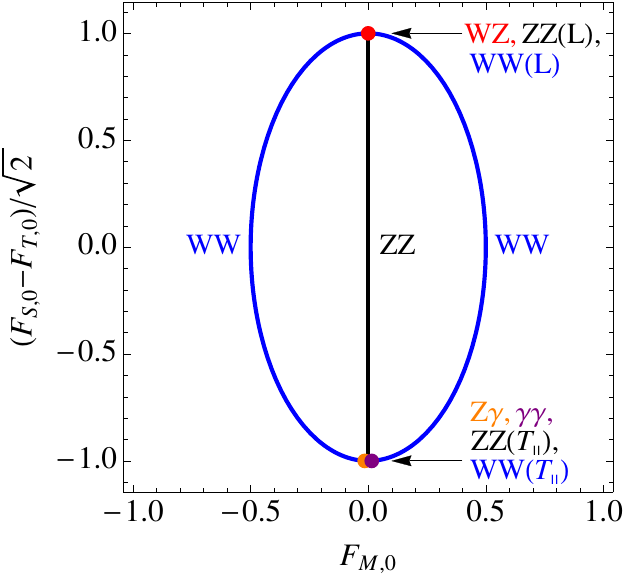}
		\hfill
	\end{center}
	\caption{Similar to Figure~\ref{fig:inter1}.
Left: the vectors $\vec{x}_i(\vec{a},\vec{b})$ intersect the plane
$\left(\frac{1+v}{\sqrt{2}},u,\frac{1-v}{\sqrt{2}}\right)$, leaving a set of
intersection points.  Right: the intersection points $\vec
P_i(\vec{a},\vec{b})=(u,v)$ on the plane form a circle.
Colors represent different scattering channels. The $T_{\parallel}$ and
$L$ in the brackets denote the corresponding polarisation
configuration in the $WW$ and $ZZ$ channels that give the top and bottom
points, see Table~\ref{tab:polar} in Section~\ref{subsec:polarisation}.
}
	\label{fig:inter2}
\end{figure}

There is still a caveat in the above approach: the boundary of the point set
$\{\vec P_i(\vec a,\vec b)\}$ could be a curve.  In that case taking
the convex hull does not help to simplify its description.
Consider a second example, where we turn on three different operators,
$F_{S,0}$, $F_{M,0}$, and $F_{T,0}$.  The corresponding $\vec{x}_i$ vectors are
\begin{flalign}
	&\vec{x}_{WW}=\left(T_a^2T_b^2,-T_aT_b\cos\phi,2\cos^2\phi\right)
	\\
	&\vec{x}_{ZZ}=\left(T_a^2T_b^2,0,4\cw^4\cos^2\phi\right)
	\\
	&\vec{x}_{WZ}=(1,0,0)
	\\
	&\vec{x}_{Z\gamma}=\vec{x}_{\gamma\gamma}=(0,0,1)
\end{flalign}
$\vec{x}_{W\gamma}$ does not give any nontrivial result.
We can see that the sum of the first and the third components are alway positive,
i.e.~these $\vec x$'s are always positive along the direction of $(1,0,1)$.
Therefore similar to the previous case, we consider the intersection points
of $\vec x_i$ on the plane
\begin{flalign}
	\vec{x}=\left(\frac{1+v}{\sqrt{2}},u,\frac{1-v}{\sqrt{2}}\right)
\end{flalign}
This is shown in Figure~\ref{fig:inter2} left.  In particular, we see that the
$WW$ channel spans a cone that intersects the plain with a circle.  Again,
denoting the intersection points by $\vec P_i(\vec{a},\vec{b})=(u,v)$, we find
\begin{flalign}
	&\vec
	P_{WW}(\vec a,\vec b)=\left(-\frac{\sqrt{2}T_aT_b\cos\phi}{T_a^2T_b^2+2\cos^2\phi},\frac{T_a^2T_b^2-2\cos^2\phi}{T_a^2T_b^2+2\cos^2\phi}\right),
	\qquad
	\vec P_{ZZ}(\vec a,\vec b)=\left(0,\frac{T_a^2T_b^2-4\cw^4\cos^2\phi}{T_a^2T_b^2+4\cw^4\cos^2\phi}\right),
	\\
	&\vec P_{WZ}(\vec a,\vec b)=(0,1),\qquad
	\vec P_{Z\gamma}(\vec a,\vec b)=P_{\gamma\gamma}(\vec a,\vec b)=(0,-1).
\end{flalign}
The intersection points are shown in Figure~\ref{fig:inter2} right. We see that
the $WW$ channel leaves a circle on the plane: if we define
$\cot\frac{\theta}{2}=-\frac{T_aT_b}{\sqrt{2}\cos\phi}$, then 
\begin{flalign}
	P_{WW}(\theta)=\left(\frac{1}{2}\sin\theta,\cos\theta\right)
\end{flalign}
 
\begin{figure}[h!]
	\begin{center}
		\includegraphics[width=.5\linewidth]{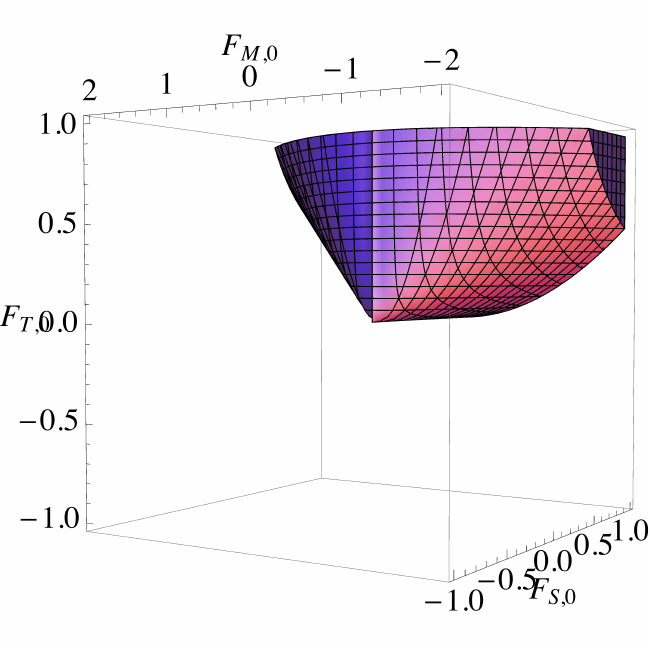}
	\end{center}
	\caption{In a subspace spanned by three coefficients, $F_{S,0}$,
	$F_{M,0}$ and $F_{T,0}$, allowed parameter space is described by a
cone.}
	\label{fig:c1}
\end{figure}

In this case, taking a convex hull allows us to keep only the $WW$ channel
while removing all other channels, but the positivity is still described by an
infinite set of vectors.  To solve the exact condition, we note that
$f_{WW}=\vec F\cdot \vec x_{WW}(\vec a,\vec b)$ is a quadratic function of
$T_aT_b/\cos\phi$:
\begin{flalign}
	f_{WW}\propto F_{S,0}(T_aT_b/\cos\phi)^2-F_{M,0}(T_aT_b/\cos\phi)+2F_{T,0}
\end{flalign}
defined on $T_aT_b/\cos\phi\in(-\infty,\infty)$. The condition for this function to be
positive is simply 
\begin{flalign}
	F_{S,0}>0 \qquad \mbox{and} \qquad 8F_{S,0}F_{T,0}-F_{M,0}^2>0
\end{flalign}
So even though the original inequality is a linear function of $\vec F$,
by removing the $\vec a$ and $\vec b$ dependence the solution is upgraded to a
quadratic function.  This is because the most constraining value of $T_aT_b/\cos\phi$
is a function of the $F_i$ coefficients, which in this case is
\begin{flalign}
	\frac{T_aT_b}{\cos\phi}\to \frac{F_{M,0}}{2F_{S,0}}
\end{flalign}
and so $f_{WW}=\vec F\cdot \vec x_{WW}(T_aT_b)$ becomes a quadratic function
once the above value is plugged in.
This quadratic function describes a cone in the parameter space spanned
by $F_{S,0}$, $F_{M,0}$ and $F_{T,0}$. If we rotate the $S$ and $T$ axes by
defining:
\begin{flalign}
	&F_1=\frac{F_{S,0}+F_{T,0}}{\sqrt{2}}
	\\
	&F_2=\frac{F_{S,0}-F_{T,0}}{\sqrt{2}}
\end{flalign}
Then the condition can be written as
\begin{flalign}
	F_1^2>F_2^2+(F_{M,0}/2)^2
\end{flalign}
This carves out a cone with its vertex right at the origin and the axis pointing to
the $F_1$ direction.  We show this in Figure~\ref{fig:c1}. Note that
if we allow complex polarisation, the line segment from $ZZ$ channel will
expand to an even larger circle, reducing the allowed parameter space to
$4F_{S,0}F_{T,0}-F_{M,0}^2>0$.

In more general cases, i.e.~with more operator coefficients and complex
polarisation, solving the positivity condition of each channel will lead to
several inequalities independent of polarisation. Some of them are linear
inequalities.  They are like the endpoints of the $WW$ and $WZ$ line segments
in our first toy example.  Others are quadratic or even higher order
inequalities. They are like the cones as in our second example.

\subsection{General solution}
\label{subsec:general}

Now we consider the most general case.

First of all, it is useful to rewrite $x_i$ in a different form to remove the
redundant degrees of freedom in $\vec a$ and $\vec b$.  Since they are complex
vectors,
in total they have 12 degrees of freedom. 
However, the rescaling of $\vec a$ and $\vec b$ does not change the positivity
condition. This removes
2 degrees of freedom. 
In addition, since the bounds are derived from the forward scattering amplitude,
the system has a rotational symmetry around the beam axis, so the
transversal components of the two vectors can appear only in the forms of
$a_1b_1^*+a_2b_2^*$, $a_1b_1+a_2b_2$, $|a_1|^2+|a_2|^2$ and $|b_1|^2+|b_2|^2$.
This removes 2 more degrees of freedom.
Finally, $\vec a$, $\vec b$, $\vec a^*$ and $\vec b^*$ each appear once in the 
amplitude, so adding a pure phase on all components of either $\vec a$ or $\vec b$ does not
change the result. This leads to two more redundant degrees of freedom.
As a result, if we define:
\begin{flalign}
	&|a_3|^2=T_a^2,\quad |b_3|^2=T_b^2
	\nonumber\\
	&a_3b_3^*=T_aT_b e^{i\delta},\quad
	a_3b_3=T_aT_b e^{i\Delta}
	\nonumber\\
	&a_1b_1^*+a_2b_2^*=\cos\phi e^{i(\alpha+\delta)},\quad
	a_1b_1+a_2b_2=\cos\psi e^{i(\beta+\Delta)} 
	\nonumber\\
	&|a_1|^2+|a_2|^2=1, \quad |b_1|^2+|b_2|^2=1
	\label{eq:variable1}
\end{flalign}
then $x_i(\vec a,\vec b)$ only depend on six parameters: $T_a$, $T_b$,
$\cos\phi$, $\cos\psi$, $\cos\alpha$ and $\cos\beta$.
We have rescaled the both vectors to fix $|a_1|^2+|a_2|^2$ and $|b_1|^2+|b_2|^2$.
$\delta$ and $\Delta$ are the free phases.  The six parameters can take
real values in the region
\begin{flalign}
	&T_a,T_b\in [0,+\infty)
	\nonumber\\
	&\cos\phi,\cos\psi\in [0,1] \nonumber\\
	&\cos\alpha,\cos\beta\in [-1,1]
	\label{eq:region1}
\end{flalign}
It is easy to show that these 6 parameters can independently take any values
in the above range, i.e.~the change of variable can be inverted in the region
defined above. Also, at this point one can see that the $W^\pm W^\pm$ and the
$W^\pm W^\mp$ channels give the same bounds, because
\begin{flalign}
	\vec x_{W^\pm W^\mp}\left(T_a,T_b,\phi,\psi,\alpha,\beta\right)
	=
	\vec x_{W^\pm W^\pm}\left(T_a,T_b,\phi,\psi,\pi-\beta,\alpha\right)
\end{flalign}
So we will only consider $W^\pm W^\pm$, which we simply refer to as $WW$.

The problem now is the following: under which conditions are $\sum_iF_i
x_i(\vec a,\vec b)$, as functions of the variables defined in
Eq.~(\ref{eq:variable1}), positive definite in the region defined by
Eq.~(\ref{eq:region1})?  We will show that these functions, after additional
changes of variables, are either at most quadratic functions defined in this
area, or bounded by such functions from below.  The positivity condition is
then equivalent to two conditions, 1) the minimum values for these quadratic
functions exist, and 2) the minimum values are positive.
These two conditions can be decomposed into a set of linear inequalities,
which are like the endpoints of $WW$ and $WZ$ line segments in our first
toy example, plus one or two higher-degree polynomial inequalities,
which are like the circle from $WW$ channel in our second example.
Below we consider each channel separately.

\subsubsection{$WW$}
Let us first consider $WW$ scattering. For later convenience, we rewrite
$\cos\phi$ as $W$, $\cos\psi$ as $V$, and $\cos\alpha$ as $R$.
The positivity condition implies $f_{WW}\ge0$, where $f_{WW}$ can be written as
\begin{flalign}
	f_{WW}(T_a,T_b,R,W,V)=f_1(T_a^2+T_b^2)+f_2T_a^2T_b^2+f_3W^2+f_4
	T_aT_bRW+f_5V^2+f_6
\end{flalign}
where $T_a$, $T_b$, $W$, $R$, $V$ are independent parameters in the range
\begin{flalign}
	T_a,T_b\in[0,+\infty),\quad W\in[0,1],\quad R\in[-1,1],\quad V\in[0,1]
\end{flalign}
and
\begin{flalign}
&f_1=-2 F_{M,1}+F_{M,7}
\\
&f_2=8 F_{S,0}+4 F_{S,1}+4 F_{S,2}
\\
&f_3=16 F_{T,0}+8 F_{T,1}+2 F_{T,2}
\\
&f_4=-8 F_{M,0}+2 F_{M,1}-2 F_{M,7}
\\
&f_5=16 F_{T,1}+2 F_{T,2}
\\
&f_6=6 F_{T,2}
\end{flalign}
For $f_{WW}>0$ to hold for any $T_{a,b}$, $W$, $V$, $R$, two conditions need to be satisfied:
1) the minimal value of $f_{WW}$ exists, and 2) the minimal value is positive.

Because $f_{WW}(T_a,T_b,R,W,V)$ is a polynomial, the first condition only requires that
its asymptotic values for large $T_{a,b}$ are positive.  This implies
\begin{flalign}
	f_1>0,\ f_2>0
	\label{eq:f1f2}
\end{flalign}
For the second condition we need to find its minimal value.  Using $T_a^2+T_b^2\geq2T_aT_b$
and $WT_aT_b\geq0$, and defining $Z=T_aT_b>0$, we have
\begin{flalign}
	f_{WW}(T_a,T_b,R,W,V)
	\ge f'_{WW}(Z,W,V)\equiv
	f_2Z^2+(2f_1-W|f_4|)Z+f_3W^2+f_5V^2+f_6
	\label{eq:fww}
\end{flalign}
Minimizing $f'_{WW}$ as a function of $Z$ implies
\begin{flalign}
	Z\to-\frac{2f_1-W|f_4|}{2f_2}
\end{flalign}
However $Z$ must be positive, so the at this step the minimal value of $f'_{WW}$ is
\begin{flalign}
	\min f'_{WW}(Z)=
	\left\{
		\begin{array}{ll}
			W|f_4|>2f_1:
			&W^2f_3+f_5V^2+f_6-\frac{(2f_1-W|f_4|)^2}{4f_2}
			\\
			W|f_4|<=2f_1:
			&W^2f_3+f_5V^2+f_6
		\end{array}
	\right.
\end{flalign}

Now we need to minimize this function w.r.t $W$.
If $|f_4|<=2f_1$, then $W|f_4|<=2f_1$ always holds. In this case
we have
\begin{flalign}
	&\min f'_{ZZ}(Z,W)=\min W^2f_3+f_5V^2+f_6
	\nonumber\\
	=&\left\{\begin{array}{ll}
		f_3\ge0:&f_5V^2+f_6\\
		f_3<0:&f_3+f_5V^2+f_6
	\end{array}\right.
\end{flalign}
On the other hand, if $|f_4|>2f_1$, we define $W_c=2f_1/|f_4|$. In this case $\min f'_{WW}(Z)$
as a function of $W$ is a piecewise function defined on $W\in[0,1]$:
\begin{flalign}
	&\min f'_{WW}(Z)=
	\left\{\begin{array}{ll}
		0<=W<=W_c: &W^2f_3+f_5V^2+f_6\\
		W_c<W<=1: &W^2f_3+f_5V^2+f_6-\frac{(2f_1-W|f_4|)^2}{4f_2}
	\end{array}\right.
\end{flalign}
Now
\begin{itemize}
\item 
if $f_3\ge\frac{f_4^2}{4f_2}$, the above function is a monotonically
increasing function of $W$ in $[0,1]$, so we take $W\to0$ and $\min f'_{WW}(Z,W)=f_5$.

\item
If $\frac{(2f_1-|f_4|)^2}{4f_2}<=f_3<\frac{f_4^2}{4f_2}$, the above function is
a monotonically increasing function of $W$ in $[0,W_c]$ and is a concave
function of $W$ in $(W_c,1]$.  By comparing its value at $W=0$ and $W=1$, i.e.
$f_5V^2+f_6$ and $f_5V^2+f_6+f_3-\frac{(2f_1-|f_4|)^2}{4f_2}$,
we find $\min f'_{WW}(Z,W)=f_5V^2+f_6$.

\item
Similarly, if $0<=f_3<\frac{(2f_1-|f_4|)^2}{4f_2}$,
we find $\min f'_{WW}(Z,W)=f_5V^2+f_6+f_3-\frac{(2f_1-|f_4|)^2}{4f_2}$.

\item
Finally, if $f_3<0$, the function is monotonically decreasing in $[0,1]$,
so $\min f'_{WW}(Z,W)=f_5V^2+f_6+f_3-\frac{(2f_1-|f_4|)^2}{4f_2}$.
\end{itemize}
Finally, the function is linear in $V^2$. Putting pieces together, we obtain
\begin{flalign}
	\min f'_{WW}(W,V,Z)&=
		f_6+\min\left(f_5,0\right)+\min\left(0,f_3-\frac{\max(0,-2f_1+|f_4|)^2}{4f_2}\right)
\end{flalign}
and this needs to be positive. Together with Eq.~(\ref{eq:f1f2}), $f_{WW}>0$ is
equivalent to the combination of the following linear conditions:
\begin{flalign}
	f_1>0,
	f_2>0,
	f_6>0,
	f_3+f_6>0,
	f_5+f_6>0,
	f_3+f_5+f_6>0
	\label{eq:ww1}
\end{flalign}
and the following quadratic inequalities:
\begin{flalign}
	&4f_2(f_3+f_6)>\max(0,|f_4|-2f_1)^2,
	\label{eq:ww2}
	\\
	&4f_2(f_3+f_5+f_6)>\max(0,|f_4|-2f_1)^2,
	\label{eq:ww3}
\end{flalign}

\subsubsection{$ZZ$}
For $ZZ$ and all the other scattering channels, we use a different change of variables
\begin{flalign}
	&\cos^2\phi+\cos^2\psi=2W^2
	\\
	&\cos\phi\cos\alpha-\cos\psi\cos\beta=2WR
\end{flalign}
and again, $T_a$, $T_b$, $W$, $R$ are independent parameters in the range
\begin{flalign}
	T_a,T_b\in[0,+\infty),\quad W\in[0,1],\quad R\in[-1,1]
\end{flalign}
The positivity condition from $ZZ$ channel requires $f_{ZZ}>0$, where
\begin{flalign}
	f_{ZZ}(T_a,T_b,R,W)=g_1(T_a^2+T_b^2)+g_2T_a^2T_b^2+g_3W^2+g_4
	T_aT_bRW+g_5
\end{flalign}
and
\begin{flalign}
&g_1=-2 c_W^6 F_{M,1}+c_W^6 F_{M,7}+c_W^4 s_W^2 F_{M,5}-c_W^2
s_W^4 F_{M,3}\\
&g_2=8 c_W^4 F_{S,0}+8 c_W^4 F_{S,1}+8 c_W^4 F_{S,2}\\
&g_3=32 c_W^8 F_{T,0}+32 c_W^8 F_{T,1}+16 c_W^4 s_W^4
F_{T,5}+16 c_W^4 s_W^4 F_{T,6}+4 c_W^4 s_W^4 F_{T,7}\nonumber\\
&+8 {c_W^8}
 F_{T,2}+8 s_W^8 F_{T,8}+2 s_W^8 F_{T,9}
\\
&g_4=-16 c_W^6 F_{M,0}+4 c_W^6 F_{M,1}-2 c_W^6 F_{M,7}-8 c_W^4
s_W^2 F_{M,4}-2 c_W^4 s_W^2 F_{M,5}\nonumber\\&
-8 c_W^2 s_W^4 F_{M,2}+2 c_W^2 s_W^4 F_{M,3}\\
&g_5=8 c_W^8 F_{T,2}+4 c_W^4 s_W^4 F_{T,7}+2 s_W^8 F_{T,9}
\end{flalign}
Now $f_{ZZ}$ has exactly the same form as $f_{WW}(T_a,T_b,R,W,0)$, as defined
in Eq.~(\ref{eq:fww}).  Therefore we can directly obtain the similar results:
\begin{flalign}
	g_1>0,
	g_2>0,
	g_5>0,
	g_3+g_5>0,
	\label{eq:zz1}
\end{flalign}
and one additional quadratic inequality:
\begin{flalign}
	&4g_2(g_3+g_5)>\max(0,|g_4|-2g_1)^2.
	\label{eq:zz2}
\end{flalign}

\subsubsection{$WZ$}

The positivity condition from the $WZ$ channel requires $f_{WZ}>0$, where
\begin{flalign}
	f_{WZ}(T_a,T_b,R,W)=h'_1T_a^2+h''_1T_b^2+h_2T_a^2T_b^2+h_3W^2+h_4
	T_aT_bRW+h_5
\end{flalign}
where
\begin{flalign}
&h'_1=-2 c_W^4 F_{M,1}+c_W^4 F_{M,7}-c_W^2 s_W^2 F_{M,5}-s_W^4 F_{M,3}
\\
&h''_1=c_W^2 F_{M,7}-2 c_W^2 F_{M,1}
\\
&h_2=4 c_W^2 F_{S,0}+4 c_W^2 F_{S,2}
\\
&h_3=16 c_W^4 F_{T,1}+4 s_W^4 F_{T,6}
\\
&h_4=-2 c_W^3 F_{M,7}-4 c_W s_W^2 F_{M,4}-c_W s_W^2 F_{M,5}
\\
&h_5=4 c_W^4 F_{T,2}+s_W^4 F_{T,7}
\end{flalign}
This is similar to the previous cases, and the only difference is that $T_a^2$
and $T_b^2$ have different coefficients. The minimal value exists if and only if
\begin{flalign}
	h'_1>0,\ h''_1>0,\ h_2>0
\end{flalign}
Similar to the $WW$ and $ZZ$ cases, using $h'_1T_a^2+h''_1T_b^2\ge 2\sqrt{h'_1h''_1}T_aT_b$,
$T_aT_bW\ge0$, and defining $Z\equiv T_aT_b$, $h_1\equiv \sqrt{h'_1h''_1}$, we have
\begin{flalign}
f_{WZ}(T_a,T_b,R,W)\ge f'_{WZ}(Z,W)=h_2Z^2+\left(2h_1-W|h_4|\right)Z+h_3W^2+h_5
\end{flalign}
This function has exactly the same form as $f'_{WW}(Z,W,V=0)$, so we can use directly
use the result there to obtain
\begin{flalign}
	h'_1>0,
	h''_1>0,
	h_2>0,
	h_5>0,
	h_3+h_5>0
	\label{eq:wz1}
\end{flalign}
and one addition inequality:
\begin{flalign}
	&4h_2(h_3+h_5)>\max\left(0,|h_4|-2\sqrt{h'_1h''_1}\right)^2,
	\label{eq:wz2}
\end{flalign}
Because of the square root on the r.h.s., this is a quartic inequality.

\subsubsection{$W\gamma$, $Z\gamma$ and $\gamma\gamma$}

The positivity conditions require $f_{W\gamma,Z\gamma,\gamma\gamma}\ge0$, where
\begin{flalign}
	&f_{W\gamma}(T_a,T_b,R,W)=i_1T_a^2+i_2W^2+i_3
	\\
	&f_{Z\gamma}(T_a,T_b,R,W)=j_1T_a^2+j_2W^2+j_3
	\\
	&f_{\gamma\gamma}(T_a,T_b,R,W)=k_2W^2+k_3
\end{flalign}
where
\begin{flalign}
&i_1=-2 F_{M,1}-F_{M,3}+F_{M,5}+F_{M,7}
\\
&i_2=16 F_{T,1}+4 F_{T,6}
\\
&i_3=4 F_{T,2}+F_{T,7}
\\
&j_1=-2 c_W^2 F_{M,1}-c_W^2 F_{M,3}-c_W^2 F_{M,5}+c_W^2 F_{M,7}
\\
&j_2=32 c_W^4 F_{T,0}+32 c_W^4 F_{T,1}+8 s_W^4 F_{T,8}+2 s_W^4 F_{T,9}+8
c_W^4 F_{T,2}-16 s_W^2c_W^2 F_{T,5}\nonumber\\
&+4 \left(c_W^2-s_W^2\right)^2 F_{T,6}-4 s_W^2 c_W^2 F_{T,7}
\\
&j_3=8 c_W^4 F_{T,2}+2 s_W^4 F_{T,9}+\left(c_W^2-s_W^2\right)^2 F_{T,7}
\\
&k_2=16 F_{T,0}+16 F_{T,1}+4 F_{T,2}+8 F_{T,5}+8 F_{T,6}+2 F_{T,7}+4 F_{T,8}+F_{T,9}
\\
&k_3=4 F_{T,2}+2 F_{T,7}+F_{T,9}
\end{flalign}

These functions have similar forms.  They are linear functions of
$T_a^2\in[0,+\infty)$ and $W^2\in[0,1]$, so they are positive if and only if
\begin{flalign}
	&i_1>0,\ j_1>0,
	\label{eq:a1}
	\\
	&i_3>0,\ j_3>0,\ k_3>0,
	\label{eq:a2}
	\\ 
	&i_2+i_3>0,\ j_2+j_3>0,\ k_2+k_3>0
	\label{eq:a3}
\end{flalign}
i.e.~these three channels only give rise to linear constraints in the parameter space.

\subsection{Polarisation}
\label{subsec:polarisation}

Our final result is the combination of constraints described by
Eqs.~(\ref{eq:ww1}), (\ref{eq:ww2}), (\ref{eq:ww3}), (\ref{eq:zz1}),
(\ref{eq:zz2}), (\ref{eq:wz1}), (\ref{eq:wz2}), (\ref{eq:a1}), (\ref{eq:a2})
and (\ref{eq:a3}).
However, it remains to show that these bounds can be actually
achieved.  This can be demonstrated by showing that the changes of variables we
have used in Eq.~(\ref{eq:variable1}) can be inverted, and therefore the minimum
value of $f$ can always be reached by some physical polarisation state.

Here instead, we will directly give the explicit polarisation vectors $\vec a$
and $\vec b$ that will lead to all the resulting bounds.  These are physically
the most crucial values for $\vec a$ and $\vec b$: once positivity
bounds for them are satisfied, then the same bounds are also satisfied for
arbitrary polarisation.  With them it is also more convenient to check that
all these conditions can be reached, as one can simply plug in the
corresponding $\vec a$ and $\vec b$ values into the original positivity
conditions in Eqs.~(\ref{eq:starts})-(\ref{eq:ends}), and obtain our main
results in Eqs.~(\ref{eq:ww1}), (\ref{eq:ww2}), (\ref{eq:ww3}), (\ref{eq:zz1}),
(\ref{eq:zz2}), (\ref{eq:wz1}), (\ref{eq:wz2}), (\ref{eq:a1}), (\ref{eq:a2})
and (\ref{eq:a3}) right away.  These polarisation values are obtained by
keeping track of the values of $\vec a$ and $\vec b$ when we look for the
minimal values and asymptotic values of $f_{VV'}$.  In a sense they correspond
the boundary of the polarisation space once mapped to the space of $x_{i}$.

For convenience, in Table~\ref{tab:polar} we define notations for the
polarisation cases that are relevant and useful for later presentation.
For example, $L$ and $T$ represent the cases of two longitudinal vectors
and two transversal vectors, and as we will see they naturally lead to
bounds on the $S$-type and $T$-type operators.  The $T$ case can further be
divided into four cases which lead to different constraints.  They represent
two vectors with parallel polarisation, perpendicular polarisation, and those
with $++$ (or $--$) and $+-$ helicity states.
$TL$ and $LT$ denote the scattering between one transversal vector
and one longitudinal vector, and are expected to constrain the $M$-type operators.
The $M$ and $\bar M$ polarisation are mixtures of transversal and longitudinal
states, which could give quadratic or quartic bounds.
The subscripts in this case only represent the relation of the transversal
components $(a_1,a_2)$ and $(b_1,b_2)$.  In particular the $M_I({\bar M}_I)$
is a special case of $M_\parallel({\bar M}_\parallel)$, where a relative
phase of $\pi/2$ between the longitudinal and the transversal parts is required.
Finally, the $M'$ and $\bar M'$ case requires a special relation
between $a_3/\sqrt{a_1^2+a_2^2}$ and $b_3/\sqrt{b_1^2+b_2^2}$.  This applies
to the $WZ$ channel only.

\begin{table}[h]
{\small
\begin{flalign}
\begin{array}{|l|l|l|l|}
	\hline
	&\text{Notation} & \text{Example of $a_i,b_i$ values} & \text{General form}
	\\\hline\hline
	\multirow{7}{*}{Linear}
	&L & 
	\vec{a}=(0,0,1), \vec{b}=(0,0,1) &
	\vec{a}=(0,0,m), \vec{b}=(0,0,n)
	\\\cline{2-4}
	&LT &
	\vec{a}=(0,0,1), \vec{b}=(1,0,0) &
    \vec{a}=(0,0,m_{3}),\vec{b}=(n_{1},n_{2},0)
	\\
	&TL &
	\vec{a}=(1,0,0), \vec{b}=(0,0,1) &
    \vec{a}=(m_{1},m_{2},0),\vec{b}=(0,0,n_{3})
	\\\cline{2-4}
	&T_\perp &
	\vec{a}=(1,0,0), \vec{b}=(0,1,0) &
    \vec{a}=e^{i\rho}(k_{1},k_{2},0), \vec{b}=e^{i\sigma}(k_{2},-k_{1},0)
	\\
	&T_\parallel &
	\vec{a}=(1,0,0), \vec{b}=(1,0,0) &
    \vec{a}=e^{i\rho}(k_{1},k_{2},0), \vec{b}=e^{i\sigma}(k_{1},k_{2},0)
	\\
	&T_{++} &
	\vec{a}=(1,i,0), \vec{b}=(1,-i,0) &
    \vec{a}=e^{i\rho}(k_{1},\pm ik_{1},0), \vec{b}=e^{i\sigma}(k_{1},\mp ik_{1},0)
	\\
	&T_{+-} &
	\vec{a}=(1,i,0), \vec{b}=(1,i,0) &
    \vec{a}=e^{i\rho}(k_{1},\pm ik_{1},0), \vec{b}=e^{i\sigma}(k_{1},\pm ik_{1},0)
	\\\hline\hline
	&M_{+-} &
	\vec{a}=(x,ix,1), \vec{b}=(x,ix,1) &
    \vec{a}=e^{i\rho}(k_{1},\pm ik_{1},e^{i\gamma}k_{3}), \vec{b}=e^{i\sigma}(k_{1},\pm ik_{1},e^{i\gamma}k_{3})
	\\
	&M_\parallel &
	\vec{a}=(x,0,1), \vec{b}=(x,0,1) &
    \vec{a}=e^{i\rho}(k_{1},k_{2},e^{i\gamma}k_{3}), \vec{b}=e^{i\sigma}(k_{1},k_{2},e^{i\gamma}k_{3})
	\\
	&M_I &
	\vec{a}=(ix,0,1), \vec{b}=(ix,0,1) &
    \vec{a}=e^{i\rho}(ik_{1}, ik_{2},k_{3}), \vec{b}=e^{i\sigma}(ik_{1},ik_{2}, k_{3})
	\\\cline{2-4}
	\mbox{Quadratic}&\bar{M}_{+-} &
	\vec{a}=(x,ix,1), \vec{b}=(x,ix,-1) &
    \vec{a}=e^{i\rho}(k_{1},\pm ik_{1},e^{i\gamma}k_{3}), \vec{b}=e^{i\sigma}(k_{1},\pm ik_{1},-e^{i\gamma}k_{3})
	\\
	\mbox{and}&\bar{M}_\parallel &
	\vec{a}=(x,0,1), \vec{b}=(x,0,-1) &
    \vec{a}=e^{i\rho}(k_{1},k_{2},e^{i\gamma}k_{3}), \vec{b}=e^{i\sigma}(k_{1},k_{2},-e^{i\gamma}k_{3})
	\\
	\mbox{quartic}&\bar{M}_I &
	\vec{a}=(ix,0,1), \vec{b}=(ix,0,-1) &
    \vec{a}=e^{i\rho}(ik_{1}, ik_{2},k_{3}), \vec{b}=e^{i\sigma}(ik_{1},ik_{2},-k_{3})
	\\\cline{2-4}
	&M_I' &
	\vec{a}=(ix,0,\sqrt{h_1''}), &
    \vec{a}=e^{i\rho}(ik_{1}, ik_{2},k_{3}), \vec{b}=e^{i\sigma}(ik_{1},ik_{2},k'_{3}),
    \\
    &&\vec{b}=(ix,0,\sqrt{h_1'}) &h_1'k_{3}^{2}=h_1''{k'}_{3}^{2}, 
	 k_3k'_3>0
	\\
	&\bar{M}'_I &
	\vec{a}=(ix,0,\sqrt{h_1''}), &
	\vec{a}=e^{i\rho}(ik_{1}, ik_{2},k_{3}), \vec{b}=e^{i\sigma}(ik_{1},ik_{2},k'_{3}),
	\\&&\vec{b}=(ix,0,-\sqrt{h_1'}) &h_1'k_{3}^{2}=h_1''{k'}_{3}^{2},
	 k_3k'_3<0
	\\\hline
	\hline
	\multicolumn{4}{|c|}{m_i,n_i\in\mathbb{C},\ x,k_{i},k'_{i}\in\mathbb{R},\ 
\rho,\sigma,\gamma\in[0,2\pi)}
	\\\hline
\end{array}
\nonumber
\end{flalign}
}
\caption{\label{tab:polar}
	Notation for relevant polarisation configurations.
The $L$, $LT$, $TL$ and $T_i$ cases lead to linear inequalities,
while the $M_i$ and $\bar M_i$ cases lead to higher order inequalities.
We show example values of $a_i$ and $b_i$, and in addition also
give their general form in the last column.}
\end{table}

We are now ready to collect all the solutions in the previous subsection,
keeping track of the polarisation configurations that allow the conditions to
be reached.  
First, collecting Eqs.~(\ref{eq:ww1}), (\ref{eq:ww2}), (\ref{eq:ww3}),
(\ref{eq:zz1}), (\ref{eq:zz2}), (\ref{eq:wz1}), (\ref{eq:wz2}), (\ref{eq:a1}),
(\ref{eq:a2}) and (\ref{eq:a3}), we find that the linear conditions can be
decomposed into three subspaces, spanned respectively by the three classes of
operator coefficients: the $F_{S,i}$ coefficients, the $F_{M,i}$ coefficients, and
the $F_{T,i}$ coefficients, i.e.
\begin{flalign}
	&M_{S,ij}F_{S,j}>0
	\label{eq:result1starts}
	\\
	&M_{M,ij}F_{M,j}>0
	\\
	&M_{T,ij}F_{T,j}>0
\end{flalign}
where ``>0'' means that each component is positive, and
\begin{flalign}
&F_{S,i}=(F_{S,0},F_{S,1},F_{S,2})^T
\\
&F_{M,i}=(F_{M,0},F_{M,1},F_{M,2},F_{M,3},F_{M,4},F_{M,5},F_{M,7})^T
\\
&F_{T,i}=(F_{T,0},F_{T,1},F_{T,2},F_{T,5},F_{T,6},F_{T,7},F_{T,8},F_{T,9})^T
\end{flalign}
and the $M$ matrices are given by
\begin{flalign}
	M_{S}=\left(
\begin{array}{ccc}
 2 & 1 & 1 \\
 1 & 1 & 1 \\
 1 & 0 & 1 \\
\end{array}
\right)
\begin{array}{l}
WW,L \\
ZZ,L \\
WZ,L \\
\end{array}
\label{eq:MS}
\end{flalign}
\begin{flalign}
	M_{M}=\left(
\begin{array}{cccccccc}
  0 & -2 \cw^4 & 0 & -\sw^4 & 0 & \sw^2\cw^2 & \cw^4 \\
 0 & -2 \cw^4 & 0 & -\sw^4 & 0 & -\sw^2\cw^2 & \cw^4 \\
  0 & -2 & 0 & 0 & 0 & 0 & 1 \\
  0 & -2 & 0 &
   -1 & 0 & 1 & 1 \\
  0 & -2 & 0 &
   -1 & 0 & -1 & 1 \\
\end{array}
\right)
\begin{array}{l}
ZZ,LT\&TL \\
WZ,LT  \\
WW,LT\&TL; WZ,TL\\
W\gamma,LT\\
Z\gamma,LT\\
\end{array}
\label{eq:MM}
\end{flalign}
\begin{flalign}
	M_T=\left(
\begin{array}{ccccccccc}
 0 & 0 & 1 & 0 & 0 & 0 & 0 & 0 \\
 0 & 2 & 1 & 0 & 0 & 0 & 0 & 0 \\
 2 & 1 & 1 & 0 & 0 & 0 & 0 & 0 \\
 8 & 12 & 5 & 0 & 0 & 0 & 0 & 0 \\
 8 \cw^8 & 8 \cw^8 & 4 \cw^8 & 4 \cw^4 \sw^4 & 4 \cw^4 \sw^4 & 2 \cw^4 \sw^4 & 2 \sw^8 & \sw^8 \\
 0 & 0 & 4 \cw^8 & 0 & 0 & 2 \cw^4 \sw^4 & 0 & \sw^8 \\
 0 & 0 & 4 \cw^4 & 0 & 0 & \sw^4 & 0 & 0 \\
 0 & 16 \cw^4 & 4 \cw^4 & 0 & 4 \sw^4 & \sw^4 & 0 & 0 \\
 0 & 0 & 4 & 0 & 0 & 1 & 0 & 0 \\
 0 & 16 & 4 & 0 & 4 & 1 & 0 & 0 \\
 0 & 0 & 8 \cw^4 & 0 & 0 & \left(\cw^2-\sw^2\right)^2 & 0 & 2 \sw^4 \\
 32 \cw^4 & 32 \cw^4 & 16 \cw^4 & -16 \cw^2 \sw^2 & 4 \left(\cw^2-\sw^2\right)^2 & 1-8\sw^2\cw^2 & 8 \sw^4 & 4 \sw^4 \\
 0 & 0 & 4 & 0 & 0 & 2 & 0 & 1 \\
 8 & 8 & 4 & 4 & 4 & 2 & 2 & 1 \\
\end{array}
\right)
\begin{array}{l}
WW,T_{\perp}  \\
WW,T_{++} \\
WW,T_{+-} \\
WW,T_{\parallel} \\
ZZ,T_{\parallel } \\
*ZZ,T_{\perp} \\
*WZ,T_{\perp} \\
WZ,T_{\parallel } \\
W\gamma,T_{\perp} \\
W\gamma,T_{\parallel} \\
*Z\gamma,T_{\perp} \\
Z\gamma,T_{\parallel } \\
\gamma\gamma,T_{\perp} \\
\gamma\gamma,T_{\parallel } \\
\end{array}
\label{eq:MT}
\end{flalign}
For each row, we also give the channel and the corresponding polarisation state
that leads to the very condition, to the right of the matrix.  For
example, the first row in $M_S$ simply means that the bound
$2F_{S,0}+F_{S,1}+F_{S,2}>0$ can be obtained from the $WW$ scattering channel,
by considering the $L$ polarisation, which is described by the first row in
Table~\ref{tab:polar}. One can easily check this by inserting the example
values $\vec{a}=\vec{b}=(0,0,1)$ into the positivity condition
Eq.~(\ref{eq:WW}). The same is
true for other rows and other matrices.  

Several comments are in order for these linear conditions:
\begin{itemize}
	\item Each row in these matrices is a $\vec x(\vec a, \vec b)$
		vector, with $\vec a$ and $\vec b$ taking the polarisation
		displayed to its right.  Together they describe the
		``boundaries'' of all $\vec x(\vec a, \vec b)$, just like
		the endpoints in our first toy exmaple in
		Section~\ref{subsec:toy}.
	\item 
The relevant polarisation states for linear conditions are either
purely longitudinal, or purely transversal, or one vector boson
is longitudinal and the other one transversal.
	\item 
The $T_{++}$ and $T_{+-}$ configurations require $\vec a$ and $\vec b$
to take complex values. This is one of the reasons why allowing for complex
polarisation states will further restrict the parameter space.
They give two additional constraints in the $T$ operator space.
The other bounds are not affected.
	\item 
As discussed in our first toy example, each row of these matrices corresponds
to some $\vec x_i$ vector, and once they are combined we may be able to
simplify further by taking the convex hull of them. 
For $M_S$ and $M_M$, these vectors already form a convex pyramid, so taking a
convex hull does not simplify the description.  The $M_T$ however contains
three vectors that stay inside some pyramids formed by other vectors. They are
labeled by a $*$ before the corresponding channel, and so we can further remove these
conditions.  This will be further discussed in the next section.
	\item $F_{M,0}$, $F_{M,2}$, and $F_{M,4}$ are not subject to any
		linear constraints, but we will see that they are constrained
		by the higher order ones.
\end{itemize}

In addition, there are four higher order inequalities:
\begin{enumerate}
	\item $WW$, $M_{+-}$ and $\bar{M}_{+-}$ polarisation
\begin{flalign}
	4f_2(f_3+f_6)>\max(0,|f_4|-2f_1)^2,
\end{flalign}
	\item $WW$, $M_{\parallel}$ and $\bar{M}_{\parallel}$ polarizatoin
\begin{flalign}
	4f_2(f_3+f_5+f_6)>\max(0,|f_4|-2f_1)^2,
\end{flalign}
	\item $ZZ$, $M_{I}$ and $\bar{M}_{I}$ polarisation
\begin{flalign}
	4g_2(g_3+g_5)>\max(0,|g_4|-2g_1)^2,
\end{flalign}
	\item $WZ$, $M'_{I}$ and $\bar{M}'_{I}$ polarisation
\begin{flalign}
	&4h_2(h_3+h_5)>\max\left(0,|h_4|-2\sqrt{h'_1h''_1}\right)^2,
\end{flalign}
\end{enumerate}
Each condition has two relevant polarisation configurations, corresponding to
the sign of the parameter in the absolute value function.  For example, in the
first condition, the $M_{+-}$ gives
\begin{flalign}
4f_2(f_3+f_6)>\max(0,-f_4-2f_1)^2,
\end{flalign}
while the $\bar M_{+-}$ gives
\begin{flalign}
4f_2(f_3+f_6)>\max(0,f_4-2f_1)^2,
\end{flalign}
The same is true for the other three conditions. In terms of the
original coefficients, these four conditions are given below explicitly:
\begin{flalign}
	&\fbox{$WW,\ M_{+-}\&\bar M_{+-}$}
        \nonumber\\
	&
	32 (2 F_{S,0}+F_{S,1}+F_{S,2}) (2 F_{T,0}+F_{T,1}+F_{T,2})
	\nonumber\\
	& -\max (0,4 F_{M,0}+F_{M,1},-4 F_{M,0}+3 F_{M,1}-2 F_{M,7})^2>0
	\label{eq:result2starts}
	\\
	&\fbox{$WW,\ M_{\parallel}\&\bar M_{\parallel}$}
        \nonumber\\
	& 8 (2 F_{S,0}+F_{S,1}+F_{S,2}) (8 F_{T,0}+12 F_{T,1}+5 F_{T,2})
	\nonumber\\
	&-\max (0,4 F_{M,0}+F_{M,1},-4 F_{M,0}+3 F_{M,1}-2 F_{M,7})^2>0
	\\
	&\fbox{$ZZ,\ M_{I}\&\bar M_{I}$}
        \nonumber\\
	&
8 (F_{S,0}+F_{S,1}+F_{S,2}) \left[4 \cw^8 (2 F_{T,0}+2
F_{T,1}+F_{T,2})+2 \cw^4 \sw^4 (2 F_{T,5}+2 F_{T,6}+F_{T,7})
\right.\nonumber\\&\left.
+\sw^8 (2 F_{T,8}+F_{T,9})\right]
-\max \left[0,2 \left(2 \cw^4 F_{M,0}+F_{M,2}
\sw^4-F_{M,4} \sw^4+F_{M,4} \sw^2\right),
\right.\nonumber\\&\left.
-\cw^4 (4 F_{M,0}-2 F_{M,1}+F_{M,7})-2 \cw^2 F_{M,4} \sw^2-\sw^4 (2
F_{M,2}-F_{M,3})-F_{M,5} \left(\sw^2-\sw^4\right)\right]^2>0
\\
	&\fbox{$WZ,\ M'_{I}\&\bar M'_{I}$}
        \nonumber\\
&
16 (F_{S,0}+F_{S,2}) 
\left[4 \cw^4 (4 F_{T,1}+F_{T,2})+\sw^4 (4 F_{T,6}+F_{T,7})\right]
-\max\Big[0, +2 \cw^2 F_{M,7}
\nonumber\\&
-2 \sqrt{(2 F_{M,1}-F_{M,7}) \left(\cw^4 (2 F_{M,1}-F_{M,7})+\cw^2 F_{M,5} \sw^2+F_{M,3} \sw^4\right)}+4 F_{M,4} \sw^2+F_{M,5} \sw^2,
\nonumber\\&
-2 \cw^2 F_{M,7}-2 \sqrt{(2 F_{M,1}-F_{M,7}) \left(\cw^4 (2 F_{M,1}-F_{M,7})+\cw^2 F_{M,5} \sw^2+F_{M,3} \sw^4\right)}
\nonumber\\&
-4 F_{M,4} \sw^2-F_{M,5} \sw^2\Big]^2>0
\label{eq:result2ends}
\end{flalign}
where the 2nd and the 3rd terms in the $\max$ functions correspond to
$M$ and $\bar M$ polarisations respectively.

Several comments are in order for these higher order inequalities:
\begin{itemize}
	\item These inequalities come from the convex boundaries
		of all $\vec x(\vec a,\vec b)$, e.g.~the circle in our second
		toy example in Section~\ref{subsec:toy}.
\item While $WW$ and $ZZ$ give 3 quadratic inequalities, the $WZ$ channel
	involves square roots in the form given above, so the corresponding
	inequality is a quartic one.
\item 
The polarisation states for quadratic conditions must involve
both longitudinal and transversal components.  Their ratio is only given as a
free parameter $x$.  By plugging in the example
values given in Table~\ref{tab:polar}, one will get a
quadratic function of $x^2$.  By requiring this function to be positive
definite, the corresponding quadratic constraint will be obtained.
\item The $M_{+-}$, $\bar M_{+-}$, $M_I$, $\bar M_I$, $M'_I$ and $\bar M'_I$
	polarisation states require complex values of $\vec a$ and $\vec b$.
	If restricted to real values, we will only have the 2nd quadratic
	condition from $WW$. This is another reason why complex polarisation
	could further constrain the parameter space.
\item
The quadratic constraints from $WW$ and $ZZ$ have the following form:
\begin{flalign}
	\left(\sum_i a_iF_{S,i}\right)
	\left(\sum_i b_iF_{T,i}\right)>
	\max\left(0,\sum_i c_iF_{M,i},\sum_i d_iF_{M,i}\right)^2
	\label{eq:STM}
\end{flalign}
i.e.~a linear combination of $M$ coefficients being less than the geomatic mean of a
linear combination of $S$ coefficients and another combination of $T$
coefficients.  The two combinations on the l.h.s.~have to be individually
positive: this is included already in the linear conditions.  Also, to see
the full structure of this condition, one needs to turn on all three kinds of
operators: $S$, $M$ and
$T$.

\item
If one considers a subspace spanned by $F_{S,i}$ or $F_{T,i}$ operators only,
the r.h.s.~of Eq.~(\ref{eq:STM}) vanishes, while the two sums on the l.h.s.~are
already positive as required by the linear conditions. So the quadratic
function becomes trivial in a subspace without $M$-type operators.  

\item
On the other hand, if we consider a subspace spanned by $F_{M,i}$ operators,
and setting other operators to zero, then the r.h.s.~must also be zero.  This
implies that both
\begin{flalign}
	\sum_i c_iF_{M,i}<0\quad \text{and}\quad \sum_i d_iF_{M,i}<0
	\label{eq:quadtolinear}
\end{flalign}
need to be satisfied.  So in the $F_{M,i}$ space, the quadratic condition
reduces to two linear conditions.

\item Finally
the quartic constraints from $WZ$ is also trivial in the $S$ and $T$ subspaces.
In the $M$ subspace, however, the r.h.s.~involves square roots.  It implies that
\begin{flalign}
	4h'_1h''_1>h_4^2
	\label{eq:wzm}
\end{flalign}
This is a quadratic condition that only involves $F_{M,i}$ operators. So unlike
the $S$ and $T$ subspaces that are only constrained by linear bounds,
the $M$ subspace is described by a set of linear bounds
from $M_M$, another set of linear bounds coming from the quadratic
inequalities of $WW$ and $ZZ$, and a quadratic bound coming from the quartic
inequality of $WZ$.

\end{itemize}

The intriguing fact about positivity bounds is that the parameters
of the underlying theory can take arbitrary values, while still the resulting
coefficients always satisfy the bounds.  In Appendix~\ref{sec:models} we
consider a specific simplified model with BSM resonances, to illustrate
exactly how this happens.

\section{Morphology of the parameter space}
\label{sec:morphology}
Even though the results obtained in the previous section is a complete
description of the parameter space, it is still interesting to study the
morphology of the space carved out by these bounds.  This helps to develop a
physical understanding of the parameter space, which might become useful in
future measurements. It also provides information for choosing reasonable
benchmark cases for theoretical studies.  In this section, we aim to discuss
the shapes of the bounds.  

From the previous section we see that there are basically two kinds of
constraints: the linear ones correspond to pyramids and prisms, while
the others to (approximately) cones.  We discuss them separately.

\subsection{Pyramids}
\label{subsec:pyramids}

As we have seen, linear conditions can be decomposed into three subspaces,
spanned by the $S$-, $M$- and $T$-type operators respectively,
so it is sufficient to consider them separately.

\begin{figure}[ht]
\centering
\includegraphics[width=.5\linewidth]{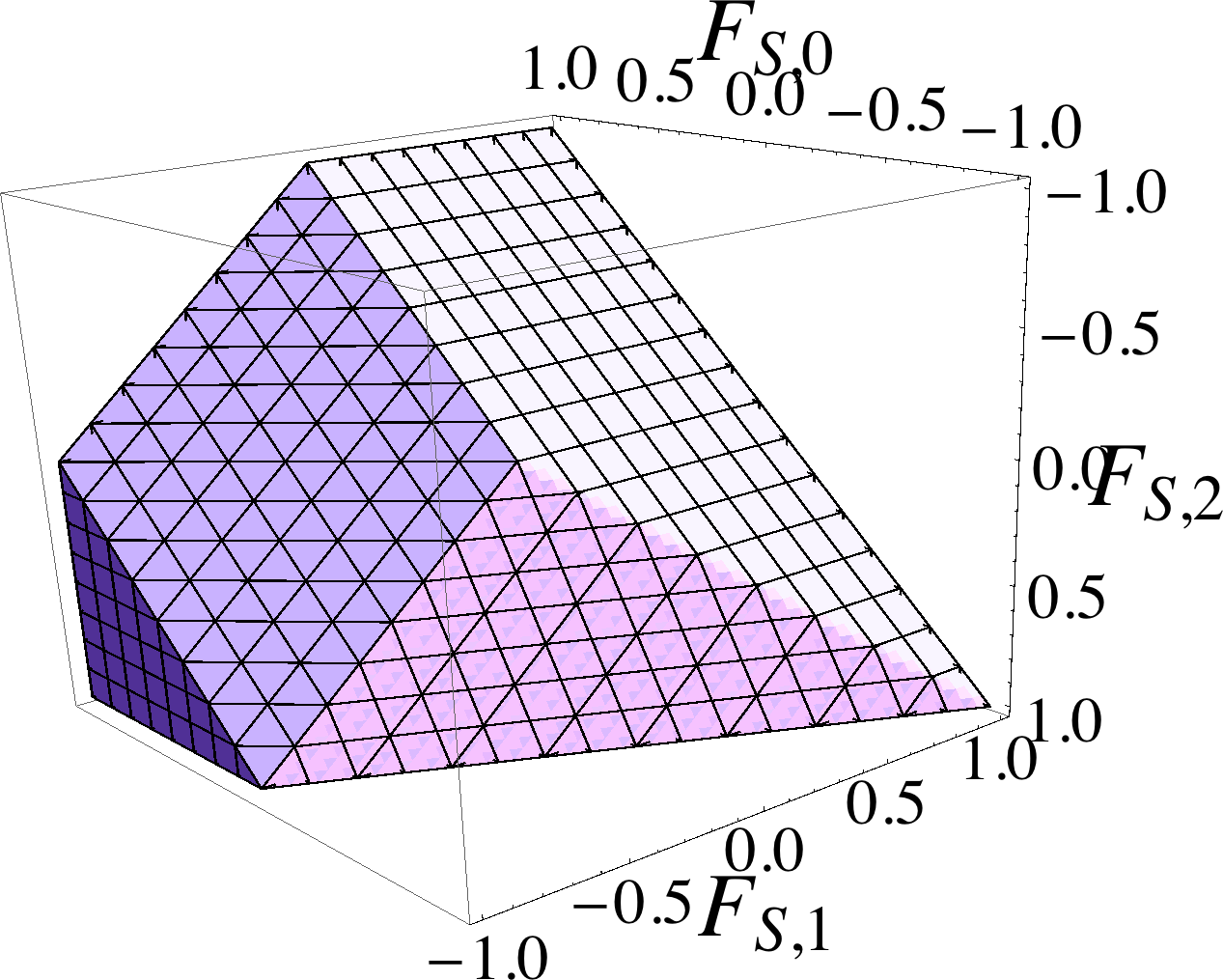}
\caption{Positivity bound described by $M_S$, the linear conditions
in the $S$-subspace.  It is also the actual bound if new degrees of freedom
only couple to longitudinal modes of SM gauge bosons.}
\label{fig:p2}
\end{figure}

The scalar operator space is the most simple one.  The $M_S$ matrix contains three
vectors, and the coefficients $\vec F_S=(F_{S,1},F_{S,2},F_{S,3})$ must be
positive upon projection onto
all three vectors. Each of these conditions describes a plane in the
3-dimensional parameter space.  All together the allowed parameter space is a
3-dimensional triangular pyramid.  This is shown in Figure~\ref{fig:p2}. 
In addition, as we have argued in the previous Section~\ref{subsec:polarisation},
if the theory only generates $S$-type operators, the quadratic and
quartic inequalities are trivially satisfied.  So in this sense
Figure~\ref{fig:p2} also represents the allowed parameter space in models where
new particles only couple to the longitudinal modes through $D_\mu H$.

\begin{figure}[ht]
\begin{minipage}[b]{0.45\linewidth}
\centering
\includegraphics[width=\textwidth]{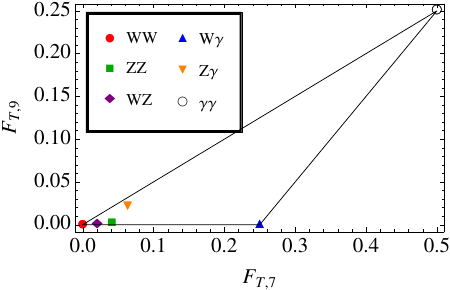}
\caption{The 6 rows in $M_T$ with a $T_\perp$ polarisation
only involve $F_{T,2}$, $F_{T,7}$ and $F_{T,9}$. The corresponding
$\vec x_i$ vectors have 3 components.  The plot shows the
intersection points $P_i=(u,v)$ of $\vec x_i$ on the plane $(1,u,v)$.
$P_{WZ}$, $P_{ZZ}$ and $P_{Z\gamma}$ are inside the triangle formed
by $P_{WW}$, $P_{W\gamma}$ and $P_{\gamma\gamma}$.}
\label{fig:279}
\end{minipage}
\hfill
\begin{minipage}[b]{0.45\linewidth}
\centering
\includegraphics[width=\textwidth]{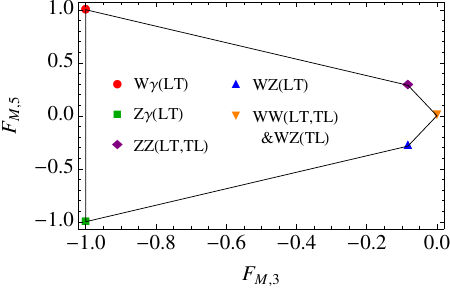}
\caption{The 5 rows in $M_M$ only involve three parameters:
$-2F_{M,1}+F_{M,7}$, $F_{M,3}$ and $F_{M,5}$.  The corresponding $\vec
x_i$ vectors have 3 components.  The plot shows the intersection points
$P_i=(u,v)$ of $\vec x_i$ on the plane $(1,u,v)$.
We see that $P_i=(u,v)$ form a convex pentagon.
}
\label{fig:M}
\end{minipage}
\end{figure}

The $M_T$ matrix contains 14 vectors.  However, 3 of them are redundant
as they can be written as positive linear combinations of the others.
This can be seen by taking the 6 rows in $M_T$ with a $T_\perp$ polarisation.
The corresponding conditions only involve the $F_{T,2}$, $F_{T,7}$
and $F_{T,9}$ coefficients.  Similar to the first toy example discussed in
Section 2, the $F_{T,2}$ component of these vectors is always positive, so we
rescale these vectors so that the $F_{T,2}$ component is always equal to 1,
and plot the 2nd and the 3rd components on a 2D plane, as in Figure~\ref{fig:279}.
We see that the three channels involving a $Z$ boson correspond to points
staying inside the triangle formed by the other three channels.  Therefore,
similar to the discussion in Section~\ref{subsec:toy}, the three vectors from
$ZZ$, $WZ$ and $Z\gamma$ channels with a $T_{\perp}$ polarisation cannot lead
to additional exclusion, and so we can remove them from $M_{T}$. 
They are labeled with $*$ in Eq.~(\ref{eq:MT}).
One can check that the remaining 11 vectors form a convex pyramid, so it is not
possible to further simplify $M_T$.  The resulting constraint is then described
by a pyramid with 11 edges in an 8-dimensional space.

\begin{figure}[ht]
\centering
\includegraphics[width=.6\linewidth]{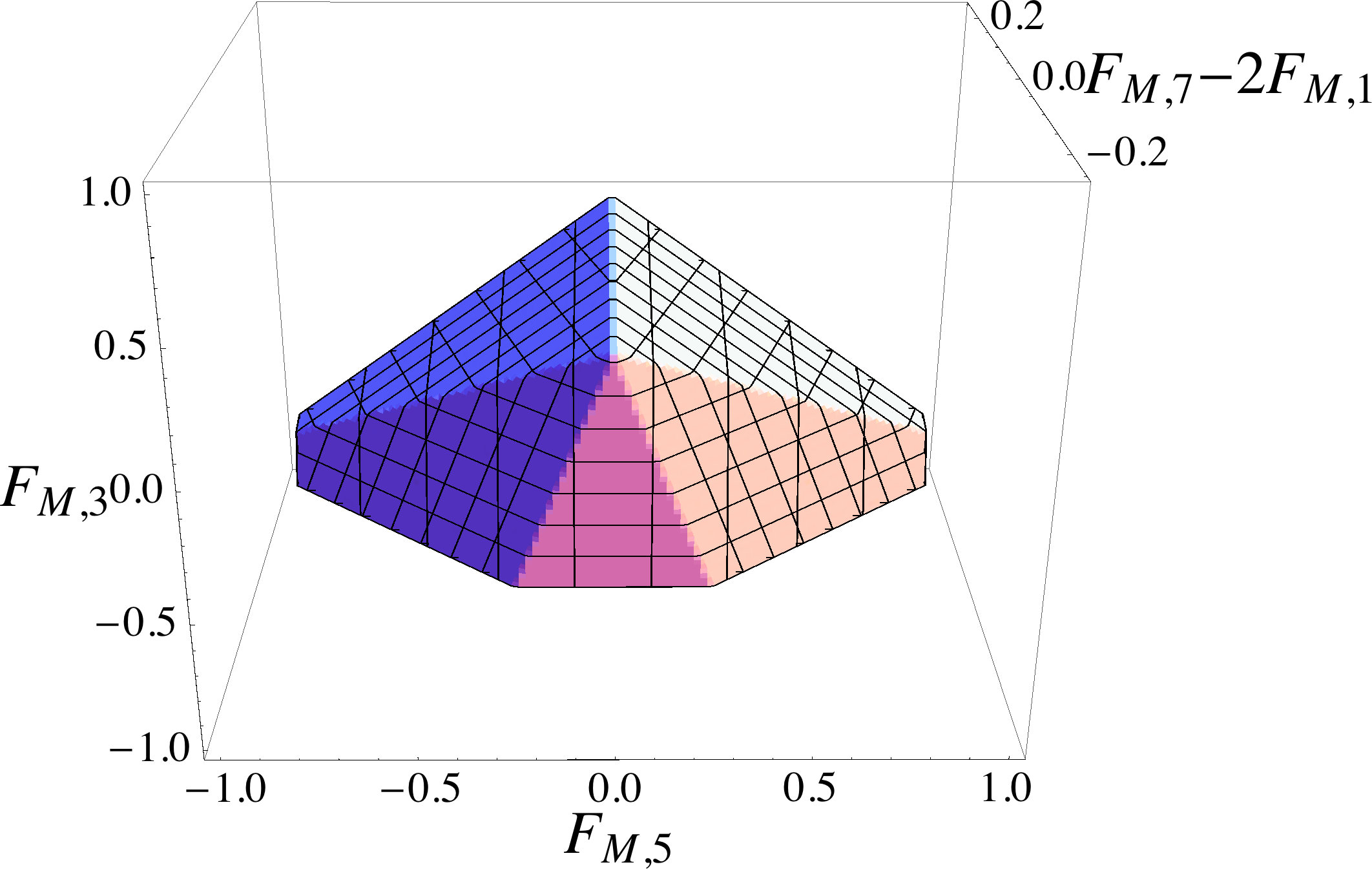}
\caption{Positivity bound described by $M_M$, the linear conditions in the
$M$-subspace.}
\label{fig:M2}
\end{figure}

The $M_M$ matrix contains 5 vectors, but only in a 3-dimensional space, spanned
by $-2F_{M,1}+F_{M,7}$, $F_{M,3}$ and $F_{M,5}$.  They already form a convex
pyramid. Similar to Section 2, this can be seen by rescaling the vectors
so that the $-2F_{M,1}+F_{M,7}$ component is always one, and plotting the
remaining two components on a plane. We show this in Figure~\ref{fig:M}.
Further simplification is not possible. The allowed parameter space
is a pentagonal pyramid, shown in Figure~\ref{fig:M2}.

As we have mentioned, if we consider a case where only the $M$-type operators
are allowed, then quadratic conditions from $WW$ and $ZZ$ apply but they
will be reduced to an additional set of linear conditions in the $M$ subspace,
as shown in Eq.~(\ref{eq:quadtolinear}).
For example, this is the case if the theory has a new vector field 
$\V$ which is a $SU(2)_L$ doublet, see Appendix~\ref{sec:models}.  In these
special cases, we should also consider the additional linear conditions, which
are described by the following matrix
\begin{flalign}
\left(
\begin{array}{ccccccc}
4 & -3 & 0 & 0 & 0 & 0 & 2 \\
-4 & -1 & 0 & 0 & 0 & 0 & 0 \\
4 \cw^4 & -2 \cw^4 & 2 \sw^4 & -\sw^4 & 2 \sw^2 \cw^2 & \sw^2\cw^2 & \cw^4 \\
-2 \cw^4 & 0 & -\sw^4 & 0 & -\sw^2 \cw^2 & 0 & 0 \\
\end{array}
\right)
\begin{array}{l}
	 WW,\bar M_{+-}\&\bar M_{\parallel}
	\\
	 WW,M_{+-}\&M_{\parallel}
	\\
	 ZZ,\bar M_I
	\\
	 ZZ,M_I
\end{array}
\end{flalign}
With this matrix, two rows from $M_M$ can be removed: the $\vec x$ vector from
$ZZ$ with $LT,TL$ polarisation is proportional to the sum of
the $\vec x$ from the same channel but the $M_I$ and the $\bar M_I$
polarisation; the $\vec x$ vector from $WW$($ZZ$) channel with $LT,TL$($TL$)
polarisation is proportional to the sum of the $\vec x$ from the $WW$
channel but with $M_\parallel$ and $\bar M_\parallel$ polarisation.  Combining
the two matrices and removing the two redundant rows gives a new matrix with 7
rows, which involve all $F_{M,i}$ operators.  In addition, the quartic
inequality from $WZ$ also applies, and it is reduced to a quadratic condition
in the $M$ subspace, as in Eq.~(\ref{eq:wzm}).  To sum up, if the theory only
generates $M$-type operators, the coefficients are constrained by 7 linear
inequalities and 1 quadratic inequality.

To better understand the constraints in the $M$ subspace, it is useful to perform
the following change of basis. Defining a set of new coefficients $C_{M,i}$ such
that
\begin{flalign}
	M_{M,i}=R_{ij}C_{M,j}
\end{flalign}
where $R$ is given by
\begin{flalign}
	R=\left(
\begin{array}{ccccccc}
 \frac{1}{24 \cw^2} & -\frac{1}{24 \cw^2} & 0 & 1 & -\frac{\sw^4}{2 \cw^4} & \frac{1}{24 \cw^2} & \frac{1}{48 \cw^2} \\
 -\frac{1}{6 \cw^2} & \frac{1}{6 \cw^2} & -\frac{1}{2 \cw^2} & -4 & 0 & -\frac{1}{6 \cw^2} & -\frac{1}{12 \cw^2} \\
 0 & 0 & 0 & 0 & 1 & 0 & -\frac{\cw^2}{8 \sw^4} \\
 0 & -\frac{\cw^2}{\sw^4} & \frac{\cw^2}{\sw^4} & 0 & 0 & 0 & 0 \\
 \frac{5}{12 \sw^2} & -\frac{5}{12 \sw^2} & 0 & \frac{4 \cw^2}{\sw^2} & 0 & -\frac{1}{12 \sw^2} & \frac{1}{12 \sw^2} \\
 -\frac{1}{\sw^2} & \frac{1}{\sw^2} & 0 & 0 & 0 & 0 & 0 \\
 -\frac{1}{3 \cw^2} & \frac{1}{3 \cw^2} & 0 & -8 & 0 & -\frac{1}{3 \cw^2} & -\frac{1}{6 \cw^2} \\
\end{array}
\right)
\end{flalign}
Then the 7 linear conditions in the $M$ space can be written as
\begin{flalign}
	&M'_{M,ij}C_{M,j}>0,
	\\
	&M'_M=\left(
\begin{array}{cccccccc}
  1 & 0 & 0 & 0 & 0 & 0 & 0 \\
  -\sw^2 & 1 & \sw^2/\cw^2-1 & 0 & 0 & 0 & 0 \\
  \sw^2 & \cw^2-\sw^2 & \sw^2/\cw^2-1 & 0 & 0 & 0 & 0 \\
  0 & 0 & 3 \cw^2 & 0 & -4 \sw^4 & 0 & 0 \\
  0 & 0 & \cw^2 & 0 & 4 \sw^4 & 0 & 0 \\
  0 & 1 & 0 & 12 \cw^2 & 0 & 0 & 0 \\
  -1 & 1 & 0 & -12 \cw^2 & 0 & 0 & 0 \\
\end{array}
\right)
\begin{array}{l}
	WZ,LT\\
	W\gamma,LT\\
	Z\gamma,LT\\
	WW,\bar M_{+-}\&\bar M_{\parallel} \\
	WW,M_{+-}\&M_{\parallel} \\
	ZZ,\bar M_I \\
	ZZ,M_I
\end{array}
\end{flalign}
and the quadratic condition from $WZ$ is simply $4C_{M,1}C_{M,3}>C_{M,6}^2$.
We can see that $C_{M,7}$ is a free parameter that is not constrained.
This is the only degree of freedom that is free in the full 18-dimensional
parameter space.  For the rest of the conditions, the 7 linear ones form a
pyramid with 7 edges in a 5-dimensional subspace, while the quadratic
condition carves out a cone in a 3-dimensional subspace that involves the 6th
direction and is partially orthogonal to the pyramid. It is difficult to
display the shape, but a parameter scan can be easily performed in this basis.
Let
\begin{flalign}
	C_{M,1},C_{M,3}\in [0,+\infty)
\end{flalign}
vary freely.  Then $C_{M,2}$ is bounded from below by the 2nd, the 3rd, and
the combination of the last two linear conditions:
\begin{flalign}
	&C_{M,2}>\max\left(\sw^2C_{M,1}+\frac{\cw^2-\sw^2}{\cw^2}C_{M,3},
	-\frac{\sw^2}{\cw^2-\sw^2}C_{M,1}+\frac{1}{\cw^2}C_{M,3},
	\frac{1}{2}C_{M,1}\right)
\end{flalign}
Then the rest coefficients are bounded by the above three from both sides:
\begin{flalign}
	&C_{M,4}\in \left(-\frac{1}{12\cw^2}C_{M,2},\frac{1}{12\cw^2}\left( C_{M,2}-C_{M,1} \right)\right)
	\\
	&C_{M,5}\in \left(-\frac{\cw^2}{4\sw^4}C_{M,3},\frac{3\cw^2}{4\sw^4}C_{M,3}\right)
	\\
	&C_{M,6}\in \left(-2\sqrt{C_{M,1}C_{M,3}},2\sqrt{C_{M,1}C_{M,3}}\right)
\end{flalign}

\subsection{Cones}
\label{subsec:cones}

\begin{figure}[ht]
\begin{minipage}[b]{0.5\linewidth}
\centering
\includegraphics[width=\textwidth]{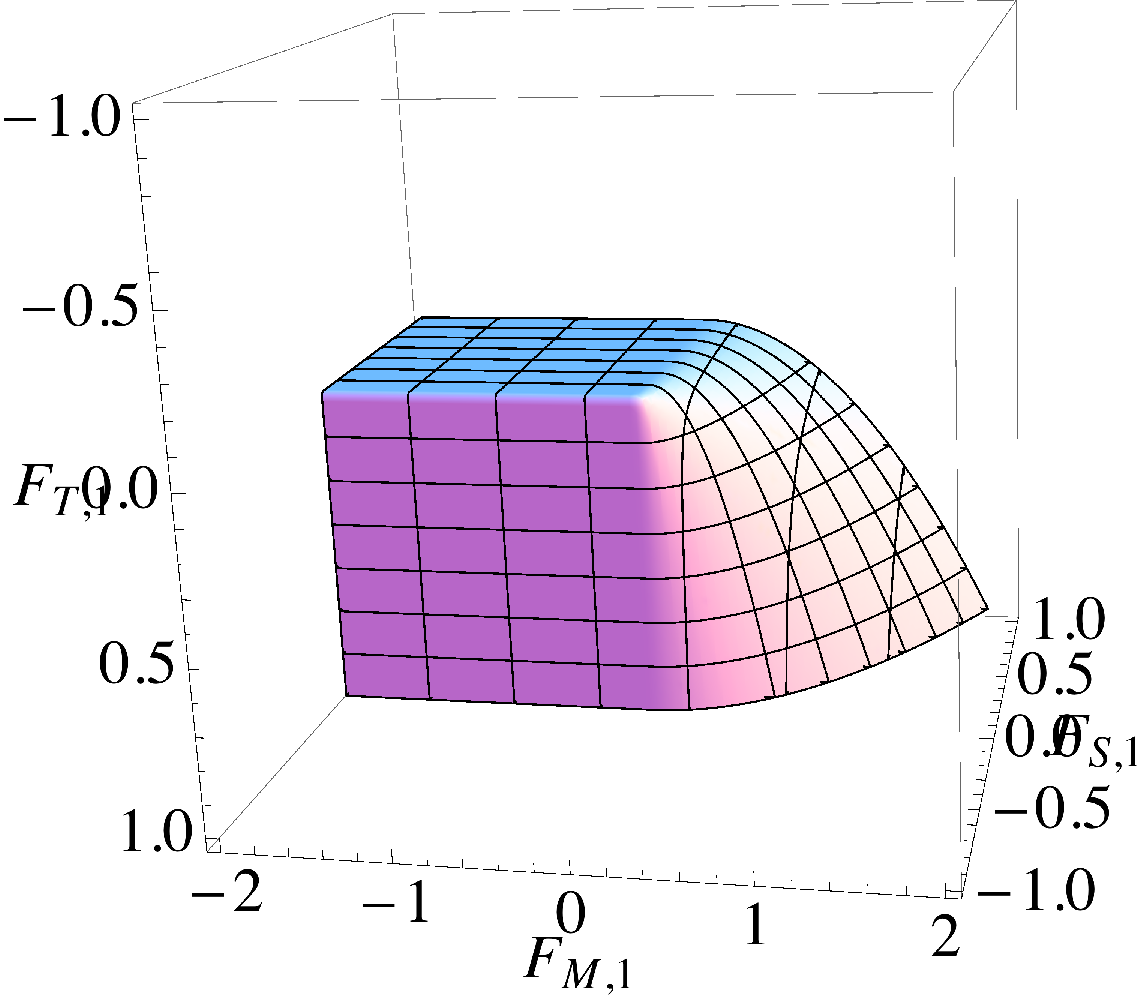}
\caption{Positivity bound from the quadratic inequality in the $WW$ channel,
	in the subspace spanned by $F_{S,1}$, $F_{M,1}$ and $F_{T,1}$.}
\label{fig:c2}
\end{minipage}
\hfill
\begin{minipage}[b]{0.44\linewidth}
\centering
\includegraphics[width=\textwidth]{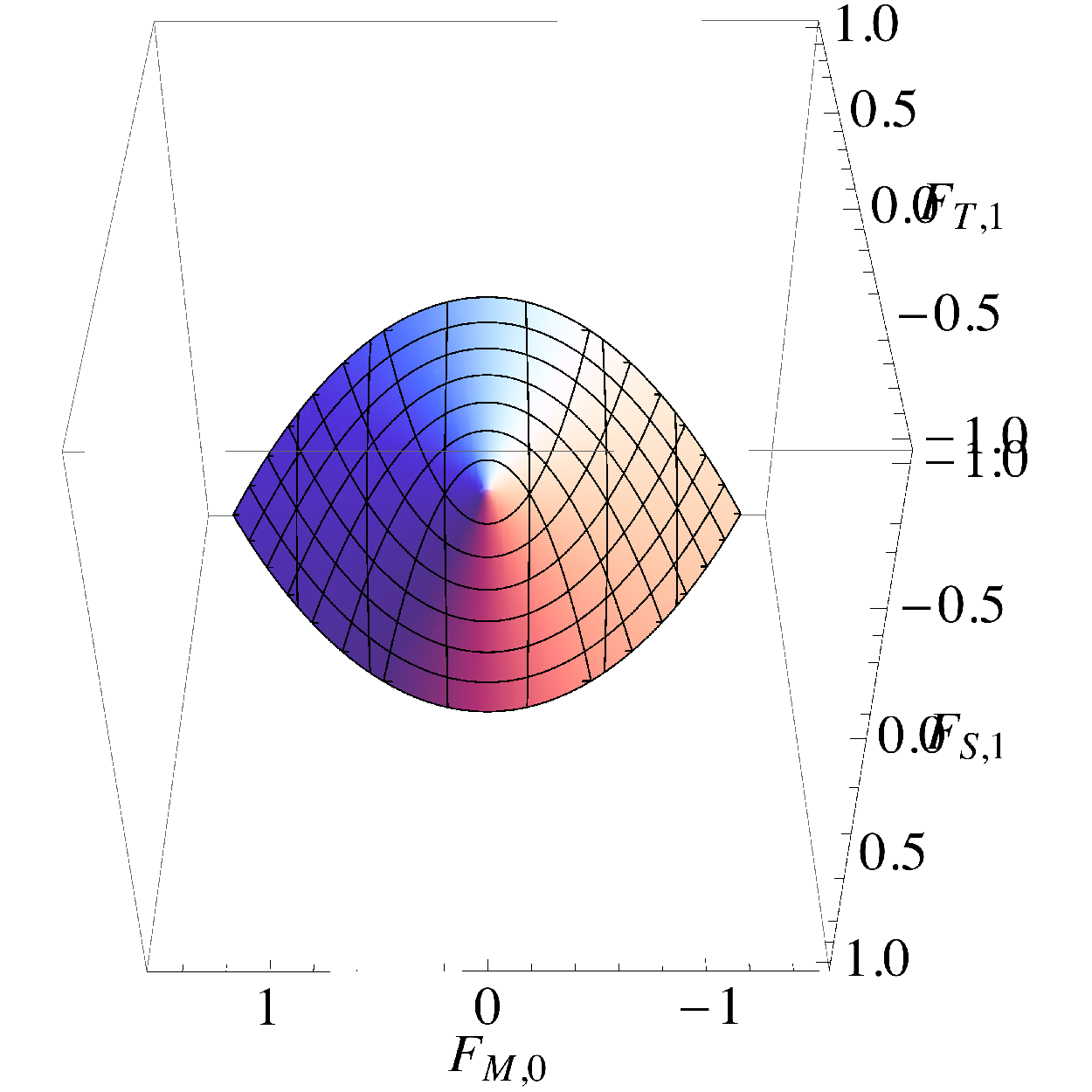}
\caption{Same as Figure~\ref{fig:c2}, but
	in the subspace spanned by $F_{S,1}$, $F_{M,0}$ and $F_{T,1}$.
	\vspace{13pt}
}
\label{fig:c3}
\end{minipage}
\end{figure}

Now let us focus on the quadratic conditions.  These conditions
typically have the following form
\begin{flalign}
	4xy>\max(0,z_1,z_2)^2
	\label{eq:quadraticgeneral}
\end{flalign}
where $x$ and $y$ are linear combinations of $S$ and $T$ operator coefficients
respectively, and $z_{1,2}$ are linear combinations of $M$ coefficients.  The
shape of this constraint can be fully seen only when we turn on three different
types of operators simultaneously.  Since $x>0$ and $y>0$ are always included
in the linear conditions, this condition can be considered as the intersection
of two constraints:
\begin{flalign}
	2\sqrt{xy}>z_1\quad \mbox{and}\quad 2\sqrt{xy}>z_2
\end{flalign}
A constraint like $2\sqrt{xy}>z$ is trivial for a negative $z$, while for
a positive $z$ it gives a half cone:
\begin{flalign}
	&u^2-v^2>z^2,\ z>0
	\\
	&\mbox{where}\ u\equiv x+y\ \mbox{and}\ v\equiv x-y
\end{flalign}
As an example, consider the first quadratic condition from $WW$,
in the subspace spanned by $F_{S,1}$, $F_{M,1}$ and $F_{T,1}$:
\begin{flalign}
	32F_{S,1}F_{T,1}>\max(0,3F_{M,1},F_{M,1})^2
	=\left\{\begin{array}{ll}0 &
		F_{M,1}<0\\9F_{M,1}^2&F_{M,1}>=0\end{array}\right.
\end{flalign}
We show this constraint in Figure~\ref{fig:c2}. It is a combination
of a triangular pyramid and a half cone.

\begin{figure}[ht]
\begin{minipage}[b]{0.47\linewidth}
\centering
\includegraphics[width=\textwidth]{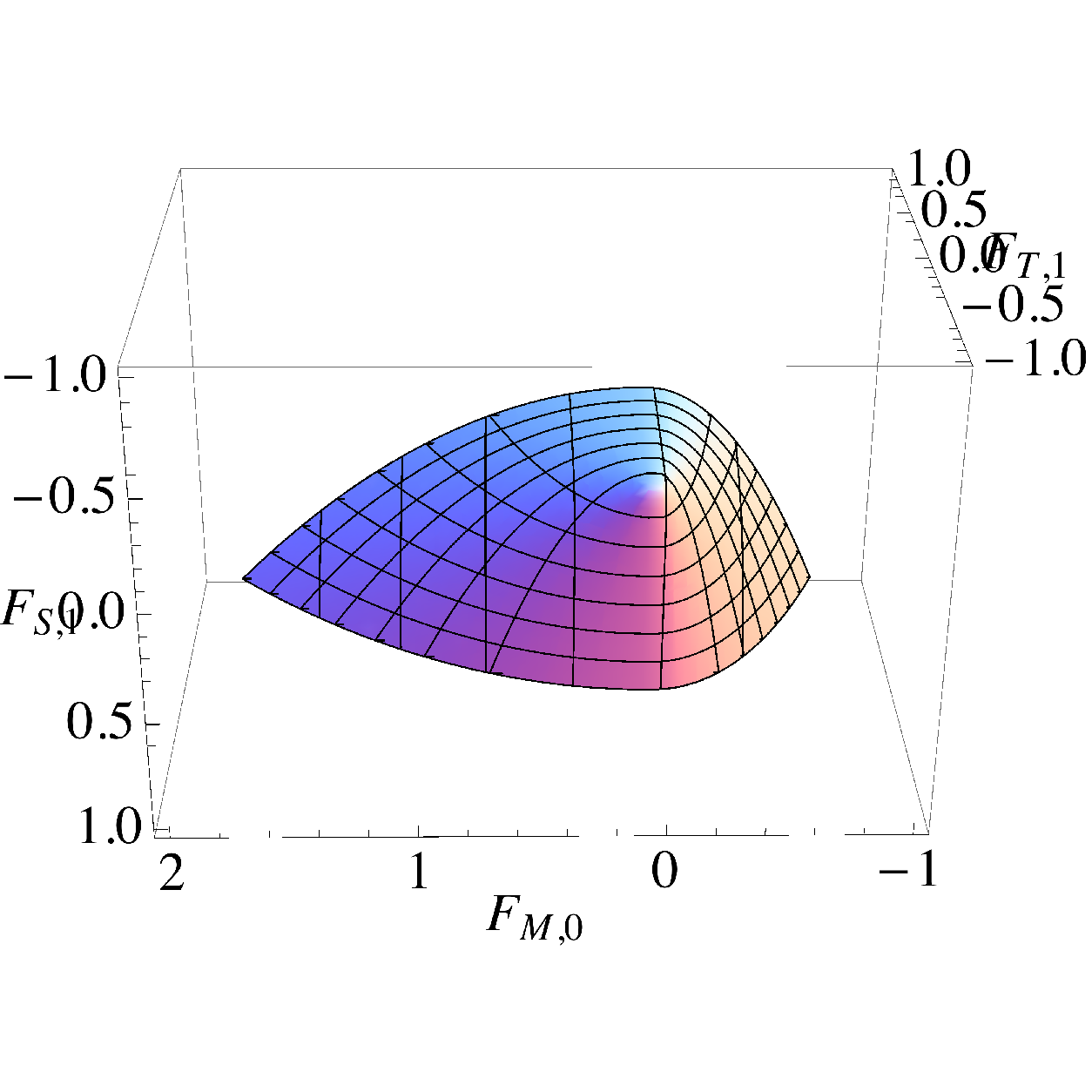}
\caption{Similar to Figure~\ref{fig:c3}, but imposing
a restriction $F_{M,0}+F_{M,1}=0$.
}
\label{fig:c4}
\end{minipage}
\hfill
\begin{minipage}[b]{0.47\linewidth}
\centering
\includegraphics[width=\textwidth]{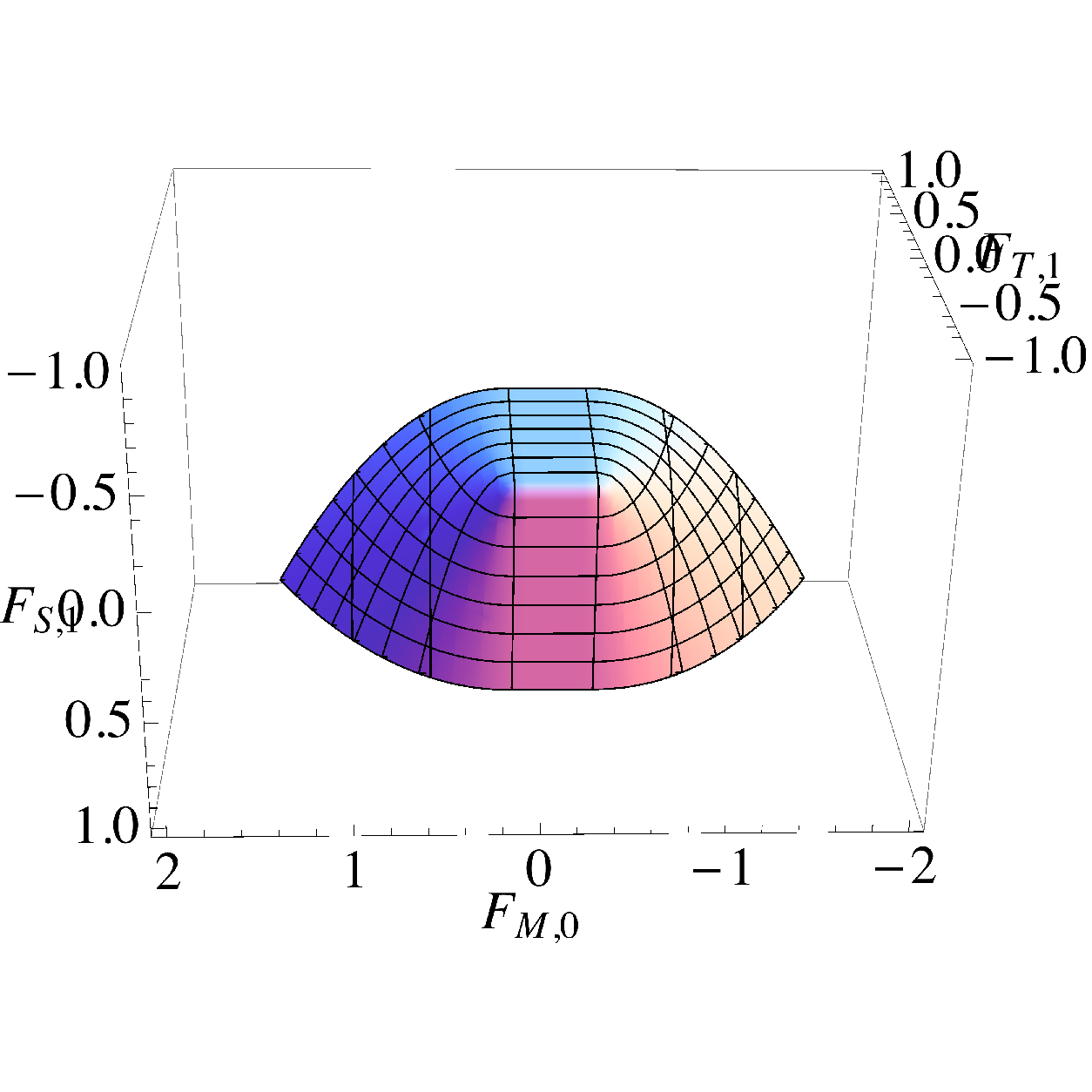}
\caption{Similar to Figure~\ref{fig:c3}, but imposing
a restriction $F_{M,1}=-1/2$.}
\label{fig:c5}
\end{minipage}
\end{figure}

Sometimes the bounds from $z_1$ and $z_2$ combine to form a full cone.
This is the case for e.g.~$F_{S,1}$, $F_{M,0}$ and $F_{T,1}$,
\begin{flalign}
	2F_{S,1}F_{T,1}>\max(0,-F_{M,0},F_{M,0})^2
	=F_{M,0}^2
\end{flalign}
The corresponding bound is shown in Figure~\ref{fig:c3}, which is
now symmetric on both sides.
It is also possible that the two sides of the cone are not symmetric.
Consider a case where we impose a restriction $F_{M,0}+F_{M,1}=0$,
then the quadratic condition becomes:
\begin{flalign}
	32F_{S,1}F_{T,1}>\max(0,3F_{M,0},-7F_{M,0})^2
\end{flalign}
We show this ``asymmetric cone'' in Figure~\ref{fig:c4}.

\begin{figure}[ht]
\centering
\includegraphics[width=.49\linewidth]{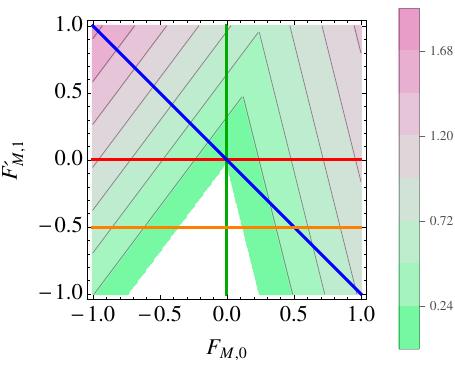}
\includegraphics[width=.49\linewidth]{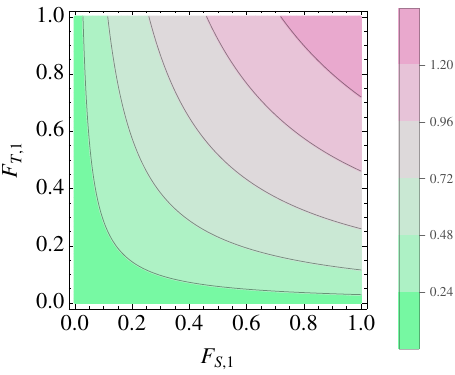}
\caption{Contour plots for Eqs.~(\ref{eq:4dcone1}) and (\ref{eq:4dcone2}),
which describes the quadratic inequality from the $WW$ channel,
in the subspace of $F_{S,1}$, $F_{T,1}$, $F_{M,0}$ and $F_{M,1}$.
}
\label{fig:contour1}
\end{figure}

The full structure of the bound, described Eq.~(\ref{eq:quadraticgeneral}),
become manifest only in a 4-dimensional space, because in general $z_1$ and
$z_2$ are linearly independent. Consider the subspace spanned
by $F_{S,0}$, $F_{M,0}$, $F_{M,1}$ and $F_{T,1}$.  The condition becomes
\begin{flalign}
	32F_{S,1}F_{T,1}>\max(0,4F_{M,0}+F_{M,1},-4F_{M,0}+3F_{M,1})^2
\end{flalign}
It is not possible to make a 4D plot.  However, to visualize this,
we can write the boundary as:
\begin{flalign}
	&\max(0,4F_{M,0}+F_{M,1},-4F_{M,0}+3F_{M,1})=4w
	\label{eq:4dcone1}
	\\
	&\sqrt{2F_{S,1}F_{T,1}}=w
	\label{eq:4dcone2}
\end{flalign}
and make contour plots for $w$ in the $F_{M,0}-F_{M,1}$ space and the
$F_{S,1}-F_{T,1}$ space.  These two plots are shown in
Figure~\ref{fig:contour1}.  The second contour plot describes a half
cone, where $w$ is a direction perpendicular to its axis.
The first one tells us that the constraint is the intersection of two half cones,
with their $w$ directions pointing to the $F_{M,0}=4F_{M,1}$ direction and
to the $F_{M,0}=-\frac{4}{3}F_{M,1}$ direction, respectively.
Furthermore, in Figure~\ref{fig:contour1} left, if we take the green, the red
and the blue axes separately and make a 3-D plot together with $F_{S,0}$
and $F_{S,1}$, we will get the shapes of a half-cone, a symmetric cone
and an asymmetric cone, which explains Figures~\ref{fig:c2}, \ref{fig:c3}
and \ref{fig:c4}.  We could also consider the orange axis which goes off
the origin, $F_{M,1}=-1/2$.  This will lead to the shape in
Figure~\ref{fig:c5}, with a plateau between two half cones.

\begin{figure}[ht]
\centering
\includegraphics[width=\linewidth]{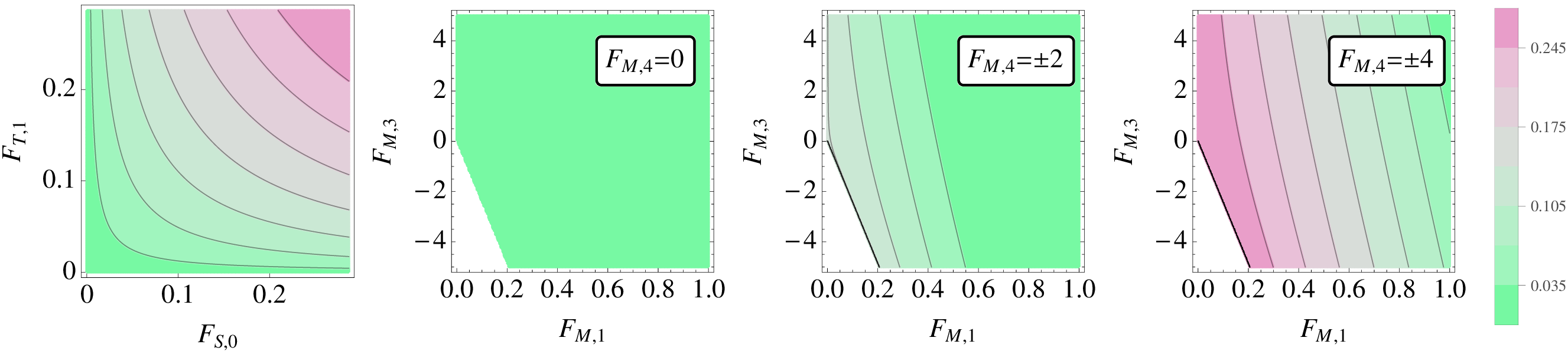}
\caption{Similar to Figure~\ref{fig:contour1}, but for the quartic inequality
from the $WZ$ channel, in the subspace of $F_{S,0}$, $F_{T,1}$,
$F_{M,1}$ and $F_{M,3}$, with different values of $F_{M,4}$.}
\label{fig:contour2}
\end{figure}

The other quadratic constraints from $WW$ and $ZZ$ have similar forms, so
we do not discuss them separately. The one from $WZ$ is more complicated,
as the terms in the $\max$ function involve additional square roots. This
inequality describes a 5-dimensional object.  In Figure~\ref{fig:contour2}, 
we show it in the space of $F_{S,0}$, $F_{T,1}$, $F_{M,1}$, $F_{M,3}$, 
and $F_{M,4}$, by plotting the first four coefficients in a similar way as in
Figure~\ref{fig:contour1}, but for different values of $F_{M,4}$, .

\section{Volumes of allowed parameter spaces}
\label{sec:volume}

While the shapes of the bounds describe the relations between different
parameters, their constraining power is more reflected by the volume of the
allowed higher-dimensional parameter space.  For example,
Ref.~\cite{Durieux:2017rsg} proposed to use the GDP parameter to assess the
overall strength of a $n$-dimensional constraint, which is proportional to
the $n$th root of the volume of the constrained parameter space.

However, this is not a useful measure for the problem at hand, because the
positivity conditions only constrain the possible directions in which the
deviation could occur, so the volume in the regular sense is always infinity.
For this reason, we define the ``volume'' of the constraint by the solid angle
spanned by the allowed parameter space.  They can be easily computed in a Monte
Carlo approach.  A random vector of coefficients
$\vec{F}=(F_{S,0},F_{S,1},\cdots)$ can be
generated by combining 18 independent random numbers under the Gaussian
distribution, and normalizing them so that the magnitude is equal to one.
Vectors generated in this way are uniformly distributed on a unit sphere in the
18-D parameter space. With a random vector $\vec{F}$ we can check explicitly if a
certain positivity condition is satisfied.  By generating many of these vectors,
we can count the fraction of the ones that satisfy this condition.  In the
rest of this section, we will refer to this fraction as the volume of the
parameter space allowed by a given positivity condition, and we denote it by $V$.
Of course, the volume defined in this way depends on the relative rescaling of
operators, and needs to be interpreted with a grain of salt.

The linear conditions form 3 pyramids in 3 subspaces, $S$, $M$ and $T$, that
are completely orthogonal to each other, so their volumes can be computed
separately.  We find that
\begin{flalign}
	V_S=0.38,\quad V_M=0.34,\quad V_T=0.16
\end{flalign}
and all together they constrain the full parameter space to
\begin{flalign}
	V_SV_MV_T=0.022
\end{flalign}
i.e.~2.2\% of the full parameter space.
In other words, the linear conditions solely are sufficient to reduce the
full parameter space by two orders of magnitude.

We have mentioned that there are two linear inequalities derived
from $WW$ channel that require complex polarisation.  If we restrict to only
real polarisation, the $T$ subspace will have a slightly weaker constraining
power, which leads to $V_{T,real}=0.17$.  The size of the total constraint
will become 0.023.

\begin{table}[ht]
	\centering
	\begin{tabular}{l|ccccc}
& $WW$ 1st & $WW$ 2nd & $ZZ$ & $WZ$ & All
\\ \hline
Volume & 0.20&0.22&0.20&0.12&0.05
\\
Improvement &12\% & $\approx$0\% & 10\% & 64\% & 94.5\%
	\end{tabular}
	\caption{The volume of parameter space allowed by each quadratic and
		quartic positivity condition, together
		with the corresponding improvement when all other bounds
	are turned on.  ``All'' is the combination of all quadratic and quartic
conditions.}
	\label{tab:quad}
\end{table}

Now let us consider the quadratic and quartic conditions.  There are two from
$WW$, one from $ZZ$, and one from $WZ$.
In Table~\ref{tab:quad}, the first row gives the volume of each condition, while
the second line gives the corresponding improvement due to this condition.  The
latter is defined by comparing two volumes: $V_{total}$ which is volume of the
combination of all quadratic and quartic conditions, and $V_{remove}$ which is the volume
of the combination of all conditions except for one that is removed.
The improvement is defined as $1-V_{total}/V_{remove}$, which gives us an idea
of how useful a specific bound is, in the presence of all other bounds.
From the table we see that the constraining power of each quadratic or quartic one
is quite good, but they are not orthogonal to each other, so the volume
of all conditions combined are weaker than the linear case, at about 5\% of the
total parameter space.
In particular, from the second row we can see that once three conditions are
taken into account, the last one does not improve much except for $WZ$.  The
second condition from $WW$ almost leads to no improvement.  We checked that
this improvement is not strictly zero, so the condition is still an independent
one, but the size of the improvement only shows up in the second digit after
the decimal point.  This means that most part of the parameter space allowed by
the other three constraints automatically satisfies this condition.

\begin{table}[ht]
	\centering
	\begin{tabular}{l|cccccccc}
& S & M & T & $WW$ 1st & $WW$ 2nd & $ZZ$ & $WZ$ & All
\\ \hline
Volume & 0.38 & 0.34 & 0.16 & 0.20&0.22&0.20&0.12&0.021
\\
Improvement &0 & 27\% & 49\% &1.1\% & $\approx$0\% & 2.5\% & 0.1\% & 97.9\%
	\end{tabular}
	\caption{Similar to Table~\ref{tab:quad}, but for all linear, quadratic
and quartic conditions. The S, M, and T column represent the linear conditions
in $S$-, $M$- and $T$-subspaces, while the next four columns represent quadratic
and quartic conditions as in Table~\ref{tab:quad}.}
	\label{tab:volume}
\end{table}

Finally in Table~\ref{tab:volume}, we show a similar table but including
both linear and quadratic conditions.  The first line compares the volumes
of each condition.  The volume of all conditions together is 2.1\% of the total,
only slightly better than the case of linear conditions only, which is 2.2\%.
So even though the quadratic conditions themselves could carve out a parameter
space with a volume of only 5\%, the impact is tiny once the linear conditions
are already taken into account.  In the second row we show the improvements of
each condition. As expected we see that the improvements due to the quadratic
and quartic
conditions are small.  The linear conditions in the $S$ subspace lead to no
improvement, because the three corresponding linear inequalities are required
to define the quadratic/quartic conditions from $WW$, $ZZ$ and $WZ$ respectively.
Finally, if we only use real polarisation, the volume of the total constraint
is 2.3\%.

\section{Positivity bounds for one, two, and three operators}
\label{sec:benchmark}

From the SMEFT point of view, the most useful and correct bounds are the global
ones, obtained with all relevant EFT operators switched on. 
However, studying lower dimensional subspace of the full parameter space
is always illustrative.  It could help to understand for example the constraining
power on certain directions, or correlations between different parameters.
In addition, in VBS studies, experimental limits have been presented only in
terms of individual or at most two-dimensional bounds, even though
more global analyses can be foreseen in the future.

To better understand the positivity constraints in lower dimensional parameter
space, and to facilitate a quick comparison with the current experimental results,
in this section we will explicitly list the positivity bounds with
one, two or three operators switched on at a time. These results could also be
used as a guidance for future searches.

\subsection{One operator at a time}

We first consider the case where only one operator is nonzero. This is the easiest way to see the impact of positivity,
as it can be directly confronted with experimental limits, most of which are
presented as individual limits \cite{qgclimits}. In Table \ref{tab:oneatatime},
we list the positivity constraints on all individual operators. In particular,
$F_{T,5}, F_{T,7}, F_{M,0}, F_{M,2}, F_{M,4}, F_{M,5}$ cannot individually take
any nonzero value. Individual limits on these operators do not have a clear
meaning, because there is no UV completion that can generate any of these
operators alone.  All other coefficients are constrained to have a certain
sign, either positive or negative, and so their individual bounds could have
been presented as one-sided limits.

\begin{table}
\label{tab:oneatatime}
    \centering
	\begin{tabular}{ccccccccc}
		\hline
	  $F_{S,0}$
	& $F_{S,1}$
	& $F_{S,2}$
	& $F_{M,0}$
	& $F_{M,1}$
	& $F_{M,2}$
	& $F_{M,3}$
	& $F_{M,4}$
	& $F_{M,5}$
		\\\hline
		+ & + & + & \ding{55} & $-$ & \ding{55} & $-$ &\ding{55} & \ding{55}
		\\\hline\hline
	  $F_{M,7}$
	& $F_{T,0}$
	& $F_{T,1}$
	& $F_{T,2}$
	& $F_{T,5}$
	& $F_{T,6}$
	& $F_{T,7}$
	& $F_{T,8}$
	& $F_{T,9}$
		\\\hline
		+ & + & + & + & \ding{55}  & + & \ding{55} &+ & +
		\\\hline
	\end{tabular}
	\caption{\label{tab:one}
Positivity constraints on individual VBS operator coefficients.
$+/-$ means the coefficient must be non-negative or non-positive.
\ding{55} means that nonzero value is not allowed.  }
\end{table}

\subsection{Two operators at a time}

Two-operator constraints have been presented for example in
Refs.~\cite{CMS:2018ysc,Khachatryan:2014sta,Aad:2014zda}. 
Since all positivity inequalities are homogeneous polynomials,
they always carve out triangle areas in the 2D parameter spaces.

Consider the three types of operators: $S$-type, $M$-type and $T$-type.  When
only two operators are turned on, there is no correlation between operators
from two different categories. This is obvious for the linear conditions
in Eqs.~(\ref{eq:MS}), (\ref{eq:MM}) and (\ref{eq:MT}).
To see that this is also true for higher order inequalities, we note that
both quadratic and quartic inequalities can be written in the form of
\begin{flalign}
	f(F_{S,i})g(F_{T,i})>\max(0,h_1(F_{M,i}),h_2(F_{M,i}))^2
\end{flalign}
where $f$, $g$ and $h_{1,2}$ are homogeneous functions of degree 1.
If $M$-type operator is not turned on, the r.h.s.~is 0 and so the inequality is trivial.
If a $M$-type operator is turned on with another $S$-type or $T$-type operator,
the l.h.s.~is 0 and so the inequality reduces to $h_{1,2}(F_{M,i})<0$.  In either case
there is no correlation across different types.
This means that we only need to show bounds for operator pairs from the same category.
For those from different ones, one can simply refer to Table~\ref{tab:one}.

These 2D bounds are shown in Figures~\ref{fig:2DS}, \ref{fig:2DM}, and \ref{fig:2DT}.
The exact description of the bounds are given in Appendix~\ref{subsec:A11},
together with the corresponding polarisation configuration.

\begin{figure}[ht]
\centering
\includegraphics[width=0.3\textwidth]{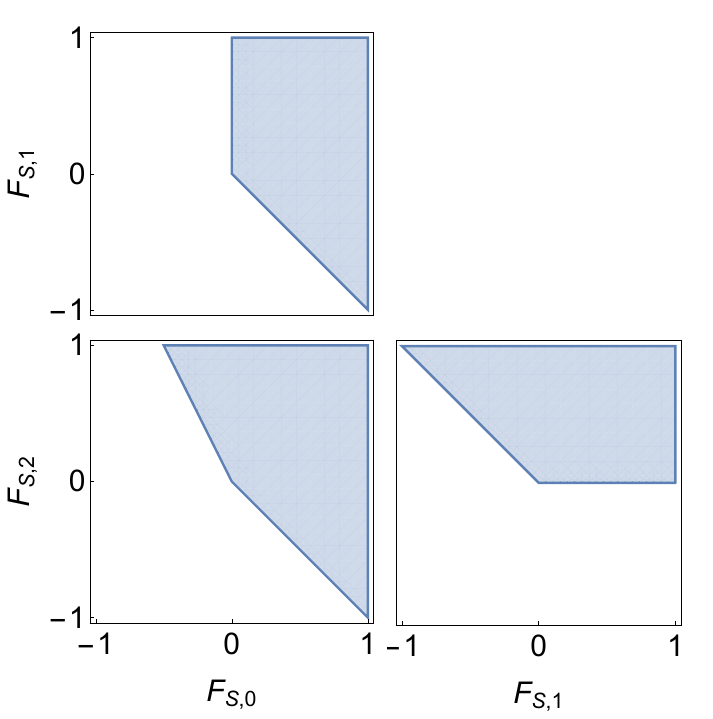}
\caption{Positivity constraints on all $F_{S,i}$ pairs.
\label{fig:2DS}}
\end{figure}

\begin{figure}
\centering
\includegraphics[width=.8\textwidth]{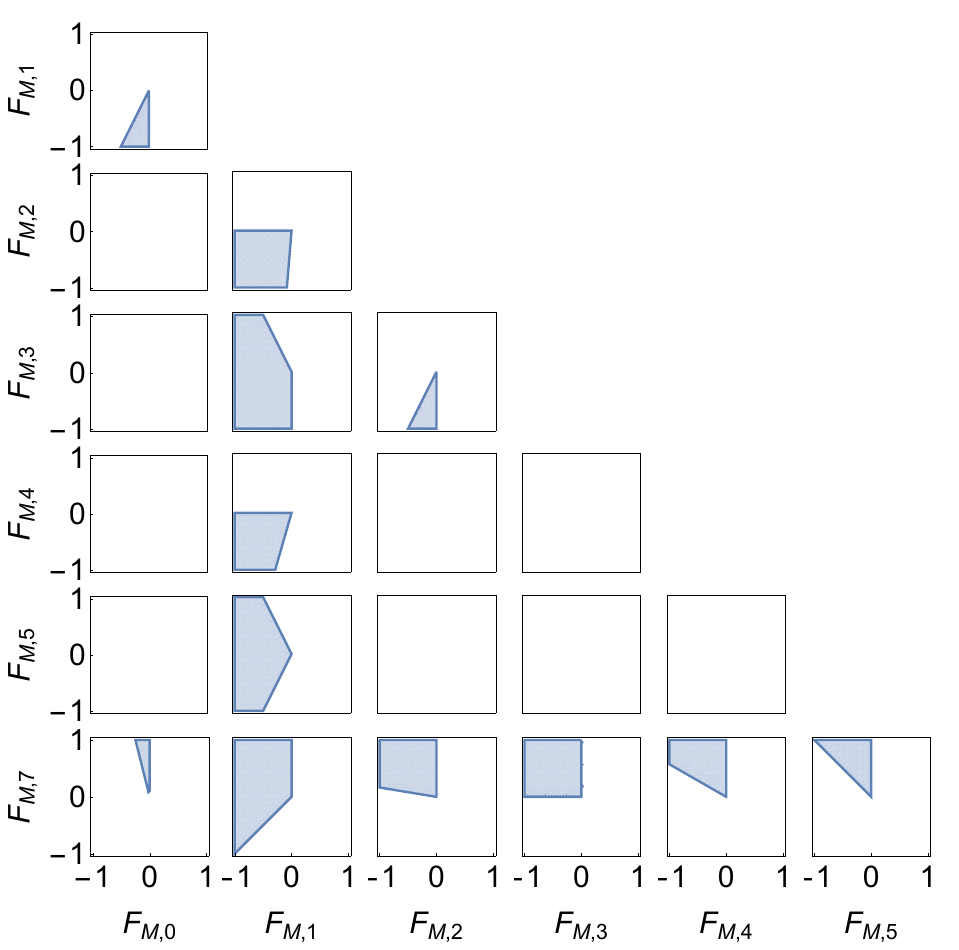}
\caption{Positivity constraints on all $F_{M,i}$ pairs.
\label{fig:2DM}}
\end{figure}

\begin{figure}[ht]
\centering
\includegraphics[width=.8\textwidth]{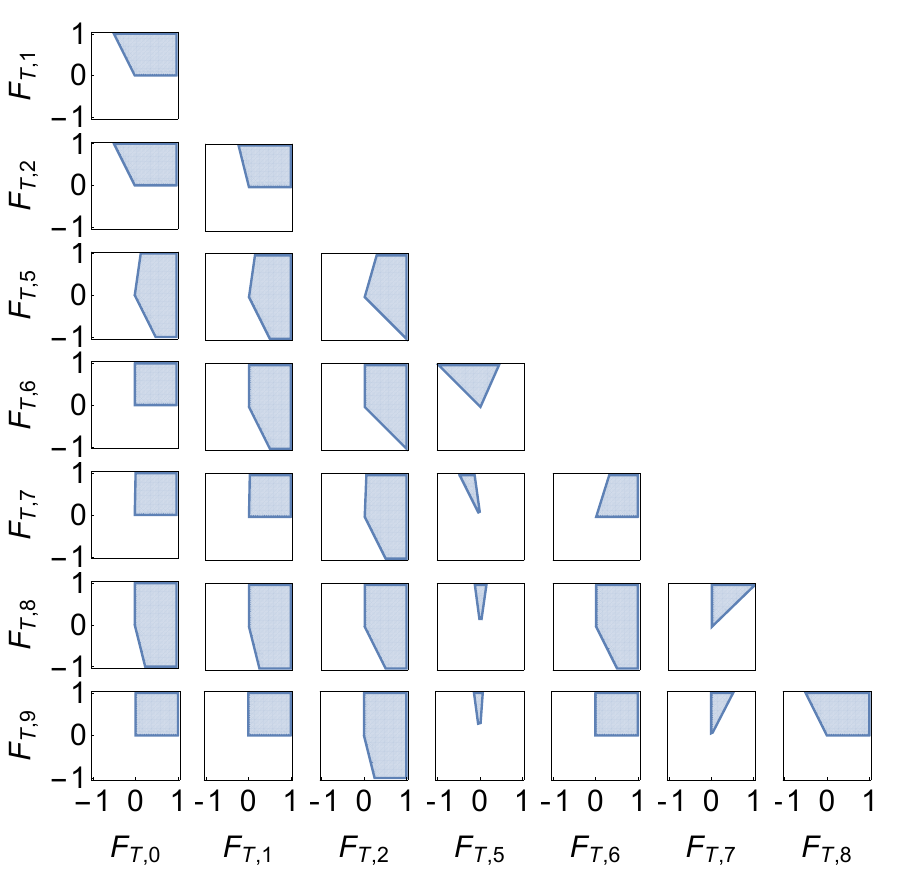}
\caption{Positivity constraints on all $F_{T,i}$ pairs.
\label{fig:2DT}}
\end{figure}

\subsection{Three operators at a time}

When three operators are turned on simultaneously, the linear conditions
continue to carve out pyramids in each $S$-, $M$-, and $T$-subspaces.
The higher order ones could potentially carve out cones or half cones,
as discussed in Section~\ref{subsec:cones}.  
The correlation across different types of operators could appear,
but only if one turns on one $S$, one $M$, and one $T$ operator simultaneously.
The shape of bound depends on which types of operators are turned on.
\begin{itemize}
\item If we turn on three operators in the same category, the situation has
	been discussed already in Section~\ref{subsec:pyramids}.
        The $S$ and $T$-subspaces are constrained by pyramids or prisms,
	while the $M$-subspace by a pyramid and/or a cone, depending
	on whether the chosen operators can probe the cone in the $C_{M,1}$,
	$C_{M,3}$ and $C_{M,6}$ subspace. For example,
	choosing $(F_{M,0}, F_{M,1}, F_{M,2})$ gives a pyramid, while
	$(F_{M,3}, F_{M,5}, F_{M,7})$ gives a cone.
\item If we turn on operators that belong to two different categories, similar
	to the two-operator case, there is again no correlation between the two
	categories. The bounds are then reduced to individual bound in
	one category and 2D bounds in another, which have all been
	discussed in the previous two subsections.
\item Finally, if we consider three operators that belong to three different
	types, the quadratic and quartic bounds cannot be decomposed into
	individual category.  They will potentially carve out cones or half cones
	as discussed in Section~\ref{subsec:cones}.
\end{itemize}

Due to the large number of possible combinations, here we only list some
representative cases
in Figures~\ref{fig:3D1}, \ref{fig:3D2}, \ref{fig:3D3}, and \ref{fig:3D4}. The
corresponding boundaries are described in Appendix~\ref{subsec:A12}, together
with polarisation conditions.

\begin{figure}[ht]
\centering
\includegraphics[width=.3\linewidth]{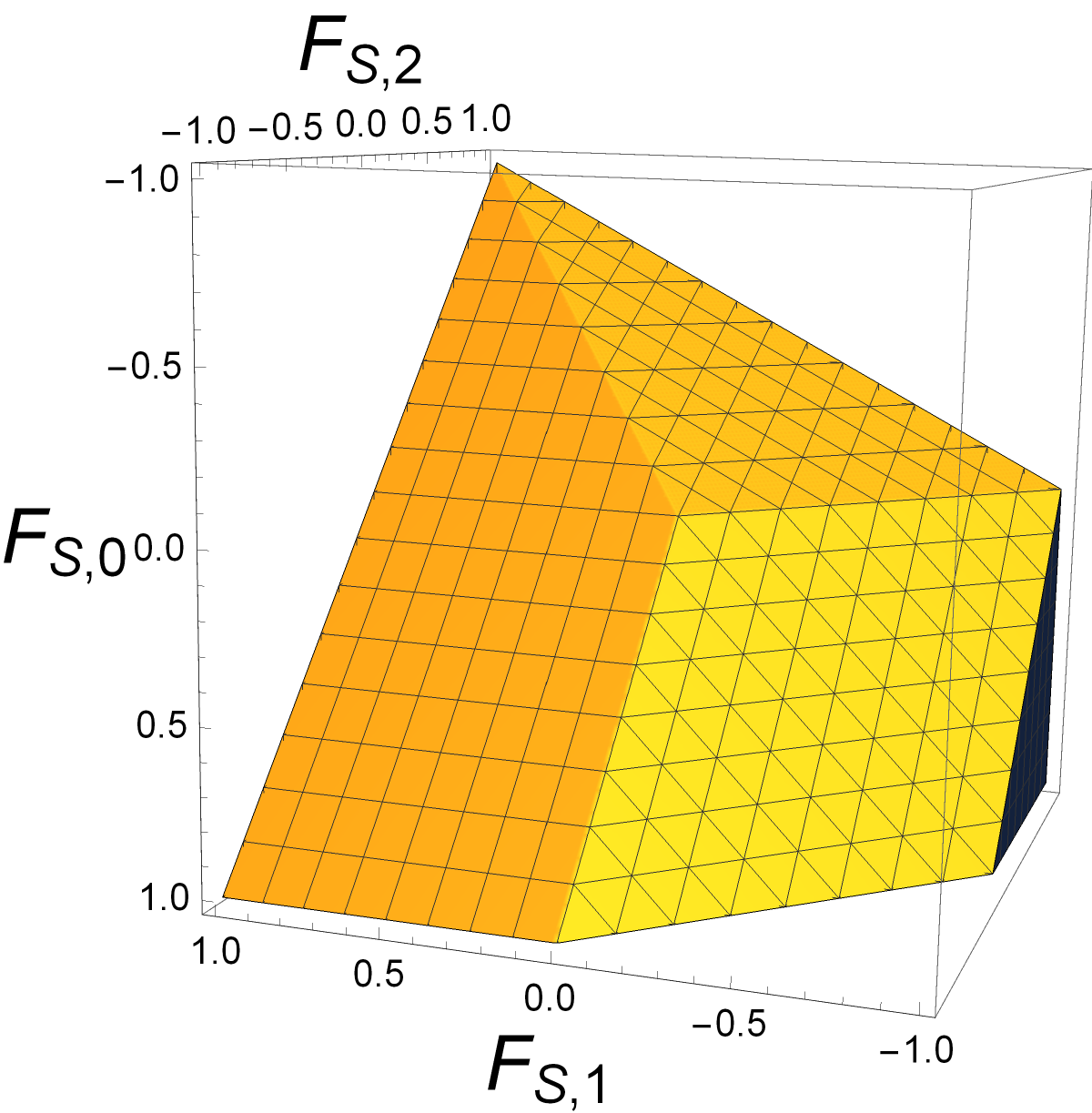}
\includegraphics[width=.3\linewidth]{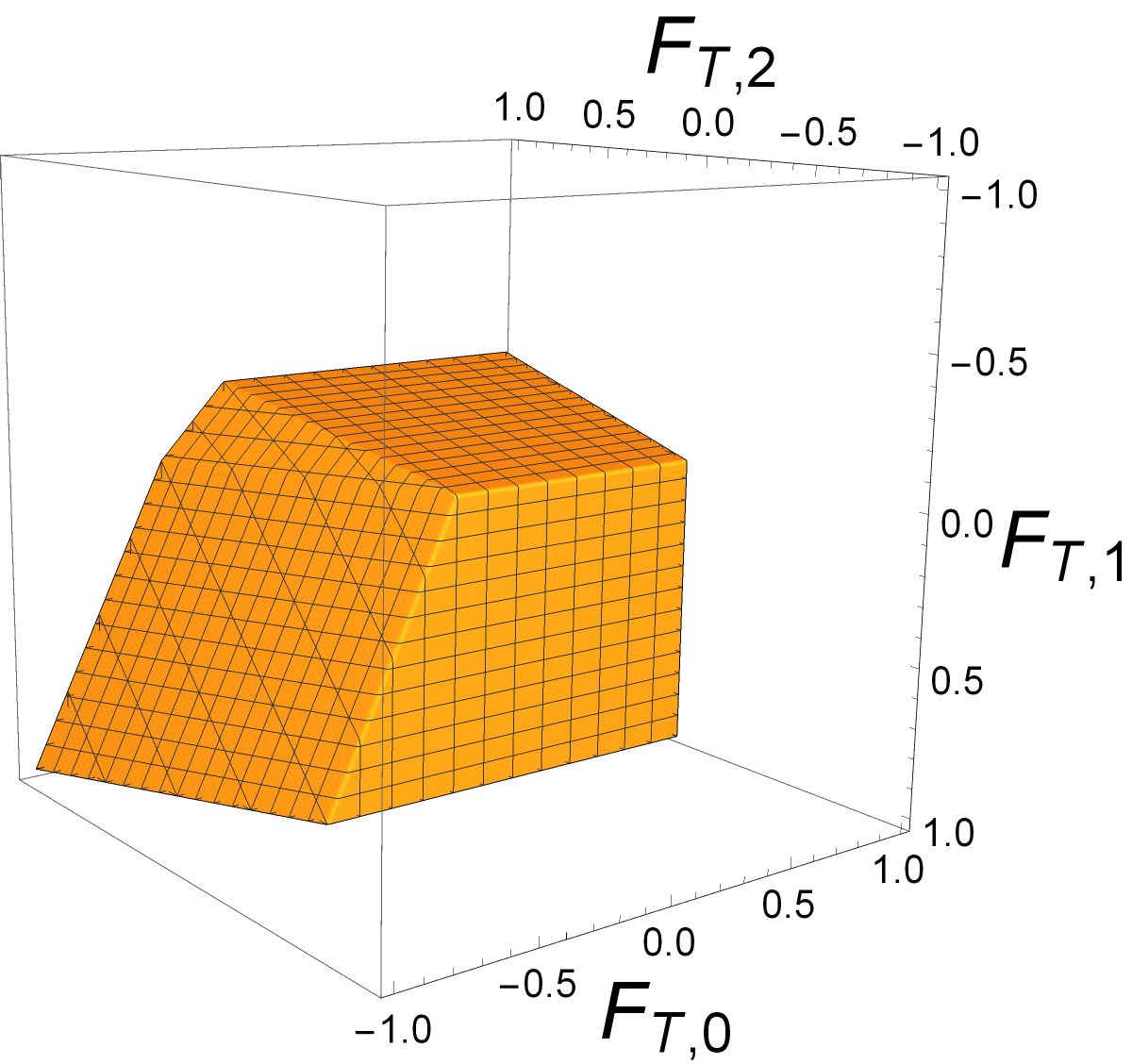}
\includegraphics[width=.3\linewidth]{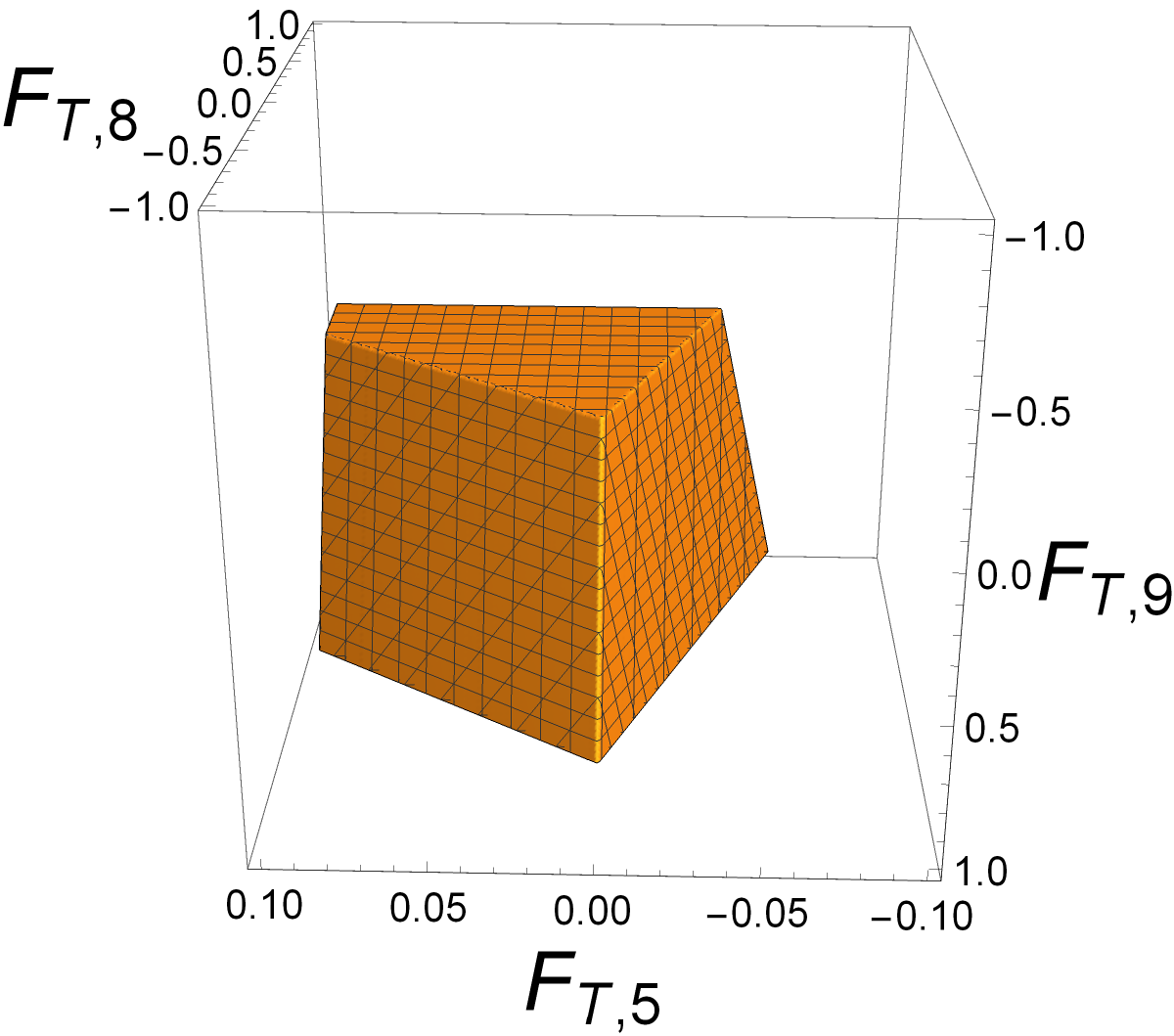}
\caption{Representative cases for three $S$-type or three $T$-type
	operators. Allowed parameter spaces are described by pyramids.
\label{fig:3D1}}
\end{figure}

\begin{figure}[ht]
\centering
\includegraphics[width=.3\linewidth]{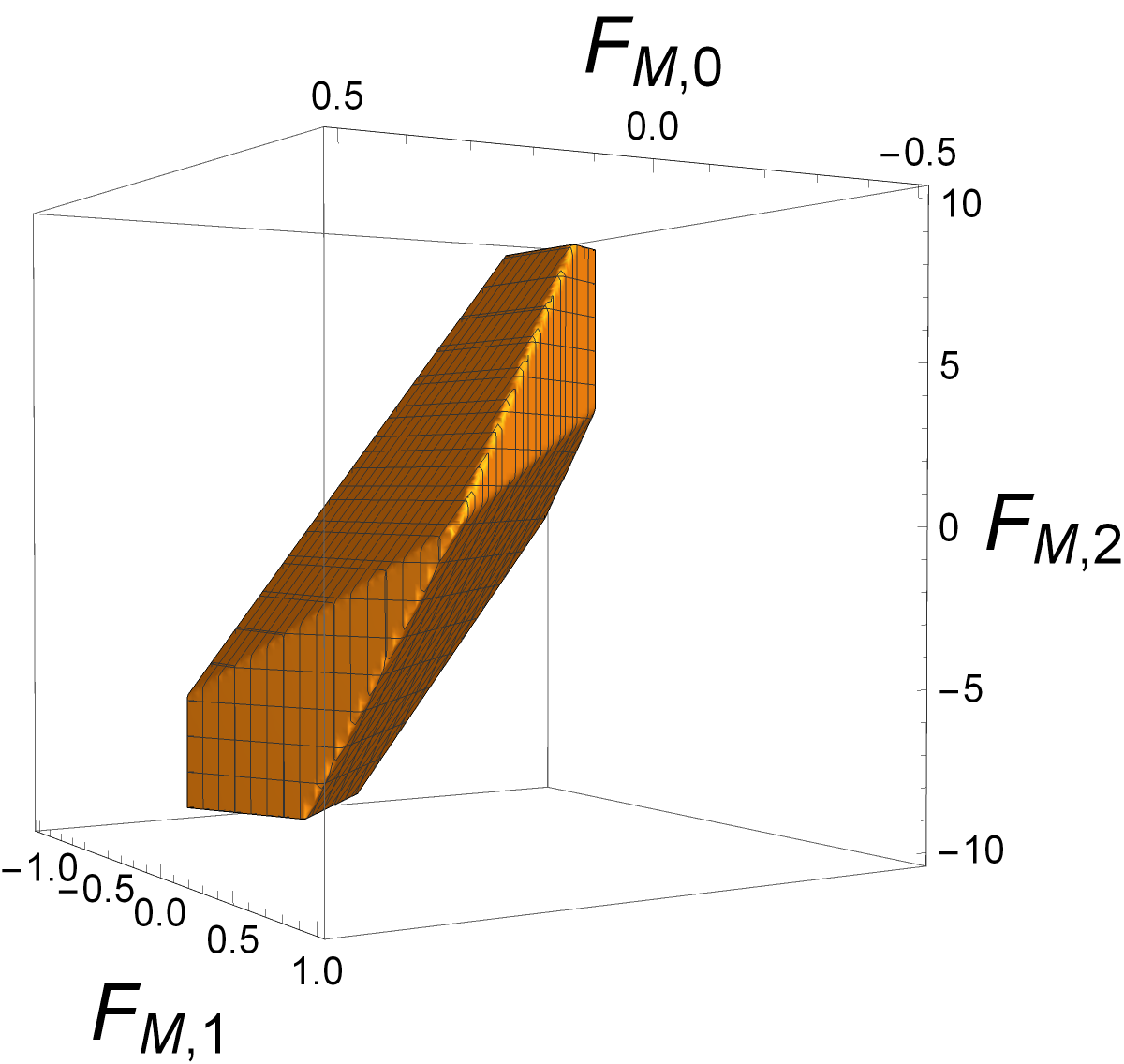}
\includegraphics[width=.3\linewidth]{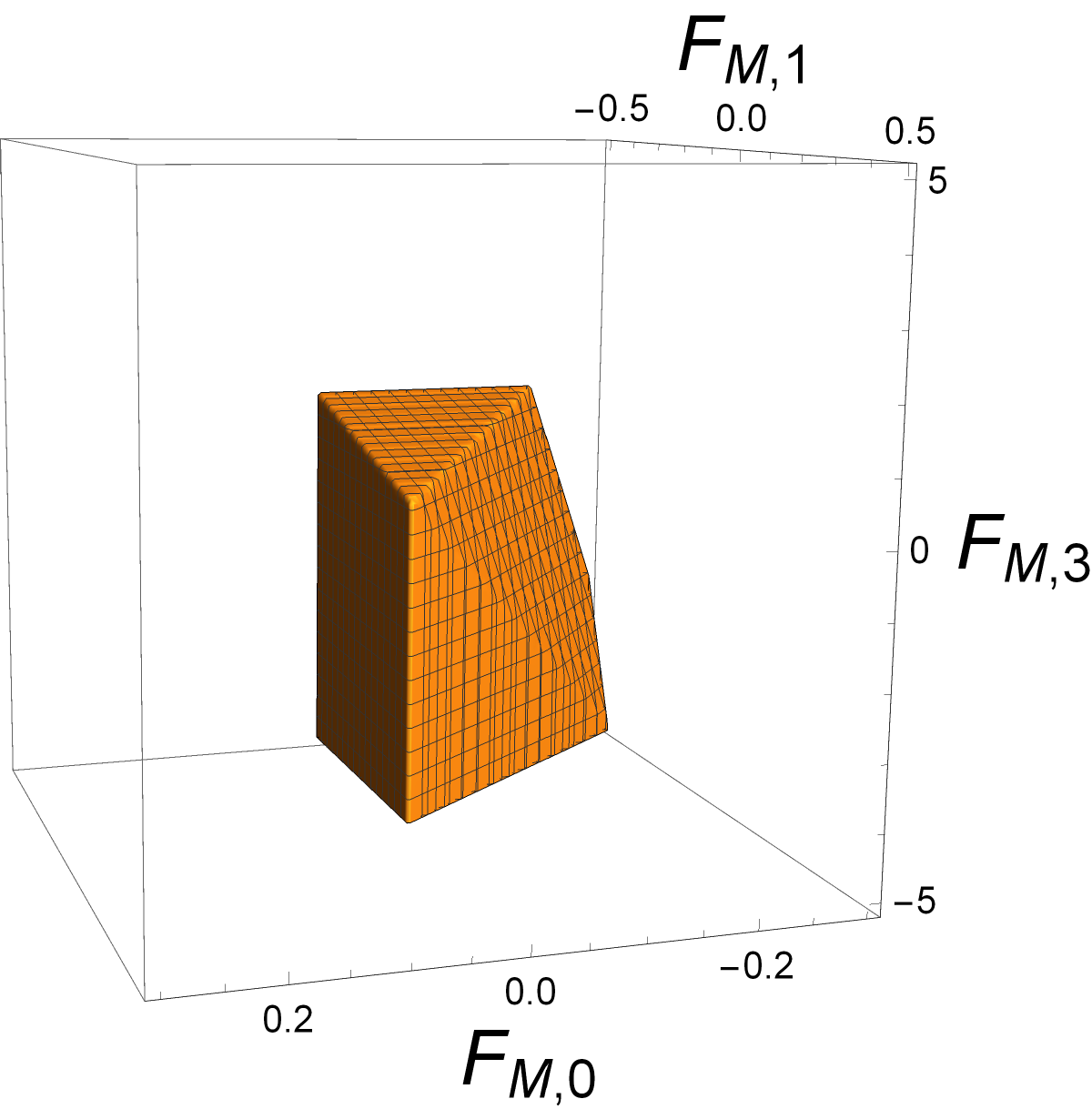}
\caption{Representative cases for three $M$-type operators.
	Allowed parameter spaces are described by pyramids.
\label{fig:3D2}}
\end{figure}

\begin{figure}[ht]
\centering
\includegraphics[width=.3\linewidth]{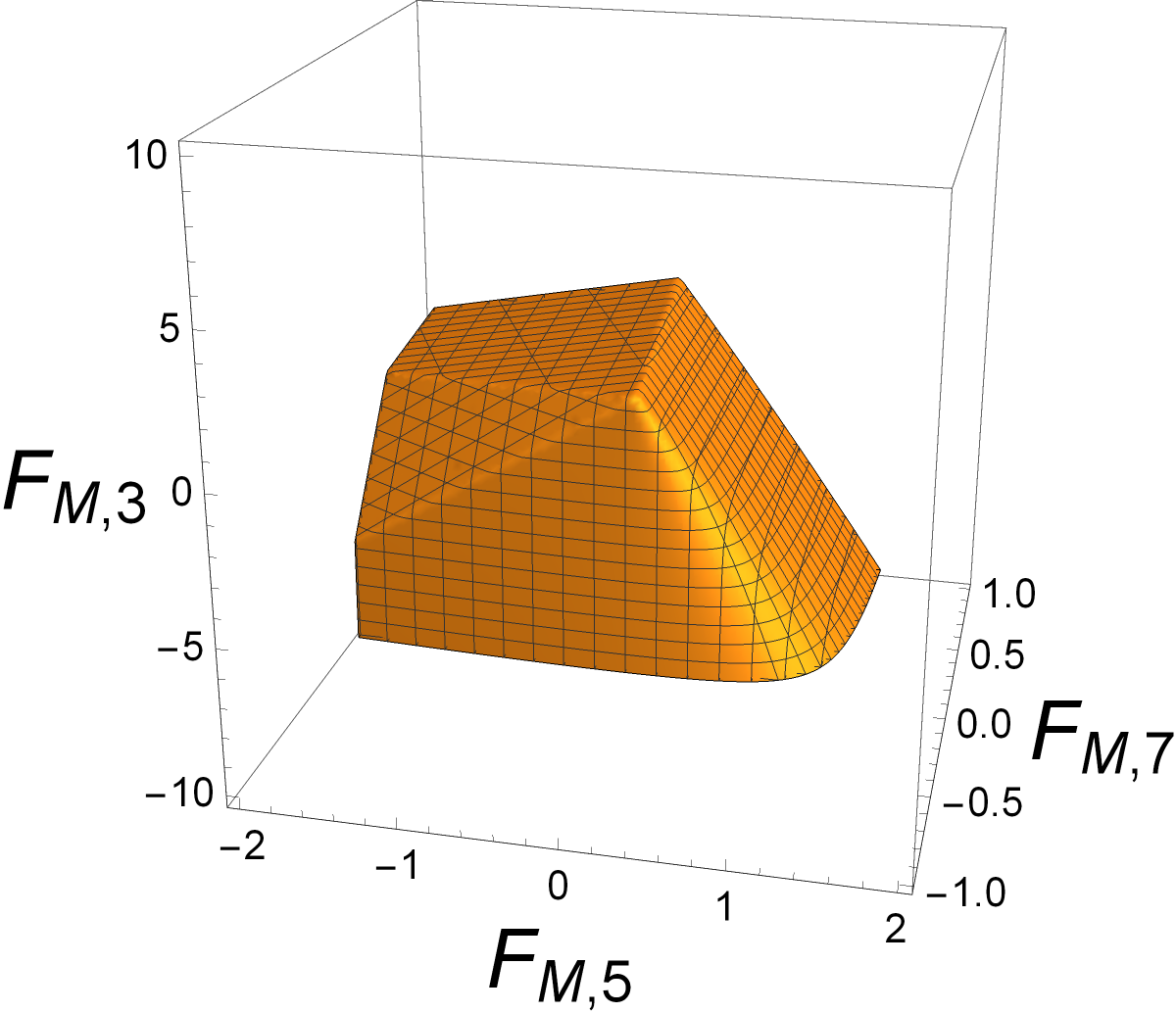}
\includegraphics[width=.3\linewidth]{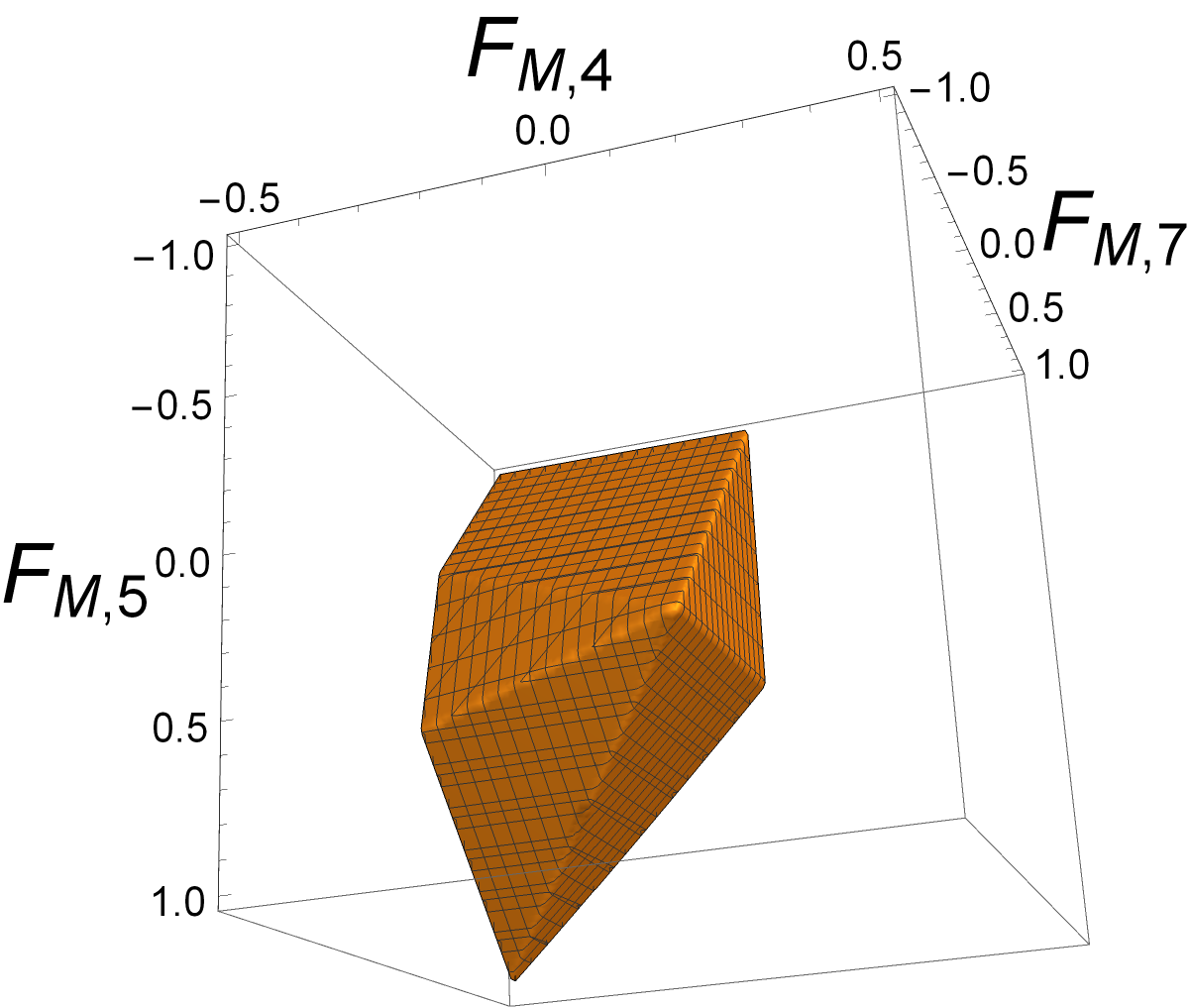}
\includegraphics[width=.3\linewidth]{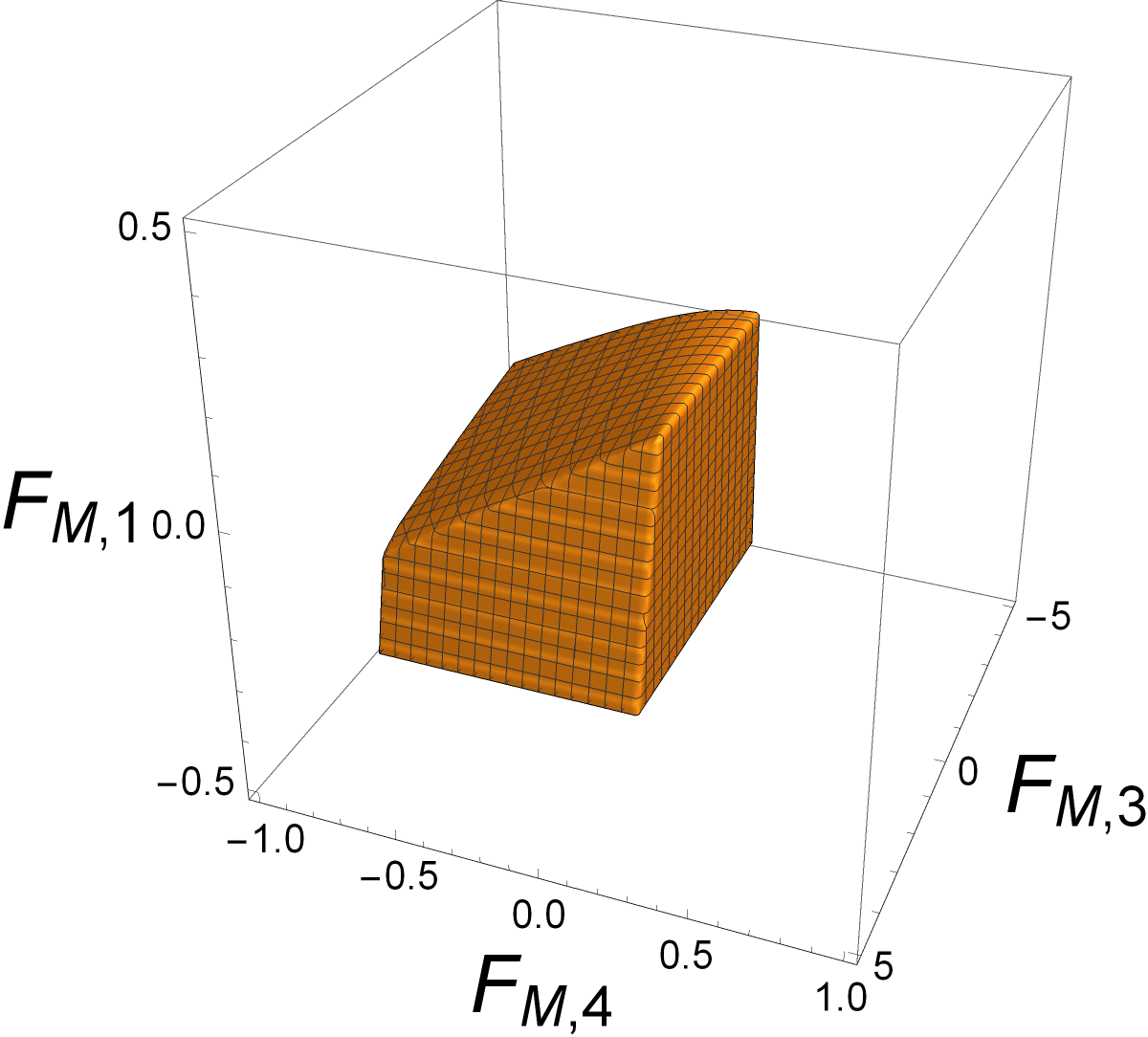}
\caption{Representative cases for three $M$-type operators. Allowed parameter
	spaces are described by combining a pyramid and a cone.
\label{fig:3D3}}
\end{figure}

\begin{figure}[ht]
\centering
\includegraphics[width=.3\linewidth]{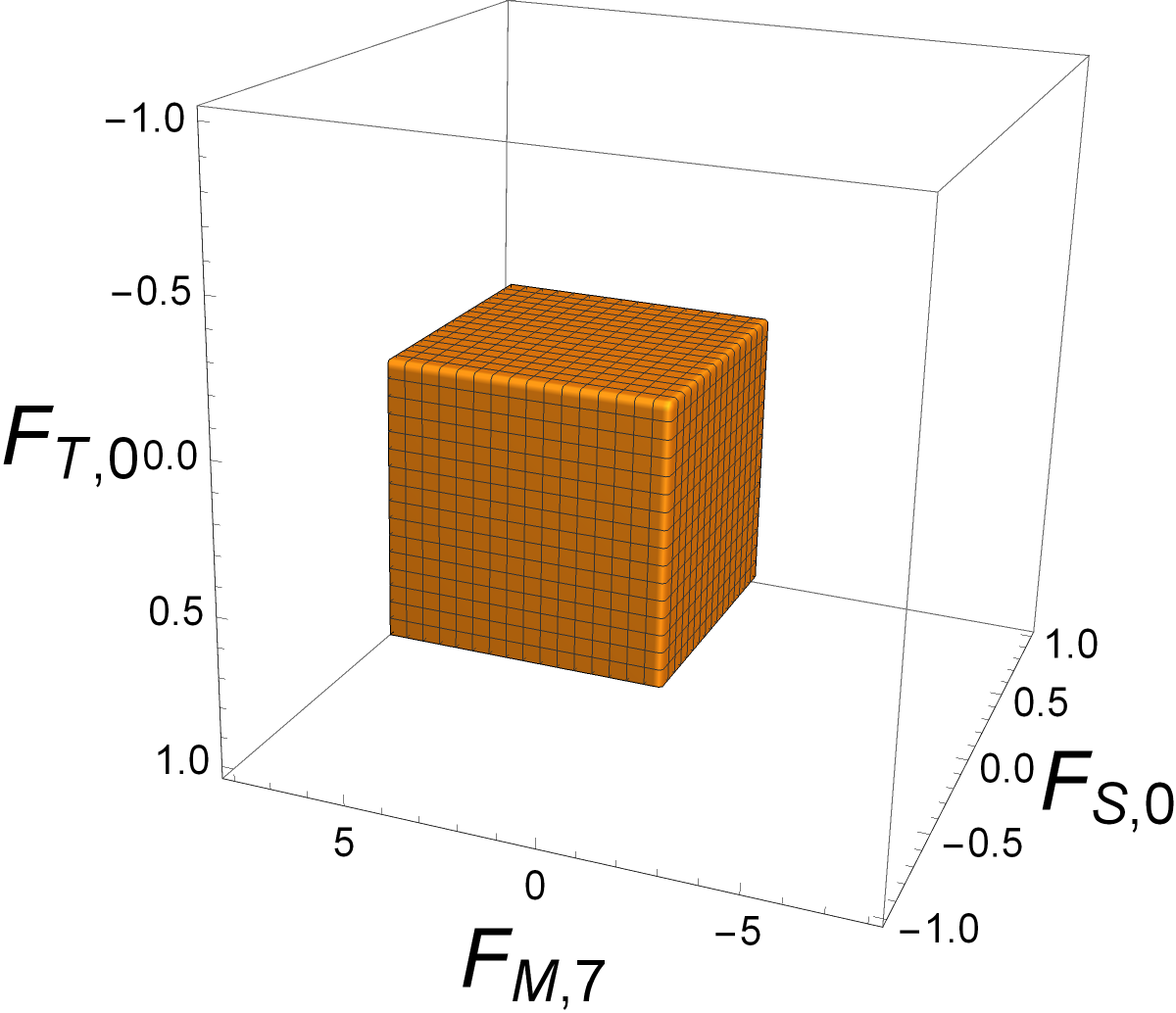}
\includegraphics[width=.3\linewidth]{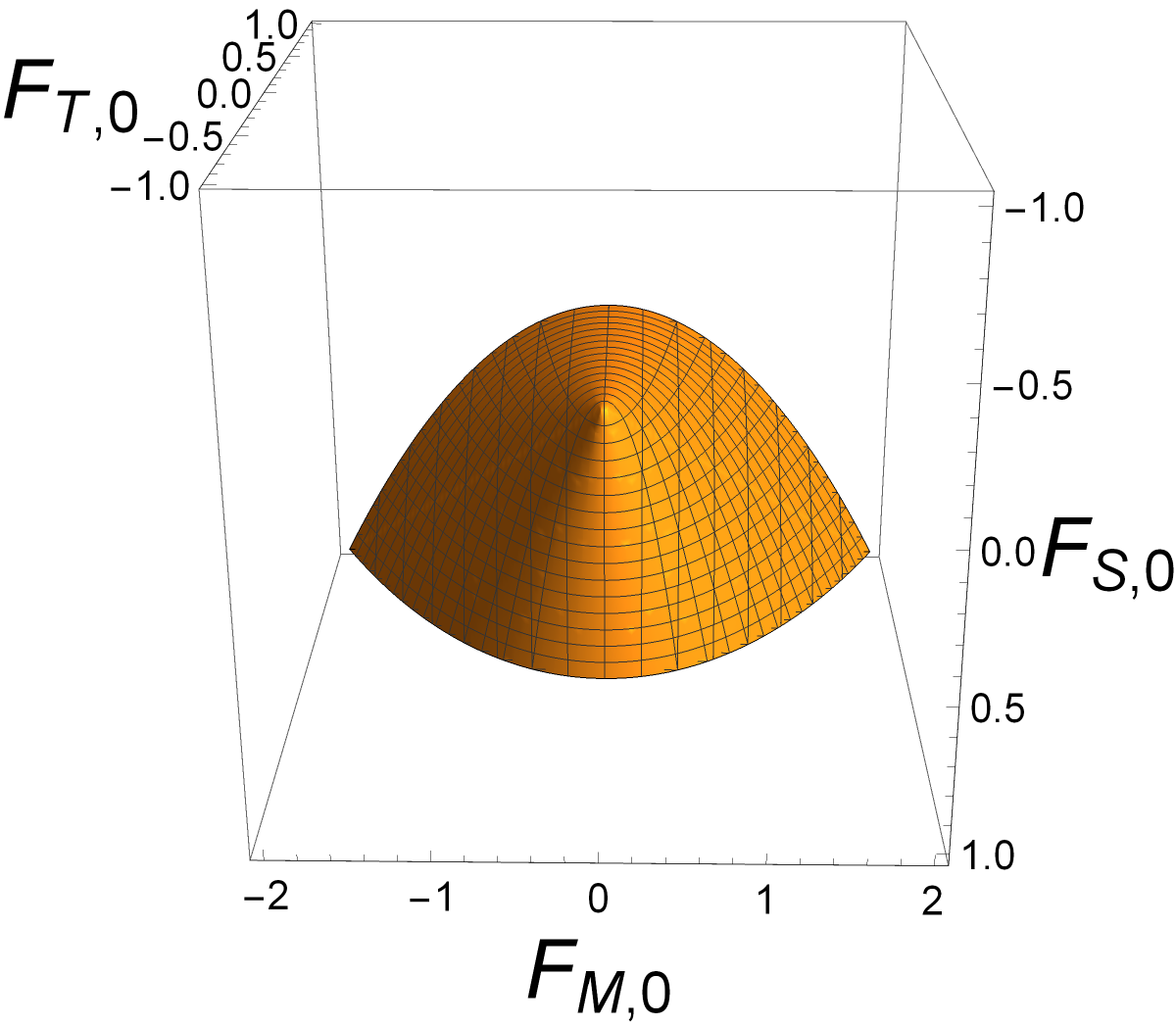}
\caption{Representative cases for operators from three different categories.
Allowed parameter spaces are described by a pyramid or a cone.
\label{fig:3D4}}
\end{figure}

\section{Summary}
\label{sec:summary}

Searching for deviations in QGC couplings is one of the main goals of the
electroweak program at the LHC.  We have shown that a set of theoretical
bounds on anomalous QGC couplings can be derived, by requiring that the
scattering amplitudes of specific VBS channels with specific polarisation
states satisfy the fundamental principles of the quantum field theory.
While these bounds do not constrain the magnitude of possible deviations from
the SM, they constrain the possible directions in the 18-dimensional QGC
parameter space in which SM deviations may be observed at all.  Therefore
they have important implications on future studies.

We have solved these positivity constraints and removed the dependence on the
polarisation of external particles.  
The exact parameter space which satisfies all positivity conditions is
described by a set of most crucial
polarisation states: once positivity bounds are satisfied for these states,
they are guaranteed to be satisfied for any other polarisation.  
The bounds derived from these polarisation states can be easily
used to guide future experimental searches, to optimise the parameter scan, and
to improve the presentation of resulting limits.  They also help future
theoretical studies based on the EFT formalism in a bottom-up
way, to choose reasonable benchmark scenarios.  We have then studied the shapes
and the volumes of these bounds.  To understand how these constraints may arise
from the underlying model point of view, we have considered a simplified model
with several new resonant states coupled to SM electroweak gauge bosons, and
showed that the positivity bounds are indeed satisfied by the corresponding EFT.

The most important results are summarised below:
\begin{itemize}
	\item The positivity condition on coefficients $\vec F$, from channel
		$i$ and external polarisation $\vec a$ and $\vec b$, is
		described by 
		$\vec F\cdot \vec x_i(\vec a,\vec b)>0$,
		where the vector $\vec x_i(\vec a,\vec b)$ can be computed
		from $f^{q_1q_2}_{ab,\epi\Lambda}(s)$.
		For arbitrary complex $\vec a$ and
		$\vec b$, we identify the boundaries
		of the set of $\vec x_i(\vec a,\vec b)$, which fully characterise
		the allowed parameter space independent of $\vec a$ and $\vec b$.
		They are described by 19 linear inequalities, 3 quadratic
		inequalities, and 1 quartic inequality, given in
		Eqs.~(\ref{eq:ww1}), (\ref{eq:ww2}), (\ref{eq:ww3}),
		(\ref{eq:zz1}), (\ref{eq:zz2}), (\ref{eq:wz1}), (\ref{eq:wz2}),
		(\ref{eq:a1}), (\ref{eq:a2}) and (\ref{eq:a3}).
	\item The $S$-type, $M$-type of $T$-type operator coefficients form 3
		subspaces.
		The 19 linear positivity
		constraints are decoupled into these 3 subspaces, with
		no correlation across.
		These linear constraints then describe the direct
		product of a pyramid in the $S$-space, one in the $T$-space,
		and a prism in the $M$-space.  Altogether, this reduces the
		allowed parameter space to 2.2\% of the total, in terms of
		solid angle.
	\item The quadratic and quartic bounds always involve all three
		subspaces.  They can be written in the form of
\begin{flalign}
	f(F_{S,i})g(F_{T,i})>\max(0,h_1(F_{M,i}),h_2(F_{M,i}))^2
\end{flalign}
		where $f$, $g$ and $h_{1,2}$ are homogeneous functions of degree
		1.  They carve out (approximately) half cones in the parameter
		space.  Together, they can reduce the parameter space to 5\% of
		the total.  When combined with the linear constraints, they
		only slightly improve the parameter space to 2.1\%.
	\item In UV completions where new degrees of freedom dominantly couple
		to the longitudinal modes of electroweak gauge bosons,
		positivity bounds carve out a 3-dimensional triangular pyramid
		in the $S$-space.
	\item In UV completions where new degrees of freedom dominantly couple
		to the transversal modes of electroweak gauge bosons, the
		constraint is described by a pyramid with 11 edges in the
		8-dimensional $T$-space.
	\item In UV completions where new degrees of freedom couple to both
		modes, but only generate $M$-type operators, the constraint
		in the 7-dimensional $M$-space is the intersection of a
		pyramid with 7 edges in a 5-dimensional subspace and a cone
		in a 3-dimensional subspace that involves the 6th direction and is partially
		orthogonal to the pyramid.  The 7th direction is unconstrained.
	\item For the convenience of future QGC studies, we have presented
		the complete descriptions of all 1-D and 2-D subspaces,
		i.e.~constraints on all individual operators and all pairs of
		operators.  We have also shown examples of 3-D subspaces.  
\end{itemize}

\acknowledgments
We would like to thank Ken Mimasu, Andrew J.~Tolley, Yu-Sheng Wu, and Xiao-Ran
Zhao for helpful discussions.  
QB is supported by IHEP under Contract No.~Y6515580U1.
CZ is supported by IHEP under Contract
No.~Y7515540U1 and the National 1000 Young Talents Program of China.  SYZ
acknowledges support from the starting grant from University of Science and
Technology of China (KY2030000089) and the National 1000 Young Talents Program
of China (GG2030040375).

\appendix

\section{Simplified models}
\label{sec:models}

Inspired by the simplified model in Ref.~\cite{Brass:2018hfw}, we extend the SM
by adding new particles that coupled to the longitudinal and/or the transversal
modes of the electroweak gauge bosons.  Following the notation of
Ref.~\cite{deBlas:2017xtg}, we consider four possible new fields.  Their names
and irreducible representations under $SU(3)_C\times
SU(2)_L\times U(1)_Y$ are given in Table~\ref{tab:newfields}.
\begin{table}[ht]
	\centering
\begin{tabular}{lcccc}
Names & scalar $\S$ & scalar $\X$ & scalar $\XX$ & vector $\V$
\\
Irrep & $(1,1)_0$ & $(1,3)_0$ & $(1,3)_1$ & $(1,2)_{\frac{1}{2}}$
\end{tabular}
\caption{New particles in the simplified model that we consider as examples of
UV completion.}
\label{tab:newfields}
\end{table}

The Lagrangian of the simplified model can be split into two pieces
\begin{flalign}
	\mathcal{L}_{BSM}=\mathcal{L}_{kin}+\mathcal{L}_{int}
\end{flalign}
The kinematic terms are standard:
\begin{flalign}
	\mathcal{L}_{kin}=&
	\frac{1}{2}D_\mu\S D^\mu \S-\frac{1}{2}M_\S^2 \S^2
	+\frac{1}{2}D_\mu\X^i D^\mu \X^i-\frac{1}{2}M_\X^2 \X^i\X^i
	\nonumber\\
	&
	+\left(D_\mu\XX^i\right)^\dagger D^\mu \XX^i-M_\XX^2 {\XX^i}^\dagger\XX^i
	\nonumber\\
	&+\left(D_\mu\V_\nu\right)^\dagger D^\nu\V^\mu
	-\left(D_\mu\V_\nu\right)^\dagger D^\mu\V^\nu
	+M_\V^2 \V^{\mu\dagger}\V_\mu\,.
\end{flalign}
The interaction terms can be written as
\begin{flalign}
	\mathcal{L}_{int}=&
	\S \J_\S + \X^i\J_\X^i + \left({\XX^i}^\dagger \J_\XX^i+h.c.\right)
	+ \left({\V^\mu}^\dagger {\J_\V}_\mu+h.c.\right)
	\label{eq:bsminteraction}
\end{flalign}
and the currents are
\begin{flalign}
	&\J_\S=
	a_H \left(D_\mu\phi\right)^\dagger D^\mu\phi
	+a_W W_{\mu\nu}^i W^{i\mu\nu}
	+a_B B_{\mu\nu} B^{\mu\nu}
	\\
	&\J_\X^i=
	b_H \left(D_\mu\phi\right)^\dagger \sigma^i D^\mu\phi
	+b_{WB}W^i_{\mu\nu}B^{\mu\nu}
	\\
	&\J_\XX^i=
	c_H\left( D_\mu\tilde\phi \right)^\dagger\sigma^i D_\mu\phi
	\\
	&\J_\V^\mu=
	d_{HB}D_\nu\phi B^{\nu\mu}
	+d_{HW}\sigma^iD_\nu\phi W^{i\nu\mu}
\end{flalign}
where $i$ is the $SU(2)_L$ triplet index. 
$\tilde\phi\equiv i\sigma^2\phi^*$.
$a_i,b_i,c_i,d_i$ are coupling strengths.
By solving the equation of motion for the heavy fields, and plugging them
back into the Lagrangian, we find that up to the dimension-8, the effective
Lagrangian can be written as
\begin{flalign}
	&\mathcal{L}=
	\frac{1}{2}M_\S^{-2}\J_\S^2
	+\frac{1}{2}M_\X^{-2}\J_\X^i\J_\X^i
	+M_\XX^{-2}{\J_\XX^i}^\dagger \J_\XX^i
	-M_\V^{-2}{\J_\V^\mu}^\dagger\J_{\V\mu}
\end{flalign}
With this we can write down the coefficients of the 18 QGC operators.
\begin{flalign}
	\begin{aligned}
		&F_{S,0}=2\frac{c_H^2}{M_\XX^2}
		\\
		&F_{S,1}=\frac{a_H^2}{2M_\S^2}-\frac{b_H^2}{2M_\X^2}
		\\
		&F_{S,2}=\frac{b_H^2}{M_\X^2}
		\\
		&F_{T,0}=\frac{\bar a_W^2}{2M_\S^2}
		\\
		&F_{T,5}=\frac{\bar a_W\bar a_B}{M_\S^2}
		\\
		&F_{T,6}=\frac{\bar b_{WB}^2}{M_\X^2}
		\\
		&F_{T,8}=\frac{\bar a_B^2}{2M_\S^2}
	\end{aligned}
	\quad
	\begin{aligned}
		&F_{M,0}=\frac{a_H\bar a_W}{M_\S^2}
		\\
		&F_{M,1}=-\frac{\bar d_{HW}^2}{M_\V^2}
		\\
		&F_{M,2}=\frac{a_H\bar a_B}{M_\S^2}
		\\
		&F_{M,3}=-\frac{\bar d_{HB}^2}{M_\V^2}
		\\
		&F_{M,4}=2\frac{b_H\bar b_{WB}}{M_\X^2}
		\\
		&F_{M,5}=2\frac{\bar d_{HW}\bar d_{HB}}{M_\V^2}
		\\
		&F_{M,7}=-\frac{\bar d_{HW}^2}{M_\V^2}
	\end{aligned}
	\label{eq:matched}
\end{flalign}
where for convenience we have defined:
\begin{flalign}
	&\bar a_W=-2g^{-2}a_W,\qquad
	\bar a_B=-4g'^{-2}a_B,
	\\
	&\bar b_{WB}=-2g^{-1}g'^{-1}b_{WB},
	\qquad
	\bar d_{HW}=2g^{-1}d_{HW},\qquad
	\bar d_{HB}=2g'^{-1}d_{HB}\,.
\end{flalign}
The coefficients which are not included above vanish in this model.

Note that this simplified model itself is not a UV completed theory.  The
interaction terms in Eq.~(\ref{eq:bsminteraction}) are of mass dimension five, so
they need to be UV completed above some energy scale.  However, at
the tree level we expect the positivity condition to be satisfied.  This is
because for dim-5 interactions we expect the amplitude to grow like
$\mathcal{O}(s)$ at large $s$, which already satisfies the Froissart-Martin
bound.  This is one of the main requirements to derive positivity bounds.
The others such as analyticity and Lorentz invariance
are trivially satisfied in this model.

To see explicitly how the positivity is satisfied, we first check the linear
conditions. Plugging in Eq.~(\ref{eq:matched}) to the linear inequalities
(\ref{eq:MS})--(\ref{eq:MT}), we find
\begin{flalign}
	&M_S\cdot F_S=
\left(
\begin{array}{c}
 \frac{1}{2} \left(\frac{a_H^2}{M_\S^2}+\frac{b_H^2}{M_\X^2}+\frac{8 c_H^2}{M_\XX^2}\right) \\
 \frac{1}{2} \left(\frac{a_H^2}{M_\S^2}+\frac{b_H^2}{M_\X^2}+\frac{4 c_H^2}{M_\XX^2}\right) \\
 \frac{b_H^2}{M_\X^2}+\frac{2 c_H^2}{M_\XX^2} \\
\end{array}
\right)\,,
\\
&M_T\cdot F_T=
\left(
\begin{array}{c}
 0 \\
 0 \\
 \frac{a_W^2}{M_\S^2} \\
 \frac{4 a_W^2}{M_\S^2} \\
 \frac{(a_B\sw^4+2a_W\cw^4)^2}{M_\S^2}+
\frac{4 b_{WB}^2 \sw^4 \cw^4}{M_\X^2}
\\
 0 \\
 0 \\
 \frac{4 b_{WB}^2 \sw^4}{M_\X^2} \\
 0 \\
 \frac{4 b_{WB}^2}{M_\X^2} \\
 0 \\
 \frac{4 b_{WB}^2 \left(\cw^2-\sw^2\right)^2}{ M_\X^2} +
 \frac{4  \left(a_B \sw^2-2 a_W \cw^2\right)^2}{M_\S^2} \\
 0 \\
 \frac{(a_B+2 a_W)^2}{M_\S^2}+\frac{4 b_{WB}^2}{M_\X^2} \\
\end{array}
\right)\,,
\\
&M_M\cdot F_M=
\left(
\begin{array}{c}
 \frac{2 d_{HW}^2}{M_\V^2} \\
 \frac{ d_{HW}^2 \cw^4+\left(d_{HW}\cw^2+d_{HB}\sw^2\right)^2}{M_\V^2} \\
 \frac{ d_{HW}^2 \cw^4+\left(d_{HW}\cw^2-d_{HB}\sw^2\right)^2}{M_\V^2} \\
 \frac{2 d_{HW}^2}{M_\V^2} \\
 \frac{\left(d_{HB}+d_{HW}\right)^2+d_{HW}^2}{M_\V^2} \\
 \frac{\left(d_{HB}-d_{HW}\right)^2+d_{HW}^2}{M_\V^2} \\
\end{array}
\right)\,.
\end{flalign}
The r.h.s.~are always sums of squared terms, which are positive definite, so the
linear positivity conditions are indeed satisfied.

\begin{figure}[ht]
\centering
\includegraphics[width=.42\linewidth]{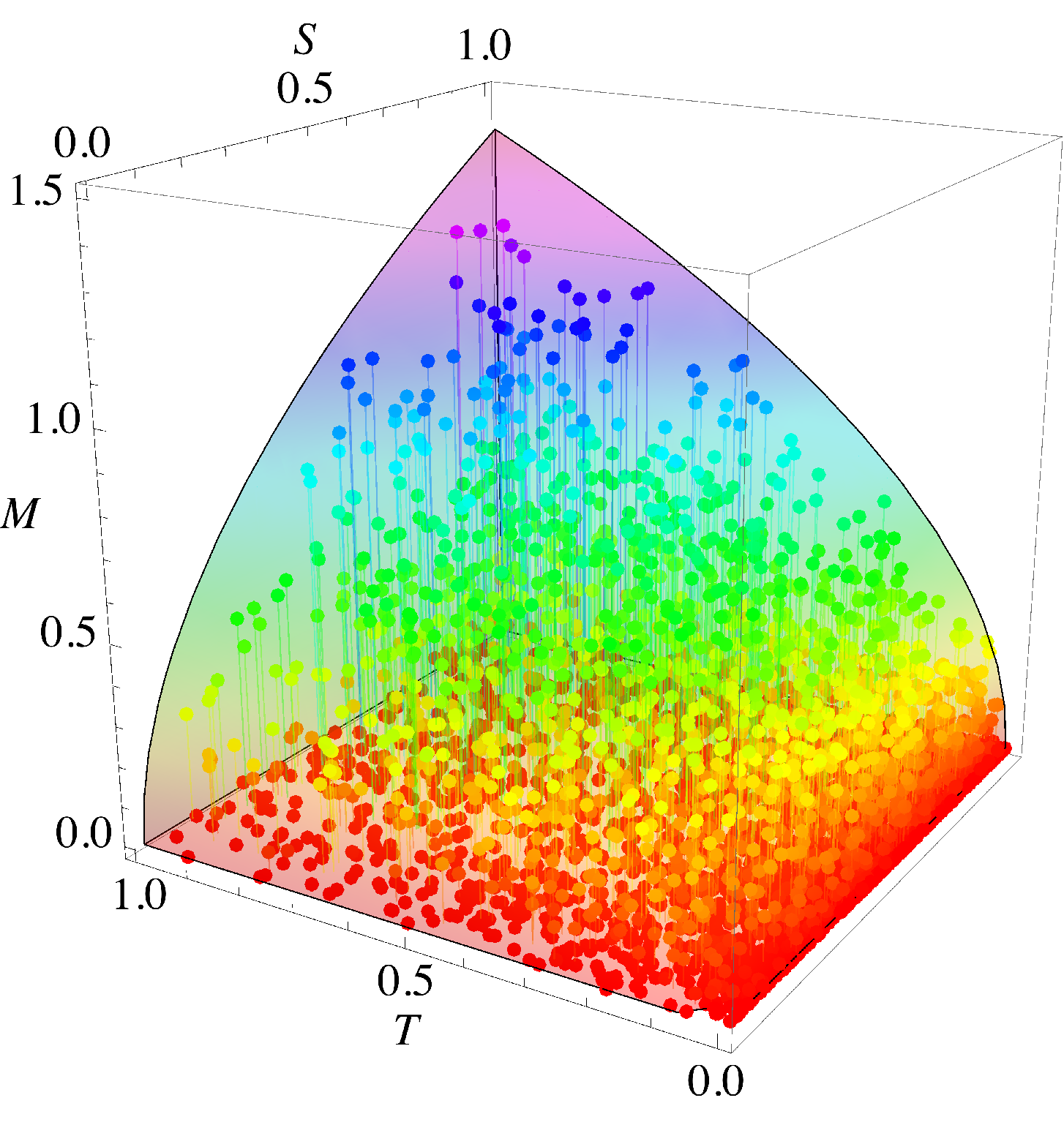}
\includegraphics[width=.49\linewidth]{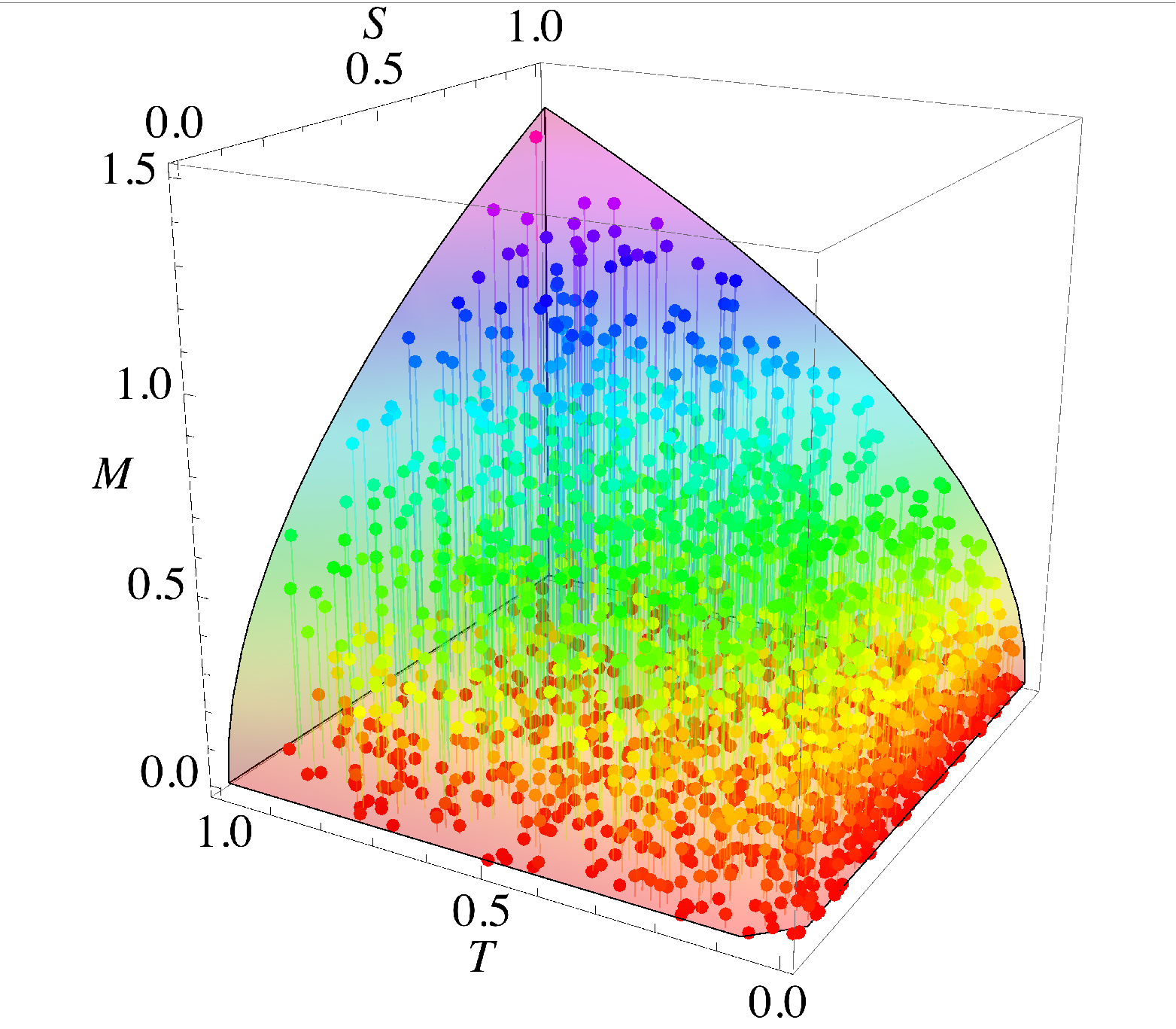}
\includegraphics[width=.45\linewidth]{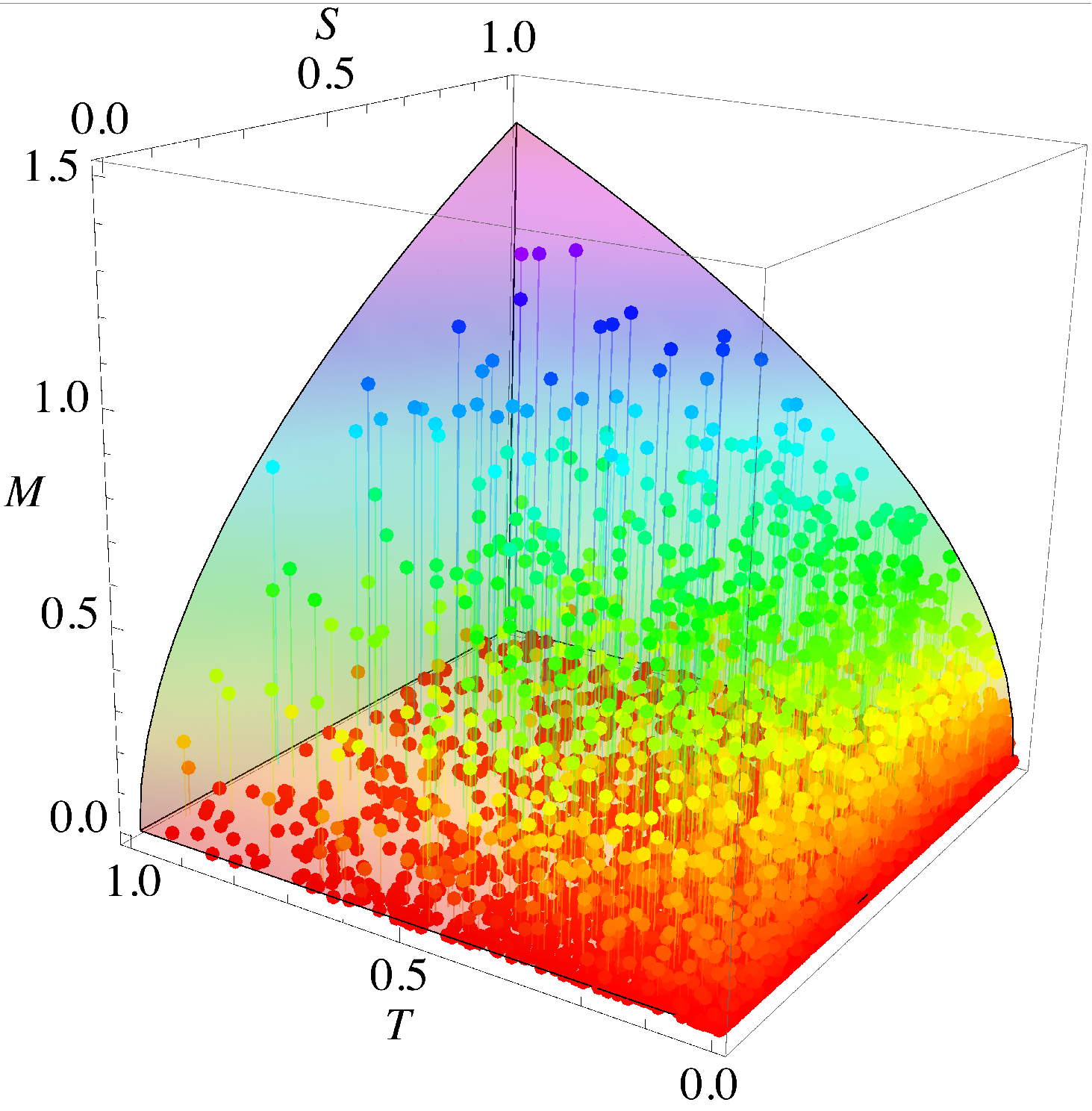}
\caption{Random points generated in the parameter space of the simplified model,
compared with quadratic and quartic positivity bounds in the same space.
Top left: quadratic condition from $WW$;
Top right: quadratic condition from $ZZ$;
Bottom: quadratic condition from $WZ$.
}
\label{fig:scatterwwzz}
\end{figure}

To see that the quadratic conditions are satisfied, we randomly generate the
coupling strengths in the range $-1\sim1$ TeV$^{-1}$ and masses in the range
$1\sim2$ TeV, and plot the corresponding parameter space.  For example,
the first quadratic condition from the $WW$ channel is
$4f_2(f_3+f_6)>\max(0,|f_4|-2f_1)^2$, so we choose three axes:
\begin{flalign}
	&S: \frac{f_2}{4}=2F_{S,0}+F_{S,1}+F_{S,2}
	\\
	&T: \frac{f_3+f_6}{8}=2F_{T,0}+F_{T,1}+F_{T,2}
	\\
	&M: \max\left(0,\frac{|f_4|-2f_1}{8}\right)=
	\max\left(
	0,F_{M,0}+\frac{F_{M,1}}{4},-F_{M,0}+\frac{3F_{M,1}}{4}-\frac{F_{M,7}}{2}
	\right)
\end{flalign}
and plot the corresponding $S,T,M$ for each set of parameters, together with the
bound $2ST>M^2$.  This is shown in Figure~\ref{fig:scatterwwzz} up left, and
we can see that all the points representing different values of model
parameters are wrapped by the surface which represents the quadratic positivity
condition, $2ST>M^2$.  The second condition from $WW$ does not give new
information in our model.  We make the same plots for $ZZ$ and $WZ$ channels in
Figure~\ref{fig:scatterwwzz} up right and down. The axes for $ZZ$ scattering
are
\begin{flalign}
	&S:F_{S0}+F_{S1}+F_{S2}
	\\
	&T:4 \cw^8 (2 F_{T0}+2 F_{T1}+F_{T2})+2 \cw^4 \sw^4 (2 F_{T5}+2
	F_{T6}+F_{T7})+\sw^8 (2 F_{T8}+F_{T9})
	\\
	&M:\max \Big(0,\frac{1}{2} \left(\cw^4 (-(4 F_{M0}-2 F_{M1}+F_{M7}))-\cw^2 \sw^2 (2 F_{M4}+F_{M5})+\sw^4 (F_{M3}-2 F_{M2})\right),
	\nonumber\\
	&\qquad2 \cw^4 F_{M0}+F_{M2} \sw^4-F_{M4} \sw^4+F_{M4} \sw^2\Big)
\end{flalign}
and the axes for $ZZ$ scattering are
\begin{flalign}
	&S:2 (F_{S0}+F_{S2})
	\\
	&T:4 \cw^4 (4 F_{T1}+F_{T2})+\sw^4 (4 F_{T6}+F_{T7})
	\\
	&M:\max \left(0,\cw^2 \sqrt{(2 F_{M1}-F_{M7}) \left(\cw^4 (2
	F_{M1}-F_{M7})+\cw^2 F_{M5} \sw^2+F_{M3} \sw^4\right)}
	\right.  \nonumber\\ &\qquad\left. 
	-2 F_{M4} \sw^2-\frac{F_{M5} \sw^2}{2}-\cw^2F_{M7},
	\right.  \nonumber\\ &\qquad\left. 
	-\sqrt{(2 F_{M1}-F_{M7}) \left(\cw^4 (2 F_{M1}-F_{M7})+\cw^2 F_{M5}
	\sw^2+F_{M3} \sw^4\right)}
	\right.  \nonumber\\ &\qquad\left. 
	+2 F_{M4} \sw^2+\frac{F_{M5} \sw^2}{2}+\cw^2 F_{M7}\right)
\end{flalign}
The results show how the model space is completely covered by the surfaces
of the quadratic and quartic bounds.

\section{Positivity bounds on two operators}
\label{subsec:A11}

Here we list positivity bounds on all two-operator pairs.
For $F_{S,i}$ operators:
\begin{flalign}
\fbox{$F_{S,0},F_{S,1}$} && \nonumber\\
&\quad F_{S,0}+F_{S,1}>0 \quad (ZZ,L) & \\
&\quad F_{S,0}>0 \quad (WZ,L) & \\
\fbox{$F_{S,0},F_{S,2}$} && \nonumber\\
&\quad F_{S,0}+F_{S,2}>0 \quad (ZZ,L;WZ,L) & \\
&\quad 2F_{S,0}+F_{S,2}>0 \quad (WW,L) & \\
\fbox{$F_{S,1},F_{S,2}$} && \nonumber\\
&\quad F_{S,1}+F_{S,2}>0 \quad (WW,L;ZZ,L) &\\
&\quad F_{S,2}>0 \quad (WZ,L)   . &
\end{flalign}
For $F_{T,i}$ operators:
\begin{flalign}
\fbox{$F_{T,0},F_{T,1}$} && \nonumber\\
&\quad 2F_{T,0}+F_{T,1}>0 \quad (WW,T_{+-}) & \\
&\quad F_{T,1}>0 \quad (WW,T_{++};WZ,T_{\parallel};W\gamma,T_{\parallel}) & \\
\fbox{$F_{T,0},F_{T,2}$} && \nonumber\\
&\quad 2F_{T,0}+F_{T,2}>0 \quad (WW,T_{+-};ZZ,T_{\parallel};Z\gamma,T_{\parallel}) & \\
&\quad F_{T,2}>0 \quad (WW,T_{\perp};WW,T_{++};ZZ,T_{\perp};
WZ,T_{\perp};WZ,T_{\parallel};  & \nonumber\\
&\quad \quad W\gamma,T_{\perp}; W\gamma,T_{\parallel};Z\gamma,T_{\perp};\gamma\gamma,T_{\perp}) & \\
\fbox{$F_{T,0},F_{T,5}$} && \nonumber\\
&\quad 2F_{T,0}-t_{w}^2F_{T,5}>0 \quad (Z\gamma,T_{\parallel}) & \\
&\quad 2F_{T,0}+F_{T,5}>0 \quad (\gamma\gamma,T_{\parallel}) &\\
\fbox{$F_{T,0},F_{T,6}$} && \nonumber\\
&\quad F_{T,0}>0 \quad (WW,T_{+-};WW,T_{\parallel}) &\\
&\quad F_{T,6}>0 \quad (WZ,T_{\parallel}) & \\
\fbox{$F_{T,0},F_{T,7}$} && \nonumber\\
&\quad F_{T,7}>0 \quad (ZZ,T_{\perp};WZ,T_{\perp};WZ,T_{\parallel}; W\gamma,T_{\perp};W\gamma,T_{\parallel};Z\gamma,T_{\perp}; \gamma\gamma,T_{\perp}) & \\
&\quad 32c_{w}^4F_{T,0} + (8s_{w}^4-8s_{w}^2+1)F_{T,7}>0 \quad (Z\gamma,T_{\parallel}) & \\
\fbox{$F_{T,0},F_{T,8}$} && \nonumber\\
&\quad F_{T,0}>0 \quad (WW,T_{+-};WW,T_{\parallel}) & \\
&\quad 4T_{T,0}+F_{T,8}>0 \quad (\gamma\gamma,T_{\parallel}) & \\
\fbox{$F_{T,0},F_{T,9}$} && \nonumber\\
&\quad F_{T,0}>0 \quad (WW,T_{+-};WW,T_{\parallel}) & \\
&\quad F_{T,9}>0 \quad (ZZ,T_{\perp};Z\gamma,T_{\perp}; \gamma\gamma,T_{\perp}) & \\
\fbox{$F_{T,1},F_{T,2}$} && \nonumber\\
&\quad F_{T,2}>0 \quad (WW,T_{\perp};ZZ,T_{\perp};WZ,T_{\perp};
W\gamma,T_{\perp};Z\gamma,T_{\perp}; \gamma\gamma,T_{\perp}) & \\
&\quad 4F_{T,1}+F_{T,2}>0 \quad (WZ,T_{\parallel};W\gamma,T_{\parallel}) & \\
\fbox{$F_{T,1},F_{T,5}$} && \nonumber\\
&\quad 2F_{T,1}-t_{w}^2F_{T,5}>0 \quad (Z\gamma,T_{\parallel}) & \\
&\quad 2F_{T,1}+F_{T,5}>0  \quad (\gamma\gamma,T_{\parallel}) & \\
\fbox{$F_{T,1},F_{T,6}$} && \nonumber\\
&\quad F_{T,1}>0 \quad (WW,T_{++};WW,T_{+-};WW,T_{\parallel}) & \\
&\quad 2F_{T,1}+F_{T,6}>0  \quad (\gamma\gamma,T_{\parallel}) & \\
\fbox{$F_{T,1},F_{T,7}$} && \nonumber\\
&\quad F_{T,7}>0 \quad (ZZ,T_{\perp};WZ,T_{\perp};W\gamma,T_{\perp};Z\gamma,T_{\perp};\gamma\gamma,T_{\perp}) & \\
&\quad 32c_{w}^4F_{T,1}+(8s_{w}^4-8s_{w}^2+1)F_{T,7}>0 \quad (Z\gamma,T_{\parallel}) & \\
\fbox{$F_{T,1},F_{T,8}$} && \nonumber\\
&\quad F_{T,1}>0 \quad (WW,T_{++};WW,T_{+-};WW,T_{\parallel}; WZ,T_{\parallel};W\gamma,T_{\parallel}) & \\
&\quad 4F_{T,1}+F_{T,8}>0 \quad (\gamma\gamma,T_{\parallel}) & \\
\fbox{$F_{T,1},F_{T,9}$} && \nonumber\\
&\quad F_{T,1}>0 \quad (WW,T_{++};WW,T_{+-};WW,T_{\parallel}; WZ,T_{\parallel};W\gamma,T_{\parallel}) & \\
&\quad F_{T,9}>0 \quad (ZZ,T_{\perp};Z\gamma,T_{\perp}; \gamma\gamma,T_{\perp}) & \\
\fbox{$F_{T,2},F_{T,5}$} && \nonumber\\
&\quad F_{T,2}-t_{w}^2F_{T,5}>0 \quad (Z\gamma,T_{\parallel}) & \\
&\quad F_{T,2}+F_{T,5}>0 \quad (\gamma\gamma,T_{\parallel}) & \\
\fbox{$F_{T,2},F_{T,6}$} && \nonumber\\
&\quad F_{T,2}>0 \quad (WW,T_{\perp};WW,T_{++};WW,T_{+-}; WW,T_{\parallel};ZZ,T_{\perp}; & \nonumber\\
&\quad\quad\quad\quad WZ,T_{\perp};W\gamma,T_{\perp};Z\gamma,T_{\perp}; \gamma\gamma,T_{\perp}) & \\
&\quad F_{T,2}+F_{T,6}>0 \quad (\gamma\gamma,T_{\parallel}) & \\
\fbox{$F_{T,2},F_{T,7}$} && \nonumber\\
&\quad 16c_{w}^4F_{T,2}+(8s_{w}^4-8s_{w}^2+1)F_{T,7}>0 \quad (Z\gamma,T_{\parallel}) & \\
&\quad 2F_{T,2}+F_{T,7}>0 \quad (\gamma\gamma,T_{\perp}; \gamma\gamma,T_{\parallel}) & \\
\fbox{$F_{T,2},F_{T,8}$} && \nonumber\\
&\quad F_{T,2}>0 \quad (WW,T_{\perp};WW,T_{++};WW,T_{+-}; WW,T_{\parallel};ZZ,T_{\perp}; & \nonumber\\
&\quad \quad\quad\quad WZ,T_{\perp};WZ,T_{\parallel};W\gamma,T_{\perp}; W\gamma,T_{\parallel};Z\gamma,T_{\perp};\gamma\gamma,T_{\perp}) & \\
&\quad 2F_{T,2}+F_{T,8}>0 \quad (\gamma\gamma,T_{\parallel}) & \\
\fbox{$F_{T,2},F_{T,9}$} && \nonumber\\
&\quad F_{T,2}>0 \quad (WW,T_{\perp};WW,T_{++};WW,T_{+-}; WW,T_{\parallel}; & \nonumber\\
&\quad \quad\quad\quad WZ,T_{\perp};WZ,T_{\parallel};W\gamma,T_{\perp}; W\gamma,T_{\parallel}) & \\
&\quad 4F_{T,2}+F_{T,9}>0 \quad (\gamma\gamma,T_{\perp}; \gamma\gamma,T_{\parallel}) & \\
\fbox{$F_{T,5},F_{T,6}$} && \nonumber\\
&\quad -4c_{w}^2s_{w}^2F_{T,5}+(c_{w}^2-s_{w}^2)^2F_{T,6}>0 \quad (Z\gamma,T_{\parallel}) & \\
&\quad F_{T,5}+F_{T,6}>0 \quad (ZZ,T_{\parallel};\gamma\gamma,T_{\parallel}) & \\
\fbox{$F_{T,5},F_{T,7}$} && \nonumber\\
&\quad -16c_{w}^2s_{w}^2F_{T,5}+(8s_{w}^4-8s_{w}^2+1)F_{T,7}>0 \quad (Z\gamma,T_{\parallel}) & \\
&\quad 2F_{T,5}+F_{T,7}>0 \quad (ZZ,T_{\parallel}; \gamma\gamma,T_{\parallel}) & \\
\fbox{$F_{T,5},F_{T,8}$} && \nonumber\\
&\quad -2F_{T,5}+t_{w}^2F_{T,8}>0 \quad (Z\gamma,T_{\parallel}) & \\
&\quad 2F_{T,5}+t_{w}^4F_{T,8}>0  \quad (ZZ,T_{\parallel}) & \\
\fbox{$F_{T,5},F_{T,9}$} && \nonumber\\
&\quad -4F_{T,5}+t_{w}^2F_{T,9}>0 \quad (Z\gamma,T_{\parallel}) & \\
&\quad 4F_{T,5}+t_{w}^4F_{T,9}>0 \quad (ZZ,T_{\parallel}) & \\
\fbox{$F_{T,6},F_{T,7}$} && \nonumber\\
&\quad F_{T,7}>0 \quad (ZZ,T_{\perp};WZ,T_{\perp};W\gamma,T_{\perp};Z\gamma,T_{\perp};\gamma\gamma,T_{\perp}) & \\
&\quad 4(c_{w}^2-s_{w}^2)^2F_{T,6}+(8s_{w}^4-8s_{w}^2+1)F_{T,7}>0  \quad (Z\gamma,T_{\parallel}) & \\
\fbox{$F_{T,6},F_{T,8}$} && \nonumber\\
&\quad F_{T,6}>0 \quad (WZ,T_{\parallel}; W\gamma,T_{\parallel}) & \\
&\quad 2F_{T,6}+F_{T,8}>0  \quad (\gamma\gamma,T_{\parallel}) & \\
\fbox{$F_{T,6},F_{T,9}$} && \nonumber\\
&\quad F_{T,9}>0 \quad (ZZ,T_{\perp};Z\gamma,T_{\perp}; \gamma\gamma,T_{\perp}) & \\
&\quad F_{T,6}>0 \quad (WZ,T_{\parallel};W\gamma,T_{\parallel}) & \\
\fbox{$F_{T,7},F_{T,8}$} && \nonumber\\
&\quad F_{T,7}>0 \quad (ZZ,T_{\perp}; WZ,T_{\perp};WZ,T_{\parallel}; W\gamma,T_{\perp};W\gamma,T_{\parallel};
Z\gamma,T_{\perp};\gamma\gamma,T_{\perp}) & \\
&\quad (8s_{w}^4-8s_{w}^2+1)F_{T,7}+8s_{w}^4F_{T,8}>0  \quad (Z\gamma,T_{\parallel}) & \\
\fbox{$F_{T,7},F_{T,9}$} && \nonumber\\
&\quad F_{T,7}>0 \quad (WZ,T_{\perp}; WZ,T_{\parallel}; W\gamma,T_{\perp};W\gamma,T_{\parallel}) & \\
&\quad (8s_{w}^4-8s_{w}^2+1)F_{T,7}+4s_{w}^4F_{T,9}>0 \quad (Z\gamma,T_{\parallel}) & \\
\fbox{$F_{T,8},F_{T,9}$} && \nonumber\\
&\quad F_{T,9}>0 \quad (ZZ,T_{\perp};Z\gamma,T_{\perp};\gamma\gamma,T_{\perp}) & \\
&\quad 2F_{T,8}+F_{T,9}>0  \quad (ZZ,T_{\parallel};Z\gamma,T_{\parallel};\gamma\gamma,T_{\parallel})   . &
\end{flalign}
For $F_{M,i}$ operators,
\begin{flalign}
\fbox{$F_{M,0},F_{M,1}$} && \nonumber\\
&\quad 2F_{M,0}-F_{M,1}>0 \quad (ZZ,\overline{M}_{I}) & \\
&\quad -F_{M,0}>0 \quad (ZZ,M_{I}) & \\
\fbox{$F_{M,0},F_{M,7}$} && \nonumber\\
&\quad -F_{M,0}>0 \quad (WW,M_{+-},M_{\parallel};ZZ,M_{I}) & \\
&\quad 4F_{M,0}+F_{M,7}>0 \quad (ZZ,\overline{M}_{I}) & \\
\fbox{$F_{M,1},F_{M,2}$} && \nonumber\\
&\quad -F_{M,1}+t_{w}^4F_{M,2}>0  \quad (ZZ,\overline{M}_{I}) & \\
&\quad -F_{M,2}>0 \quad (ZZ,M_{I}) & \\
\fbox{$F_{M,1},F_{M,3}$} && \nonumber\\
&\quad -F_{M,1}>0 \quad (WW,TL,LT; WZ,TL;WW,M_{+-},M_{\parallel}; &\nonumber\\
&\quad\quad\quad\quad WW,\overline{M}_{+-}, \overline{M}_{\parallel}) & \\
&\quad -2F_{M,1}-F_{M,3}>0 \quad (W\gamma,LT;Z\gamma,LT) & \\
\fbox{$F_{M,1},F_{M,4}$} && \nonumber\\
&\quad -F_{M,1}+ t_{w}^2F_{M,4}>0 \quad (ZZ,\overline{M}_{I}) & \\
&\quad -F_{M,4}>0 \quad (ZZ,M_{I}) & \\
\fbox{$F_{M,1},F_{M,5}$} && \nonumber\\
&\quad -2F_{M,1}-F_{M,5}>0 \quad (Z\gamma,LT) & \\
&\quad -2F_{M,1}+F_{M,5}>0 \quad (W\gamma,LT) & \\
\fbox{$F_{M,1},F_{M,7}$} && \nonumber\\
&\quad -F_{M,1}>0 \quad (WW,M_{+-},M_{\parallel}) & \\
&\quad -F_{M,1}+F_{M,7}>0  \quad (WZ,\overline{M}'_{I}) & \\
\fbox{$F_{M,2},F_{M,3}$} && \nonumber\\
&\quad 2F_{M,2}-F_{M,3}>0 \quad (ZZ,\overline{M}_{I}) & \\
&\quad -F_{M,2}>0 \quad (ZZ,M_{I}) & \\
\fbox{$F_{M,2},F_{M,7}$} && \nonumber\\
&\quad 2t_{w}^4F_{M,2}+F_{M,7}>0 \quad (ZZ,\overline{M}_{I}) & \\
&\quad -F_{M,2}>0  \quad (ZZ,M_{I}) & \\
\fbox{$F_{M,3},F_{M,7}$} && \nonumber\\
&\quad F_{M,7}>0 \quad (WW,TL,LT; WZ,TL;WW,\overline{M}_{+-},\overline{M}_{\parallel}) & \\
&\quad -F_{M,3}>0  \quad (WZ,M'_{I},\overline{M}'_{I}) & \\
\fbox{$F_{M,4},F_{M,7}$} && \nonumber\\
&\quad 2t_{w}^2F_{M,4}+F_{M,7}>0 \quad (ZZ,\overline{M}_{I}) & \\
&\quad -F_{M,4}>0 \quad (ZZ,M_{I}) & \\
\fbox{$F_{M,5},F_{M,7}$} && \nonumber\\
&\quad F_{M,5}+F_{M,7}>0 \quad (W\gamma,LT) & \\
&\quad -F_{M,5}>0 \quad (WZ,M'_{I}) .  & \\
\nonumber
\end{flalign}

\section{Positivity bounds on three operators}
\label{subsec:A12}

Here we list all the three operator bounds considered in
Section~\ref{sec:benchmark}.

\begin{flalign}
\fbox{$F_{S,0},F_{S,1},F_{S,2}$} && \nonumber\\
&\quad F_{S,0}+F_{S,1}+F_{S,2}>0 \quad (ZZ,L) & \\
&\quad 2F_{S,0}+F_{S,1}+F_{S,2}>0 \quad (WW,L) & \\
&\quad F_{S,0}+F_{S,2}>0 \quad (WZ,L) & \\
\fbox{$F_{T,0},F_{T,1},F_{T,2}$} && \nonumber\\
&\quad 2F_{T,0}+F_{T,1}+F_{T,2}>0 \quad (WW,T_{+-}) & \\
&\quad F_{T,2}>0 \quad (WW,T_{\perp};ZZ,T_{\perp};WZ,T_{\perp};W\gamma,T_{\perp};Z\gamma,T_{\perp};\gamma\gamma,T_{\perp}) & \\
&\quad 4F_{T,1}+F_{T,2}>0 \quad (WZ,T_{\parallel}; W\gamma,T_{\parallel}) & \\
&\quad 2F_{T,0}+2F_{T,1}+F_{T,2}>0 \quad (ZZ,T_{\parallel};Z\gamma,T_{\parallel}; \gamma\gamma,T_{\parallel}) & \\
\fbox{$F_{T,5},F_{T,8},F_{T,9}$} && \nonumber\\
&\quad F_{T,9}>0 \quad (ZZ,T_{\perp};Z\gamma,T_{\perp}; \gamma\gamma,T_{\perp}) & \\
&\quad -4F_{T,5}+2t_{w}^2F_{T,8}+t_{w}^2F_{T,9}>0 \quad (Z\gamma,T_{\parallel}) & \\
&\quad 4F_{T,5}+2t_{w}^4F_{T,8}+t_{w}^4F_{T,9}>0 \quad (ZZ,T_{\parallel}) & \\
\fbox{$F_{M,0},F_{M,1},F_{M,2}$} && \nonumber\\
&\quad 4F_{M,0}-3F_{M,1}>0 \quad  (WW,\overline{M}_{+-},\overline{M}_{\parallel}) & \\
&\quad -4F_{M,0}-F_{M,1}>0 \quad  (WW,M_{+-},M_{\parallel}) & \\
&\quad 2F_{M,0}-F_{M,1}+t_{w}^4F_{M,2}>0 \quad  (ZZ,\overline{M}_{I}) & \\
&\quad -2F_{M,0}-t_{w}^4F_{M,2}>0 \quad  (ZZ,M_{I}) & \\
\fbox{$F_{M,0},F_{M,1},F_{M,3}$} && \nonumber\\
&\quad -2F_{M,1}-F_{M,3}>0 \quad  (W\gamma,LT;Z\gamma,LT) & \\
&\quad 4F_{M,0}-3F_{M,1}>0 \quad  (WW,\overline{M}_{+-},\overline{M}_{\parallel}) & \\
&\quad 4F_{M,0}-2F_{M,1}-t_{w}^4F_{M,3}>0 \quad (ZZ,\overline{M}_{I}) & \\
&\quad -F_{M,0}>0 \quad (ZZ,M_{I}) & \\
\fbox{$F_{M,3},F_{M,5},F_{M,7}$} && \nonumber\\
&\quad F_{M,7}>0 \quad (WW,TL,LT;WZ,TL;WW,\overline{M}_{+-},\overline{M}_{\parallel}) & \\
&\quad -F_{M,3}+F_{M,5}+F_{M,7}>0 \quad (W\gamma,LT) & \\
&\quad -t_{w}^4F_{M,3}+t_{w}^2F_{M,5}+F_{M,7}>0 \quad (ZZ,TL,LT;ZZ,\overline{M}_{I}) & \\
&\quad 4F_{M,7}(c_{w}^4F_{M,7}-s_{w}^2c_{w}^2F_{M,5}-s_{w}^4F_{M,3})>(2c_{w}^2F_{M,7}+s_{w}^2F_{M,5})^2 &\nonumber\\
&\quad \quad\quad\quad (WZ,M'_{I},\overline{M}'_{I}) & \\
\fbox{$F_{M,4},F_{M,5},F_{M,7}$} && \nonumber\\
&\quad -F_{M,5}+F_{M,7}>0 \quad (Z\gamma,LT) & \\
&\quad F_{M,5}+F_{M,7}>0 \quad (W\gamma,LT) & \\
&\quad 2t_{w}^2F_{M,4}+t_{w}^2F_{M,5}+F_{M,7}>0 \quad (ZZ,\overline{M}_{I}) & \\
&\quad -F_{M,4}>0 \quad (ZZ,M_{I}) & \\
&\quad 4F_{M,7}(F_{M,7}-t_{w}^2F_{M,5})>(4t_{w}^2F_{M,4}+t_{w}^2F_{M,5}+2F_{M,7})^2 & \nonumber\\
&\quad\quad\quad\quad (WZ,M'_{I}, \overline{M}'_{I}) & \\
\fbox{$F_{M,1},F_{M,3},F_{M,4}$} && \nonumber\\
&\quad -2F_{M,1}-F_{M,3}>0 \quad (W\gamma,LT;Z\gamma,LT) & \\
&\quad -F_{M,4}>0 \quad (ZZ,M_{I}) & \\
&\quad F_{M,1}(2F_{M,1}+t_{w}^4F_{M,3})>2t_{w}^4F_{M,4}^2 \quad (WZ,M'_{I},\overline{M}'_{I})
& \\
\fbox{$F_{S,0},F_{T,0},F_{M,7}$} && \nonumber\\
&\quad F_{S,0}>0 \quad (WW,L;ZZ,L;WZ,L) & \\
&\quad F_{T,0}>0 \quad (WW,T_{+-};WW,T_{\parallel};
ZZ,T_{\parallel};Z\gamma,T_{\parallel};
\gamma\gamma,T_{\parallel}) & \\
&\quad F_{M,7}>0 \quad (ZZ,TL,LT;WZ,LT;WW,TL, LT;WZ,TL; & \nonumber\\
&\quad \quad W\gamma,LT;Z\gamma,LT;WZ,\overline{M}'_{I}) & \\
\fbox{$F_{S,0},F_{T,0},F_{M,0}$} && \nonumber\\
&\quad F_{S,0}>0 \quad (WW,L;ZZ,L;WZ,L) & \\
&\quad F_{T,0}>0 \quad (WW,T_{+-};WW,T_{\parallel};
ZZ,T_{\parallel};Z\gamma,T_{\parallel};
\gamma\gamma,T_{\parallel}) & \\
&\quad 4F_{S,0}F_{T,0}>F_{M,0}^2 \quad (ZZ,M_{I},\overline{M}_{I})   . &
\end{flalign}

\bibliography{bib.bib}
\end{document}